\documentclass[aps, prd, showpacs, superscriptaddress, preprintnumbers, nofootinbib, twocolumn]{revtex4-1}
\usepackage[pdftex]{graphicx, color}
\usepackage{amsmath, amssymb, amscd, latexsym, bm, braket}
\usepackage[colorlinks=true,pdfstartview=FitV,linkcolor=blue,citecolor=magenta,urlcolor=blue,bookmarks=true,bookmarksnumbered=true]{hyperref}

\newcommand{\Slash}[1]{{\ooalign{\hfil/\hfil\crcr$#1$}}}

\begin{document}
\title{Spin-dependent dynamically assisted Schwinger mechanism}

\author{Xu-Guang Huang}
\address{Department of Physics and Center for Field Theory and Particle Physics, Fudan University, Shanghai, 200433, China }
\address{Key Laboratory of Nuclear Physics and Ion-beam Application (MOE), Fudan University, Shanghai 200433, China }

\author{Hidetoshi Taya}
\email{{\tt h_taya@fudan.edu.cn}}
\address{Department of Physics and Center for Field Theory and Particle Physics, Fudan University, Shanghai, 200433, China }

\date{\today}

\begin{abstract}
We study electron and positron pair production from the vacuum by a strong slow electric field superimposed by a weak fast electric field pointing in an arbitrary direction as a perturbation (the dynamically assisted Schwinger mechanism).  An analytical formula for the production number is derived on the basis of the perturbation theory in the Furry picture.  The formula is found to be in good agreement with non-perturbative results obtained by numerically solving the Dirac equation if the perturbation is sufficiently weak and/or is not very slow.  We also find analytically/numerically that the Schwinger mechanism becomes spin-dependent if the perturbation has a transverse component with respect to the strong electric field.  The number difference between spin up and down particles is strongly suppressed by an exponential of the critical field strength if the frequency of the perturbation is small, while it is only weakly suppressed by powers of the critical field strength if the frequency is large enough.  We also find that the spin-imbalance exhibits non-trivial oscillating behaviors in terms of the frequency of the perturbation, the azimuthal angle, and the momentum of produced particles.
\end{abstract}


\maketitle

\section{Introduction}

Spontaneous electron and positron pair production from the vacuum in the presence of a  classical electric field (the Schwinger mechanism) is one of the most remarkable predictions of quantum electrodynamics (QED) \cite{sau31, hei36, sch51}.  Intuitively, the physical origin of the Schwinger mechanism is essentially the same as the electrical breakdown of semi-conductors (or the Landau-Zener transition \cite{lan32, zen32, stu32, maj32}).  That is, the energy band of electrons is tilted by the electric field, so that electrons filling the Dirac sea can tunnel into the positive energy band leaving holes in the Dirac sea (i.e., positrons).  The tunneling rate should be suppressed exponentially by the tunneling length, i.e., the gap energy $\sim 2m$.  Therefore, one expects that the Schwinger mechanism can be manifest only when the typical energy scale of the electric field becomes larger than the gap energy.  Indeed, the production number for a constant and homogeneous electric field was explicitly computed by Schwinger in his seminal work in 1951 \cite{sch51} as
\begin{align}
	n^{(\mp)}_{{\bm p},s} = \frac{V}{(2\pi)^3} {\rm exp}\left[ - \pi \frac{m^2 + {\bm p}_{\perp}^2}{eE} \right] ,  \label{eq1}
\end{align}
where $n^{(-/+)}$ is for electrons/positrons.  This formula implies that one has to prepare an extremely strong electric field of the order of $e E_{\rm cr} \equiv m^2 \sim \sqrt{10^{28}\;{\rm W/cm^2}}$ to test the Schwinger mechanism in laboratory experiments.  Unfortunately, it seems to be impossible to achieve this with current experimental technologies.  Indeed, HERCULES is currently the strongest laser, whose strength is $eE \sim \sqrt{10^{22}\;{\rm W/cm^2}}$ \cite{yan08}.  There are many intense laser facilities planned around the world (e.g., ELI \cite{eli} and HiPER \cite{hiper}), which are expected to reach $eE \sim \sqrt{ 10^{24}\;{\rm W/cm^2}}$ but are still weaker than the critical field strength $E_{\rm cr}$ by several orders of magnitude.

Recently, there has been an increasing interest in how to enhance the production number of the Schwinger mechanism.  One of the most reasonable ideas is to superimpose a weak fast electromagnetic field onto the original electric field (the dynamically assisted Schwinger mechanism) \cite{sch08,dun09,piz09,mon10a,mon10b}.  This idea can be understood as an analog of the Franz-Keldysh effect in semi-conductor physics \cite{fra58, kel58, tah63, cal63, tay19}.  An intuitive explanation of the idea is the following.  First, the weak fast field perturbatively interacts with electrons in the Dirac sea to kick them up into the gap.  Then, the tunneling length for the Schwinger mechanism is reduced by the kick, so that the critical field strength $E_{\rm cr}$ is effectively reduced as the frequency of the weak fast field is increased.  Notice that the perturbative kick is the essence of the dynamically assisted Schwinger mechanism.  This point was clarified in Refs.~\cite{gre17, gre19, tay19} by explicitly employing a perturbation theory in terms of the weak field in the Furry picture \cite{fur51, fra81, fra91, gre17, gre19, tay19}.  In particular, it was shown that the production process is dominated by the perturbative kick for a weak field with a frequency above the gap energy.  In that limit, the production number becomes proportional to powers of $eE/m^2$, which is free from the strong exponential suppression.

In this paper, we discuss a novel effect due to the superposition of a weak fast electric field in the Schwinger mechanism.  Namely, we show that the Schwinger mechanism becomes spin-dependent if the superimposed electric field is transverse with respect to the original strong electric field.  Notice that the original Schwinger mechanism (\ref{eq1}) is clearly spin-independent because the quantum tunneling is insensitive to spin.  The superimposition of an additional transverse electric field is essential for the spin-dependence. Intuitively, this is because a particle with transverse momentum feels a magnetic field in the longitudinal direction in its rest frame if the superimposed electric field in the observer frame has a transverse component with respect to the momentum direction.  Therefore, the mass gap is effectively reduced/enhanced through the spin-magnetic coupling depending on the charge and the momentum direction, which results in the spin-dependent Schwinger mechanism.  Note that a similar spin-dependent particle production mechanism from the vacuum was recently studied in Ref.~\cite{koh18}, in which the electron-positron pair production by a rotating electric field in the multi-photon production regime was considered.

This paper is organized as follows: In Sec.~\ref{sec2}, we analytically study the aforementioned spin-dependent Schwinger mechanism with superposition of a weak fast field based on a perturbation theory in the Furry picture \cite{fur51, fra81, fra91, gre17, gre19, tay19}.  In Sec.~\ref{sec3}, we numerically solve the Dirac equation to discuss the spin-imbalance caused by the spin-dependent Schwinger mechanism without relying on any approximations.  The numerical results are compared with the analytical formula obtained in Sec.~\ref{sec2}, and we find an excellent agreement between them.  Section~\ref{sec4} is devoted to summary and discussion.  Details of the analytical calculations in Sec.~\ref{sec2} are presented in Appendix~\ref{appA}.  More numerical results on the momentum distribution of produced electrons are presented in Appendix~\ref{appB}.

\section{Analytical discussion based on perturbation theory in the Furry picture} \label{sec2}

In this section, we analytically show that the Schwinger mechanism becomes spin-dependent if one superimposes a weak transverse electric field.  We first explain our physical setup in Sec.~\ref{sec2a}.  Next, in Sec.~\ref{sec2b}, we derive a formula for the momentum distribution of electrons and positrons produced by the Schwinger mechanism with a weak fast field pointing in an arbitrary direction by employing a perturbation theory in the Furry picture \cite{fur51, fra81, fra91, gre17, gre19, tay19}.  By using this formula, we compute the difference between the production number of spin up and down particles in Sec.~\ref{sec2c}.  We show that the difference becomes non-vanishing if the weak field is transverse, and discuss its qualitative features.

\subsection{Setup} \label{sec2a}

We consider QED in the presence of an external classical field $A_{\mu}$, whose Lagrangian is given by
\begin{align}
	{\mathcal L} = \bar{\psi} \left[ i \Slash{\partial} - e \Slash{A} - m \right] \psi.  \label{eq2}
\end{align}
In this work, we assume that the external field is purely electric and homogeneous in space.  We also adopt the temporal gauge $A_0 = 0$, so that $A_{\mu}$ depends only on time $A_{\mu}(x) = A_{\mu}(x^0)$ in the following.

We consider a situation in which $A_{\mu}$ can be separated into two parts, i.e., a strong slow field $\bar{A}_{\mu}$ and a weak fast perturbation on top of it ${\mathcal A}_{\mu}$ as
\begin{align}
	A_{\mu} = \bar{A}_{\mu} + {\mathcal A}_{\mu}. \label{eq3}
\end{align}
We also assume that the strong field $\bar{A}_{\mu}$ is sufficiently slow, so that it can be approximated by a constant electric field as
\begin{align}
	\bar{A}_{\mu} = (0, 0, 0, \bar{E}x^0), \label{eq5}
\end{align}
where $e\bar{E}>0$ and we defined the $x^3$-axis by the direction of the strong electric field.  Furthermore, in this work we focus on the simplest perturbation, i.e., a monochromatic electric field with frequency $\Omega>0$,
\begin{align}
	{\mathcal A}_{\mu} = \sin (\Omega t + \phi) \times (0,{\mathcal E}_1,{\mathcal E}_2,{\mathcal E}_3)/\Omega, \label{eq6}
\end{align}
where $\phi$ is an arbitrary phase factor.  It is straightforward to generalize the following discussions to other types of perturbations.  Note that the Keldysh parameter $\gamma_{\rm K}$ \cite{bre70,pop72,kel65,tay14} for the weak field is given by $\gamma_{\rm K} \equiv m\Omega/e|{\bm {\mathcal E}}|$.  Therefore, the effect of the weak field remains ``perturbative'' and the perturbation theory in the Furry picture is, roughly speaking\footnote{More specifically, in addition to the Keldysh parameter $\gamma_{\rm K}$, there exists one more dimensionless parameter $\nu \equiv e|{\bm {\mathcal E}}|/\omega^2$ that controls the interplay between the perturbative and non-perturbative particle production.  This is simply because the system has three dimensionful quantities $e|{\bm {\mathcal E}}|, \Omega, m$, from which one can construct two dimensionless parameters to characterize the system \cite{tay14, gel16}.  Even if one includes the new dimensionless parameter $\nu$, it is true that the perturbation theory in the Furry picture is valid for sufficiently weak $|{\bm {\mathcal E}}|$ and/or not very small $\Omega$.  }, valid for $\gamma_{\rm K} \gtrsim 1$, i.e., the perturbation should be sufficiently weak $e|{\bm {\mathcal E}}| \lesssim m\Omega$, or equivalently, the frequency should not be so small $\Omega \gtrsim e|{\bm {\mathcal E}}|/m$.  Otherwise, if $\gamma_{\rm K} \lesssim 1$, higher order corrections due to multiple scatterings by the weak field contribute to the production process non-perturbatively, which is difficult to discuss within the perturbation theory in the Furry picture.  We also note that the dynamically assisted Schwinger mechanism with a similar but different weak field configuration (i.e., on-shell dynamical photon) ${\mathcal A}_{\mu} \propto \sin (\Omega (t- {\bm n} \cdot {\bm x}) + \phi)$ with ${\bm n}$ being a unit vector $|{\bm n}| = 1$ was previously discussed, for example, in Refs.~\cite{dun09, mon10a, mon10b} with semi-classical method and Ref.~\cite{gre18} with a perturbative approach within the WKB approximation.

\subsection{Spin-dependent momentum distribution} \label{sec2b}

We analytically compute the momentum distribution of electrons $n^{(-)}$ and positrons $n^{(+)}$ produced from the vacuum in the presence of the external field $A_{\mu}$ (\ref{eq3}).  To this end, we employ a perturbation theory in the Furry picture \cite{fur51, fra81, fra91, gre17, gre19, tay19}, in which the interaction with the strong field $\bar{A}_{\mu}$ is treated non-perturbatively and that with the weak perturbation ${\mathcal A}_{\mu}$ is treated perturbatively.

In the lowest order in the perturbation ${\mathcal A}_{\mu}$, the (canonical) momentum distribution per spin $s$, $n^{(\mp)}_{{\bm p},s} \equiv d^3N^{(\mp)}_s/d{\bm p}^3$, was already given in Ref.~\cite{tay19} as
\begin{align}
	n^{(\mp)}_{\pm{\bm p},s}
		&= \sum_{s'} \int d^3{\bm p}'
		   \left|
			\!\!\!\!\!\!\!
			\parbox{5em}{\includegraphics[width=5em]{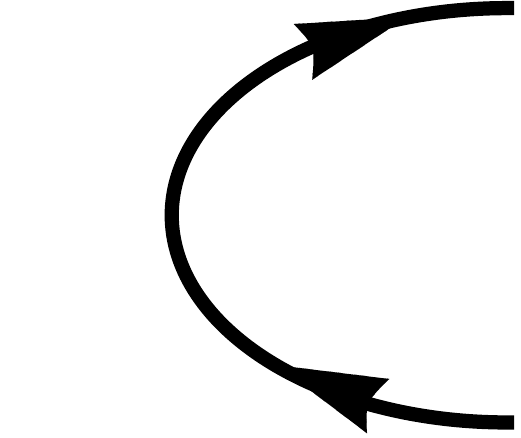}} \
			+ \
			\parbox{5em}{\includegraphics[width=5em]{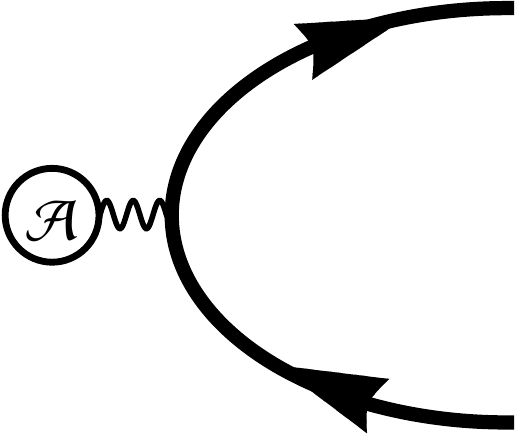}} \
		   \right|^2 \nonumber\\
		&= \sum_{s'} \int d^3{\bm p}' \left| \int d^3{\bm x} {}_{\pm} \psi^{{\rm out}\dagger}_{{\bm p},s}(x) {}_{\mp} \psi^{\rm in}_{{\bm p}',s'}(x) \right. \nonumber\\
		&\quad\quad\quad \left. - ie \int d^4x {}_{\pm} \bar{\psi}^{{\rm out}}_{{\bm p},s}(x) \Slash{\mathcal A} (x) {}_{\mp} \psi^{{\rm in}}_{{\bm p}',s'}(x) \right|^2, \label{eq7}
\end{align}
where the thick line represents the electron propagator fully dressed by the strong field $\bar{A}_{\mu}$ as
\begin{align}
	\parbox{4.5em}{\includegraphics[width=4.5em]{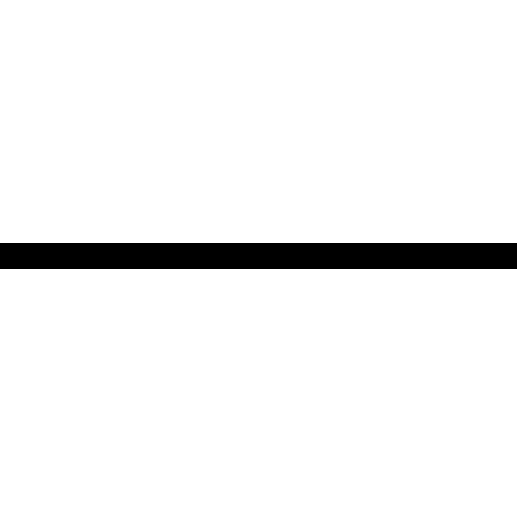}}\
	&=\
	\parbox{4.5em}{\includegraphics[width=4.5em]{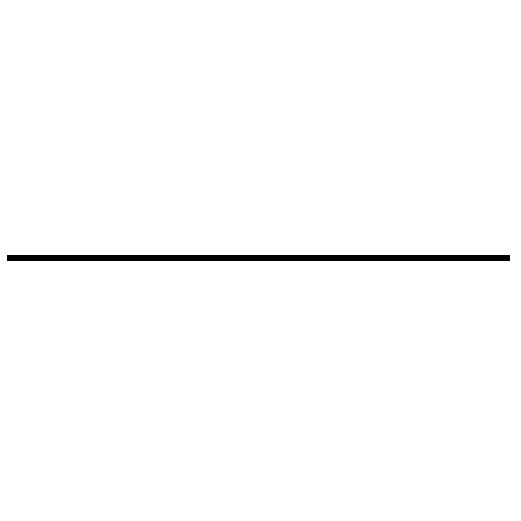}}\
	+\
	\parbox{4.5em}{\includegraphics[width=4.5em]{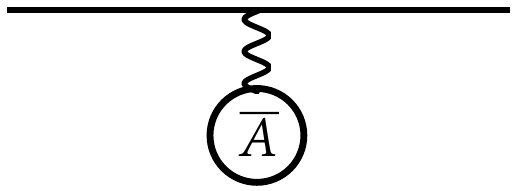}}\nonumber\\
	&\quad +\
	\parbox{4.5em}{\includegraphics[width=4.5em]{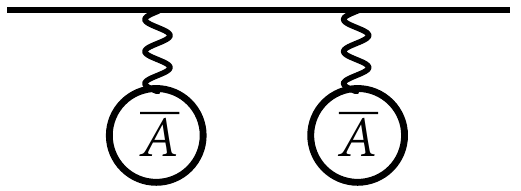}}\
	+\  \cdots. \label{eq_7}
\end{align}
Because of the dressed propagator, the positive/negative frequency mode functions defined at the asymptotic times $|x^0| \to \infty$, ${}_{\pm}\psi_{{\bm p},s}^{\rm as}$ (${\rm as}={\rm in}$ and ${\rm out}$ for $x^0 \to -\infty$ and $+\infty$, respectively), are also dressed by the strong field $\bar{A}_{\mu}$.  Namely, ${}_{\pm}\psi_{{\bm p},s}^{\rm as}$ is defined as a solution of the Dirac equation under the strong field $\bar{A}_{\mu}$,
\begin{align}
	0 = \left[ i \Slash{\partial} - e\bar{\Slash{A}} - m\right] {}_{\pm}\psi_{{\bm p},s}^{\rm as},
\end{align}
with boundary conditions
\begin{subequations}
\begin{align}
	\lim_{x^0 \to -\infty} {}_{\pm}\psi_{{\bm p},s}^{\rm in} &\propto {\rm e}^{i {\bm p}\cdot {\bm x}} {\rm e}^{ \mp i\omega_{\bm P} x^0}, \\
	\lim_{x^0 \to +\infty} {}_{\pm}\psi_{{\bm p},s}^{\rm out} &\propto {\rm e}^{i {\bm p}\cdot {\bm x}} {\rm e}^{ \mp i\omega_{\bm P} x^0},
\end{align}
\end{subequations}
(i.e., plane waves at the asymptotic times $x^0 \to \pm \infty$) and the normalization
\begin{subequations}
\begin{align}
	\int d^3{\bm x} {}_{\pm}\psi_{{\bm p},s}^{{\rm as}\dagger}(x)  {}_{\pm}\psi_{{\bm p}',s'}^{\rm as}(x) &= \delta_{ss'} \delta^3({\bm p}-{\bm p}'), \\
	\int d^3{\bm x} {}_{\pm}\psi_{{\bm p},s}^{{\rm as}\dagger}(x)  {}_{\mp}\psi_{{\bm p}',s'}^{\rm as}(x) &= 0.
\end{align}
\end{subequations}
We note that, in the presence of the strong field $\bar{A}_{\mu}$, one has to take care of the distinction between the in- and out-state mode functions ${}_{\pm}\psi_{{\bm p},s}^{\rm in/out}$.  This is because the non-perturbative interaction due to $\bar{A}_{\mu}$ mixes up the positive and negative frequency modes (i.e., particle and anti-particle modes) during the time-evolution, and thus ${}_{\pm}\psi_{{\bm p},s}^{\rm in/out}$ are no longer the same.

One can carry out the integrations in Eq.~(\ref{eq7}) for the present field configuration (i.e., the constant electric field (\ref{eq5}) superimposed by the monochromatic wave (\ref{eq6})).  Note that this calculation can be done exactly without any use of approximations such as the WKB method (e.g. Refs.~\cite{gre17, gre18, gre19}).  The detailed calculations can be found in Appendix~\ref{appA}.  We find
\begin{widetext}
\begin{align}
	n^{(\mp)}_{\pm{\bm p},s}
		&=	\frac{V}{(2\pi)^3}  \exp\left[- \pi \frac{ m^2 + p_{\perp}^2 }{ e\bar{E} } \right] \nonumber\\
			&\quad\times
			\left[
					\left|
							1
							+ i\pi {\rm e}^{i \left( \phi- \frac{\Omega p_3}{e\bar{E}} \right) }
								\Biggl\{
										\frac{p_{\perp}}{\Omega} \frac{{\mathcal E}_{\perp} }{\bar{E}} \cos(\theta_{\bm p} - \theta_{\bm {\mathcal E}}) {\rm Re}\left[ {\rm e}^{-i\frac{\Omega^2}{4e\bar{E}}} {}_1\tilde{F}_1\left( 1 - i \frac{ m^2 + p_{\perp}^2 }{ 2e\bar{E} }; 1; i \frac{\Omega^2}{2e\bar{E}} \right)  \right] \right. \right. \nonumber\\
										&\ \quad\quad\quad\quad\quad\quad\quad\quad\quad\quad\quad\quad + \sigma \frac{ p_{\perp}}{\Omega}\frac{ {\mathcal E}_{\perp} }{\bar{E}} \sin(\theta_{\bm p} - \theta_{\bm {\mathcal E}}) {\rm Im}\left[ {\rm e}^{-i\frac{\Omega^2}{4e\bar{E}}} {}_1\tilde{F}_1\left( 1 - i\frac{ m^2 + p_{\perp}^2 }{ 2e\bar{E} }; 1; i \frac{\Omega^2}{2e\bar{E}} \right)  \right] \nonumber\\
										&\ \quad\quad\quad\quad\quad\quad\quad\quad\quad\quad\quad\quad - i \frac{{\mathcal E}_3}{\bar{E}}  \frac{ m^2 + p_{\perp}^2 }{ 2e\bar{E} } {\rm Re}\left[ {\rm e}^{-i\frac{\Omega^2}{4e\bar{E}}} {}_1\tilde{F}_1\left( 1 - i\frac{ m^2 + p_{\perp}^2 }{ 2e\bar{E} }; 2; i \frac{\Omega^2}{2e\bar{E}} \right)  \right] \left.
								\Biggl\}
					\right|^2 \nonumber\\
					&\quad\quad\  + \left. \pi^2 \frac{m^2}{\Omega^2} \frac{{\mathcal E}_{\perp}^2}{\bar{E}^2} \left|  {\rm Im}\left[ {\rm e}^{-i\frac{\Omega^2}{4e\bar{E}}} {}_1\tilde{F}_1\left( 1 - i \frac{ m^2 + p_{\perp}^2 }{ 2e\bar{E} }; 1; i \frac{\Omega^2}{2e\bar{E}} \right)  \right] \right|^2
			\right], \label{eq11}
\end{align}
\end{widetext}
where $\sigma \equiv +1\  (-1)$ for spin up $s=\uparrow$ (down $s=\downarrow$) with respect to the $x^3$-axis and ${}_1\tilde{F}_1(a;b;z) \equiv {}_1 F_1(a;b;z)/\Gamma(b)$ is the regularized hypergeometric function.  We also introduced $p_{\perp}, {\mathcal E}_{\perp}, \theta_{\bm p},\theta_{\bm {\mathcal E}}$ as
\begin{align}
	&p_1 \equiv p_{\perp} \cos \theta_{\bm p},\ p_2 \equiv p_{\perp} \sin \theta_{\bm p},\nonumber\\
	&{\mathcal E}_1 \equiv {\mathcal E}_{\perp} \cos \theta_{\bm {\mathcal E}},\ {\mathcal E}_2 \equiv {\mathcal E}_{\perp} \sin \theta_{\bm {\mathcal E}}.
\end{align}

We emphasize here that the momentum distribution (\ref{eq11}) explicitly depends on spin $s$ (the second term in the curly brackets).  That is, the Schwinger mechanism becomes spin-dependent if one superimposes weak perturbations on top of a strong field.  Notice that the Schwinger mechanism without perturbations or the purely perturbative particle production from the monochromatic wave alone\footnote{In general, the perturbative particle production can be spin-dependent for general field configurations such as a rotating field \cite{koh18}.  What we would like to emphasize here is that even if the perturbation alone is insensitive to spin, the mutual assistance between the strong field and the perturbation results in the spin-dependent particle production.  } is insensitive to spin; the spin-dependent term only appears when both exist.  To see this explicitly, let us consider some limits of the momentum distribution (\ref{eq11}).  In the absence of the perturbation $e|{\bm {\mathcal E}}| \to 0$, only the first term in Eq.~(\ref{eq7}) survives and the momentum distribution (\ref{eq11}) can be simplified as
\begin{align}
	n^{(\mp)}_{\pm {\bm p},s}
	\xrightarrow{e|{\bm {\mathcal E}}| \to 0}{}
	\frac{V}{(2\pi)^3}  \exp\left[ -\pi\frac{m^2+p_{\perp}^2}{e\bar{E}}  \right] .
\end{align}
This is nothing but the Schwinger formula for a constant electric field (\ref{eq1}) \cite{sch51}, and the production is independent of spin $s$.  On the other hand, in the absence of the strong field $e\bar{E} \to 0$, the particle production process becomes purely perturbative and the perturbation theory in the Furry picture is trivially reduced to the standard perturbation theory without strong fields \cite{itz80, tay14}.  In this limit, only the second term in Eq.~(\ref{eq7}) survives and the momentum distribution (\ref{eq11}) becomes
\begin{align}
	n^{(\mp)}_{\pm {\bm p},s}
	\xrightarrow{e\bar{E}\to 0}{} &VT \frac{e^2}{16 \pi^2} \delta\left(\Omega - 2\omega_{\bm p} \right) \nonumber\\
	&\times	\Biggl[ \Biggl|
					\cos(\theta_{\bm p}-\theta_{\bm {\mathcal E}})\frac{{\mathcal E}_{\perp}}{\Omega}\frac{ p_{\perp} }{\sqrt{m^2 + p_{\perp}^2}}  \frac{p_3}{\omega_{\bm p}} \nonumber\\
					&\quad \quad - \frac{{\mathcal E}_3}{\Omega}  \frac{\sqrt{m^2 + p_{\perp}^2}}{\omega_{\bm p}}  \Biggl|^2 + \frac{ m^2 }{ m^2 + p_{\perp}^2 } \frac{{\mathcal E}_{\perp}^2}{\Omega^2} \nonumber\\
	&\quad\  + \Biggl|\sin(\theta_{\bm p}-\theta_{\bm {\mathcal E}}) \frac{{\mathcal E}_{\perp}}{\Omega}\frac{ p_{\perp} }{\sqrt{m^2 + p_{\perp}^2}} \Biggl|^2
					 \Biggl] , \label{eq14}
\end{align}
where $\omega_{\bm p} \equiv \sqrt{m^2+{\bm p}^2}$ is the on-shell energy and $T$ is the whole time interval.  Again, the production is independent of spin $s$.  Therefore, we conclude that the spin-dependence arises because of the mutual assistance between the strong field and the weak perturbation.

The spin-dependent term is proportional to ${\bf e}_{x^3}\cdot ({\bm p} \times {\bm {\mathcal E}})$, i.e., the spin-dependence appears only if the emission direction of the particle has a component perpendicular to both the strong and weak electric fields.  Intuitively, this is because a particle with transverse momentum effectively feels a magnetic field in the longitudinal direction in its rest frame if there is an electric field in the transverse direction in the observer frame.

The formula (\ref{eq11}) is divergent at $\Omega \to 0$ if the perturbation has transverse component $e{\mathcal E}_{\perp} \neq 0$.  This is not a physical behavior as we shall explicitly show in Sec.~\ref{sec3} and Appendix~\ref{appB} by comparing exact results obtained by numerically solving the Dirac equation.  This unphysical behavior originates from multiple scattering effects, which cannot be captured by the lowest order diagram in Eq.~(\ref{eq7}).  As explained below Eq.~(\ref{eq6}), the multiple scattering effects become important for small $\gamma_{\rm K} \lesssim 1$.  Hence, the formula (\ref{eq11}) becomes invalid for very small $\Omega \lesssim e|{\bm {\mathcal E}}|/m$.  Notice that we are interested in a situation such that the total field can be separated into a fast field ${\mathcal A}_{\mu}$ and a slow field $\bar{A}_{\mu}$ (see Eq.~(\ref{eq3})).  For very small $\Omega$, however, one cannot manifestly distinguish the fast field ${\mathcal A}_{\mu}$ from the slow field $\bar{A}_{\mu}$, so that the separation becomes ambiguous.  Therefore, we do not discuss very small $\Omega$ cases in detail below.  The pair production for very small $\Omega$ is rather trivial because the effect of the transverse weak field is just to change the direction of the total field, so that the production number becomes consistent with the naive Schwinger formula (\ref{eq1}) and no spin-dependence appears.

The $\phi$- and $p_3$-dependences of the distribution $n^{(\mp)}_{{\bm p},s}$ always appear with the combination $\phi \mp \Omega p_3/e\bar{E}$.  Intuitively, this is because the particle production becomes the most efficient at the instant when the longitudinal kinetic momentum $P_{3} = p_{3} \pm e\bar{E} x^0$ ($+, -$ for an electron and a positron, respectively) is vanishing, at which the energy cost of the production becomes the smallest \cite{nik70, tay19}.  Therefore, the value of the weak field at $x^0 = \mp p_{3}/e\bar{E}$ becomes important, at which the phase of the weak field (\ref{eq6}) reads $\phi \mp \Omega p_{3}/e\bar{E} $.

\subsection{Spin-imbalance} \label{sec2c}

We quantify the spin-imbalance by the difference between the production number of spin up and down particles as
\begin{align}
	\Delta n^{(\mp)}_{{\bm p}}
	\equiv n^{(\mp)}_{{\bm p},{\uparrow}}-n^{(\mp)}_{{\bm p},{\downarrow}}. \label{eq15}
\end{align}
From the momentum distribution (\ref{eq11}), the spin-imbalance $\Delta n^{(\mp)}_{{\bm p}}$ can be evaluated as
\begin{widetext}
\begin{align}
	\Delta n^{(\mp)}_{{\bm p}}
	&= \frac{V}{2\pi^2} \exp\left[ - \pi\frac{m^2+p_{\perp}^2}{e\bar{E}}  \right] \frac{ p_{\perp}}{\Omega}\frac{{\mathcal E}_{\perp}}{\bar{E}} {\rm Im}\left[ {\rm e}^{-i\frac{\Omega^2}{4e\bar{E}}} {}_1\tilde{F}_1\left( 1 - i \frac{ m^2 + p_{\perp}^2 }{ 2e\bar{E} }; 1; i \frac{\Omega^2}{2e\bar{E}} \right)  \right] \nonumber\\
	&\quad \times \Biggl\{ \mp \sin\left( \phi \mp \frac{\Omega p_3}{e\bar{E}} \right) \sin(\theta_{\bm p}-\theta_{\bm {\mathcal E}})  + \frac{\pi}{2} \frac{ p_{\perp}}{\Omega}\frac{{\mathcal E}_{\perp}}{\bar{E}}  {\rm Re}\left[ {\rm e}^{-i\frac{\Omega^2}{4e\bar{E}}} {}_1\tilde{F}_1\left( 1 - i \frac{ m^2 + p_{\perp}^2 }{ 2e\bar{E} }; 1; i \frac{\Omega^2}{2e\bar{E}} \right)  \right]   \sin\left( 2(\theta_{\bm p}-\theta_{\bm {\mathcal E}}) \right)  \Biggl\}.  \label{eq16}
\end{align}
The first term in the curly brackets appears because of the interference between the two diagrams in Eq.~(\ref{eq7}), and the second term comes solely from the second diagram.

The spin-imbalance $\Delta n^{(\mp)}_{{\bm p}}$ shows non-trivial dependence in $\Omega$.  For small $\Omega \lesssim \sqrt{e\bar{E}}, \sqrt{m^2+p_{\perp}^2}$, the magnitude of $\Delta n^{(\mp)}_{{\bm p}}$ increases monotonically with increasing $\Omega$ as
\begin{align}
	\Delta n^{(\mp)}_{{\bm p}}
	\xrightarrow{\Omega \to 0}{}& \frac{V}{2\pi^2} \exp\left[ - \pi\frac{m^2+p_{\perp}^2}{e\bar{E}}  \right] \frac{p_{\perp}}{\Omega}\frac{{\mathcal E}_{\perp}}{\bar{E}} \frac{\Omega^2}{4e\bar{E}}  \Biggl\{ \mp \sin\left( \phi \mp \frac{\Omega p_3}{e\bar{E}} \right) \sin(\theta_{\bm p}-\theta_{\bm {\mathcal E}})  + \frac{\pi}{2} \frac{ p_{\perp}}{\Omega}\frac{{\mathcal E}_{\perp}}{\bar{E}}     \sin\left( 2(\theta_{\bm p}-\theta_{\bm {\mathcal E}}) \right)  \Biggl\}.    \label{eq17}
\end{align}
On the other hand, for large $\Omega \gtrsim \sqrt{e\bar{E}}, \sqrt{m^2+p_{\perp}^2}$, the magnitude of $\Delta n^{(\mp)}_{{\bm p}}$ decreases with increasing $\Omega$ and exhibits a rapidly oscillating behavior in $\Omega$ as
\begin{align}
	\Delta n^{(\mp)}_{{\bm p}}
	\xrightarrow{\Omega \to \infty}{}
	& \frac{1}{\pi^3} \exp\left[ -\pi \frac{m^2+p_{\perp}^2}{2e\bar{E}} \right] \sqrt{1 - \exp\left[ -\pi \frac{m^2+p_{\perp}^2}{e\bar{E}} \right]} \frac{\sqrt{\pi}}{2}  \sqrt{ \frac{e\bar{E}}{m^2 + p_{\perp}^2 } } \frac{p_{\perp}}{\Omega}\frac{{\mathcal E}_{\perp}}{\bar{E}} \sin \varphi \nonumber\\
	&\times \Biggl[   \mp \sin\left( \phi \mp \frac{\Omega p_3}{e\bar{E}} \right) \sin(\theta_{\bm p}-\theta_{\bm {\mathcal E}})  \nonumber\\
	&\quad\quad + \exp\left[ + \pi \frac{m^2+p_{\perp}^2}{2e\bar{E}} \right] \sqrt{1 - \exp\left[ -\pi \frac{m^2+p_{\perp}^2}{e\bar{E}} \right]} \frac{\sqrt{\pi}}{2}  \sqrt{ \frac{e\bar{E}}{m^2 + p_{\perp}^2 } } \frac{p_{\perp}}{\Omega}\frac{{\mathcal E}_{\perp}}{\bar{E}}  \cos \varphi \sin\left(2(\theta_{\bm p}-\theta_{\bm {\mathcal E}}) \right)   \Biggl], \label{eq18}
\end{align}
where the phase factor $\varphi = \varphi(\Omega)$ determines the frequency of the oscillation as
\begin{align}
	\varphi \equiv \frac{\Omega^2}{4e\bar{E}} - \frac{m^2+p_{\perp}^2}{2e\bar{E}} {\rm ln}\frac{\Omega^2}{2e\bar{E}} - i \; {\rm ln}\left[ \frac{\Gamma\left( 1 + i \frac{m^2+p_{\perp}^2}{2e\bar{E}}  \right)}{\left|  \Gamma\left( 1 + i \frac{m^2+p_{\perp}^2}{2e\bar{E}}  \right)  \right|}   \right] .  \label{eq19}
\end{align}
\end{widetext}
The oscillation in $\Omega$ is reminiscent of the Franz-Keldysh oscillation, which originates from the oscillating distribution of electrons in the Dirac sea due to quantum reflection by the band gap \cite{tay19}.  Note that $\Delta n^{(\mp)}_{{\bm p}}$ for large $\Omega$ is no longer suppressed exponentially by $|e\bar{E}|^{-1}$ because the second term in the curly brackets in Eq.~(\ref{eq18}) is exponentially large so that it cancels with the suppression factor.  This is because the perturbative effect dominates the production for large $\Omega$ \cite{gre17, gre19, tay19}.

As the two terms (i.e., the interference term and the perturbative term) in Eq.~(\ref{eq18}) have distinct $\theta_{\bm p}$-dependences, the $\theta_{\bm p}$-dependence of the spin-imbalance $\Delta n^{(\mp)}_{{\bm p}}$ is determined by which term is dominant.  Namely, $\Delta n^{(\mp)}_{{\bm p}}$ becomes proportional to $\sin (\theta_{\bm p} - \theta_{\bm {\mathcal E}})$ ($\sin \left(2 (\theta_{\bm p} - \theta_{\bm {\mathcal E}}) \right)$) if the first (second) term dominates the production: For supercritical field strength $e\bar{E} \gtrsim m^2 + p_{\perp}^2$, the dimensionless quantity $p_{\perp}{\mathcal E}_{\perp}/\Omega\bar{E}$ (i.e., the strength of the effective magnetic field $\propto {\bm p}\times {\bm {\mathcal E}}$) determines the relative size between the two terms, and the first and the second term becomes dominant for $p_{\perp}{\mathcal E}_{\perp}/\Omega\bar{E} \lesssim 1$ and $\gtrsim 1$, respectively.  On the other hand, for subcritical field strength $e\bar{E} \lesssim m^2 + p_{\perp}^2$, $p_{\perp}{\mathcal E}_{\perp}/\Omega\bar{E}$ determines the relative size only when the frequency $\Omega$ is small.  When $\Omega$ becomes large, the exponential factor becomes important.  Thus, the second term dominates the spin-imbalance provided that $p_{\perp}{\mathcal E}_{\perp}/\Omega\bar{E}$ is not exponentially small.  

Note that the first term vanishes if one integrates $\Delta n^{(\mp)}_{{\bm p}}$ over the longitudinal momentum $p_3$ as
\begin{align}
	&\int dp_3\; \Delta n^{(\mp)}_{{\bm p}} \nonumber\\
	&= \frac{VT}{4\pi} e\bar{E} \exp\left[ - \pi\frac{m^2+p_{\perp}^2}{e\bar{E}}  \right] \left| \frac{p_{\perp}}{\Omega} \frac{{\mathcal E}_{\perp}}{\bar{E}} \right|^2 \nonumber\\
	&\quad\times {\rm Im}\left[ {\rm e}^{-i\frac{\Omega^2}{4e\bar{E}}} {}_1\tilde{F}_1\left( 1 - i \frac{ m^2 + p_{\perp}^2 }{ 2e\bar{E} }; 1; i \frac{\Omega^2}{2e\bar{E}} \right)  \right] \nonumber\\
	&\quad \times {\rm Re}\left[ {\rm e}^{-i\frac{\Omega^2}{4e\bar{E}}} {}_1\tilde{F}_1\left( 1 - i \frac{ m^2 + p_{\perp}^2 }{ 2e\bar{E} }; 1; i \frac{\Omega^2}{2e\bar{E}} \right)  \right]  \nonumber\\
	&\quad\times \sin\left( 2(\theta_{\bm p}-\theta_{\bm {\mathcal E}}) \right)  ,
\end{align}
where we used $\int dp_3 = e\bar{E}T$ \cite{nik70}.  Thus, the $\theta_{\bm p}$-dependence of the spin-imbalance in the transverse distribution $\int dp_3 (n^{(\mp)}_{{\bm p},{\uparrow}}-n^{(\mp)}_{{\bm p},{\downarrow}})$ is independent of the parameters and is always proportional to $\sin \left(2 (\theta_{\bm p} - \theta_{\bm {\mathcal E}}) \right)$.

It is evident from Eq.~(\ref{eq16}) that the spin-imbalance vanishes after $\theta_{\bm p}$-integration.  That is, the total number of spin up and down particles is the same as
\begin{align}
	\int d\theta_{\bm p} \; \Delta n^{(\mp)}_{{\bm p}}  = 0.
\end{align}
This is a natural result since the vacuum is an unpolarized state.

\section{Numerical evaluation of the spin-imbalance} \label{sec3}

In this section, we numerically study the spin-imbalance by directly solving the Dirac equation on a computer without using any approximations.  In Sec.~\ref{sec3a}, we first explain how to compute the spin-imbalance from a solution of the Dirac equation by using the Bogoliubov transformation technique.  In Sec.~\ref{sec3b}, we present our numerical results, and discuss the spin-imbalance quantitatively.  We also show that the numerical results are in excellent agreement with the analytical formula derived in Sec.~\ref{sec2} if the perturbation is sufficiently weak and/or the frequency is not so small.

\subsection{Formalism: The Bogoliubov transformation} \label{sec3a}

We explain how to compute the spin-imbalance $\Delta n^{(\mp)}_{{\bm p},s}$ on a computer without relying on any approximations.  Our formulation is based on the Bogoliubov transformation technique \cite{tan09} including spin-dependence \cite{gel16, tay17}.

We first introduce a gauge field configuration defined by
\begin{align}
	\breve{A}_{\mu}(x^0)
	\equiv \left\{
			\begin{array}{ll}
				A_{\mu}(-\tau) & (x^0<-\tau) \\
				A_{\mu}(x^0) & (-\tau<x^0<+\tau) \\
				A_{\mu}(+\tau) & (+\tau<x_0)		
			\end{array}
		  \right. , \label{eq22}
\end{align}
where $\tau>0$ and $A_{\mu}$ is the original gauge field configuration (\ref{eq3})\footnote{The electric field $\breve{\bm E} \propto \theta(t+\tau)\theta(\tau-t)$ is suddenly switched on/off at $t = \mp \tau$ for our gauge field configuration $\breve{A}_{\mu}$.  The effect of the sudden switching on/off at the boundary $t=\pm \tau$ in the production number can be safely neglected in the limit of $\tau \to \infty$ because the major production occurs not at the boundary but during $t \in (-\tau, \tau)$.  In general, the boundary effect is negligible with $\tau \to \infty$ and the final formulas (\ref{eq37a}) and (\ref{eq37b}) are insensitive to how we switch on/off the electric field $\breve{\bm E}$.}.  Notice that
\begin{align}
	\breve{A}_{\mu} \xrightarrow{\tau \to \infty}{} A_{\mu}, \label{eq23}
\end{align}
so that the production number of the original field $A_{\mu}$ can be obtained from that of $\breve{A}_{\mu}$ with sufficiently long $\tau$.

As $\breve{A}_{\mu}$ becomes a pure gauge (i.e., non-interacting with $\hat{\psi}$) at $x^0<-\tau$ and $x^0>\tau$, one can safely expand the field operator $\hat{\psi}$ by a plane wave and canonically quantize $\hat{\psi}$ to define creation/annihilation operators at $x^0<-\tau$ and $x^0>\tau$ as\footnote{Only when the gauge field becomes a pure gauge, can one define creation/annihilation operators in a well-defined manner.  In other words, one cannot uniquely define creation/annihilation operators in the presence of a non-vanishing electromagnetic field because the interaction between $\hat{\psi}$ and the field mixes up particle and anti-particle modes.  For such a case, one has to make an additional assumption (e.g., the adiabatic particle picture \cite{birrel}) to define the operators. }
\begin{widetext}
\begin{align}
	\hat{\psi}(x) 	
	= \left\{
		\begin{array}{ll}
			\displaystyle \sum_s \int d^3{\bm p}\left[ u_{{\bm P}(-\tau),s} {\rm e}^{-i \omega_{{\bm P}(-\tau)} x^0 } \hat{a}^{(-\tau)}_{{\bm p},s} + v_{{\bm P}(-\tau),s} {\rm e}^{ +i \omega_{{\bm P}(-\tau)} x^0 } \hat{b}^{(-\tau)\dagger}_{-{\bm p},s} \right] \frac{{\rm e}^{i{\bm p}\cdot {\bm x}}}{(2\pi)^{3/2}} & {\rm for}\ x^0<-\tau \\
			\displaystyle \sum_s \int d^3{\bm p}\left[ u_{{\bm P}(+\tau),s} {\rm e}^{-i \omega_{{\bm P}(+\tau)} x^0 } \hat{a}^{(+\tau)}_{{\bm p},s} + v_{{\bm P}(+\tau),s} {\rm e}^{ +i \omega_{{\bm P}(+\tau)} x^0 } \hat{b}^{(+\tau)\dagger}_{-{\bm p},s} \right] \frac{{\rm e}^{i{\bm p}\cdot {\bm x}}}{(2\pi)^{3/2}} & {\rm for}\ x^0>+\tau
		\end{array}
	  \right. , \label{eq24}
\end{align}
\end{widetext}
where, as before, ${\bm p}$ is (canonical) momentum, $s=\uparrow, \downarrow$ is spin, and $\omega_{\bm p} = \sqrt{m^2 + {\bm p}^2}$ is the on-shell energy.  We also introduced $\breve{\bm A} \equiv (\breve{A}^{1}, \breve{A}^{2}, \breve{A}^{3})$ and ${\bm P} \equiv {\bm p} - e\breve{\bm A}$, and $u_{{\bm p},s},v_{{\bm p},s}$ represent the Dirac spinors such that
\begin{subequations}
\begin{align}
	0 &= \left[ \gamma^0 \omega_{\bm p} - {\bm \gamma}\cdot {\bm p} - m \right] u_{{\bm p},s}, \\
	0 &= \left[ \gamma^0 \omega_{\bm p} + {\bm \gamma}\cdot {\bm p} + m \right] v_{{\bm p},s}
\end{align}
\end{subequations}
with ${\bm \gamma} \equiv (\gamma^1, \gamma^2, \gamma^3)$ and the normalization
\begin{align}
	u_{{\bm p},s}^{\dagger} u_{{\bm p},s'} = v_{{\bm p},s}^{\dagger} v_{{\bm p},s'} = \delta_{ss'} ,\
	u_{{\bm p},s}^{\dagger} v_{{\bm p},s'} = 0.  \label{eq26}
\end{align}
The creation/annihilation operators satisfy the following anti-commutation relations
\begin{subequations}
\begin{align}
	\{ a^{(\pm \tau)\dagger}_{{\bm p},s}, a^{(\pm \tau)}_{{\bm p}',s'}  \}
	= 	\{ b^{(\pm \tau)\dagger}_{{\bm p},s}, b^{(\pm \tau)}_{{\bm p}',s'}  \}
	&= \delta_{ss'}\delta^3({\bm p}-{\bm p}'), \\
	{\rm (others)} &= 0.
\end{align}
\end{subequations}

An important point here is that the creation/annihilation operators at $x^0 > +\tau$ and those at $x^0 < -\tau$ are inequivalent because of the interaction with the electromagnetic field $\breve{A}_{\mu}$ during the time-evolution $-\tau < x^0 < \tau$.  The inequivalence can be expressed in terms of a Bogoliubov transformation.  To see this, we first expand the field operator $\hat{\psi}$ in terms of the creation/annihilation operators at $x^0<-\tau$ as
\begin{align}
	\hat{\psi}(x) = \sum_s \int d^3{\bm p}\left[ U_{{\bm p},s}(x^0) \hat{a}^{(-\tau)}_{{\bm p},s} + V_{{\bm p},s}(x^0) \hat{b}^{(-\tau)}_{-{\bm p},s} \right] \frac{{\rm e}^{i{\bm p}\cdot {\bm x}}}{(2\pi)^{3/2}}.
\end{align}
The time-evolution of the mode functions $U,V$ is determined by the Dirac equation
\begin{align}
	0 = \left[ i \gamma^0 \partial_{x^0} - {\bm \gamma}\cdot({\bm p}- e\breve{\bm A}) - m \right] \begin{pmatrix} U_{{\bm p},s} \\ V_{{\bm p},s} \end{pmatrix}  . \label{eq29}
\end{align}
The initial condition for the mode functions $U,V$ is fixed by the first line of Eq.~(\ref{eq24}) as
\begin{subequations}
\begin{align}
	U_{{\bm p},s}(-\tau) = u_{{\bm P}(-\tau),s} {\rm e}^{-i \omega_{{\bm P}(-\tau)} x^0 }, \label{eq30a} \\
	V_{{\bm p},s}(-\tau) = v_{{\bm P}(-\tau),s} {\rm e}^{+i \omega_{{\bm P}(-\tau)} x^0 }.  \label{eq30b}
\end{align}
\end{subequations}
Note that because of this initial condition, the mode functions $U,V$ satisfy the same normalization condition for $u,v$ as
\begin{align}
	U_{{\bm p},s}^{\dagger} U_{{\bm p},s'} = V_{{\bm p},s}^{\dagger} V_{{\bm p},s'} = \delta_{ss'} ,\
	U_{{\bm p},s}^{\dagger} V_{{\bm p},s'} = 0.  \label{eq31}
\end{align}
Now, we are ready to express $\hat{a}^{(+\tau)}_{{\bm p},s} , \hat{b}^{(+\tau)\dagger}_{-{\bm p},s}$ in terms of a Bogoliubov transformation of $\hat{a}^{(-\tau)}_{{\bm p},s} , \hat{b}^{(-\tau)\dagger}_{-{\bm p},s}$.  By noting the second line of Eq.~(\ref{eq24}), we find
\begin{widetext}
\begin{align}
	\begin{pmatrix}
		\hat{a}^{(+\tau)}_{{\bm p},s} \\
		\hat{b}^{(+\tau)\dagger}_{-{\bm p},s}
	\end{pmatrix}
	&=
	\int d^3{\bm x} \frac{{\rm e}^{-i{\bm p}\cdot {\bm x}}}{(2\pi)^{3/2}}
	\begin{pmatrix}
		u^{\dagger}_{{\bm P}(+\tau),s} {\rm e}^{+ i\omega_{{\bm P}(+\tau)}\tau} \\
		v^{\dagger}_{{\bm P}(+\tau),s} {\rm e}^{- i\omega_{{\bm P}(+\tau)}\tau}
	\end{pmatrix}
	\hat{\psi}(+\tau)	\nonumber\\
	&= \sum_{s'}
	   \begin{pmatrix}
		u^{\dagger}_{{\bm P}(+\tau),s} U_{{\bm p},s'}(\tau) {\rm e}^{+ i\omega_{{\bm P}(+\tau)} \tau} & u^{\dagger}_{{\bm P}(+\tau),s} V_{{\bm p},s'}(\tau) {\rm e}^{+ i\omega_{{\bm P}(+\tau)} \tau} \\
		v^{\dagger}_{{\bm P}(+\tau),s} U_{{\bm p},s'}(\tau) {\rm e}^{- i\omega_{{\bm P}(+\tau)} \tau}  & v^{\dagger}_{{\bm P}(+\tau),s} V_{{\bm p},s'}(\tau) {\rm e}^{- i\omega_{{\bm P}(+\tau)} \tau}
	   \end{pmatrix}
	  \begin{pmatrix}
		\hat{a}^{(-\tau)}_{{\bm p},s'} \\
		\hat{b}^{(-\tau)\dagger}_{-{\bm p},s'}
	  \end{pmatrix}.  \label{eq32}
\end{align}
\end{widetext}
Notice that the matrix elements are not diagonal in the spin space $\Slash{\propto}\  \delta_{ss'}$ if the external field $\breve{A}$ affects the spin state during the time-evolution.

Now, we can evaluate the spin-dependent production number of electrons and positrons $n^{(\mp)}$ produced from the vacuum under the external field $A_{\mu}$ based on Eq.~(\ref{eq32}).  To this end, we first define a vacuum state at $x^0 < -\tau$ as
\begin{align}
	0 = \begin{pmatrix} \hat{a}^{(-\tau)}_{{\bm p},s} \\ \hat{b}^{(-\tau)}_{{\bm p},s} \end{pmatrix} \ket{{\rm vac};-\tau}\ {\rm for\ any\ }{\bm p},s.
\end{align}
Then, one can directly evaluate the vacuum expectation value of the number operator at $x^0 > \tau$ (i.e, the production number under $\breve{A}_{\mu}$) as
\begin{subequations}
\begin{align}
	\breve{n}^{(-)}_{{\bm p},s}
		&\equiv \frac{\braket{{\rm vac};-\tau| \hat{a}_{{\bm p},s}^{(+\tau)\dagger}  \hat{a}_{{\bm p},s}^{(+\tau)}|{\rm vac};-\tau}}{\braket{{\rm vac};-\tau| {\rm vac};-\tau}} \nonumber\\
		&= \frac{V}{(2\pi)^3} \sum_{s'} \left|  u^{\dagger}_{{\bm p} - e\breve{\bm A}(\tau),s} V_{{\bm p},s'}(\tau)  \right|^2, \label{eq34a} \\
	\breve{n}^{(+)}_{{\bm p},s}
		&\equiv \frac{\braket{{\rm vac};-\tau| \hat{b}_{{\bm p},s}^{(+\tau)\dagger}  \hat{b}_{{\bm p},s}^{(+\tau)}|{\rm vac};-\tau}}{\braket{{\rm vac};-\tau| {\rm vac};-\tau}} \nonumber\\
		&= \frac{V}{(2\pi)^3} \sum_{s'} \left|  v^{\dagger}_{-{\bm p} - e\breve{\bm A}(\tau),s} U_{-{\bm p},s'}(\tau) \right|^2, \label{eq34b}
\end{align}
\end{subequations}
where we used $\delta^3({\bm p}={\bm 0})=V/(2\pi)^3$.  Therefore, as noted in Eq.~(\ref{eq23}), the production number $n^{(\mp)}$ by $A_{\mu}$ can be obtained by taking the $\tau \to \infty$ limit of $\breve{n}^{(\mp)}$ as
\begin{align}
	n^{(\mp)}_{{\bm p},s} = \lim_{\tau \to \infty} \breve{n}^{(\mp)}_{{\bm p},s} .  \label{eq36}
\end{align}

From Eq.~(\ref{eq36}), we arrive at a formula for the spin-imbalance $\Delta n^{(\mp)}_{{\bm p}}$ as
\begin{subequations}
\begin{align}
	\Delta n^{(-)}_{{\bm p}}
		&= \lim_{\tau \to \infty} \sum_{s} \left[ \left|  u^{\dagger}_{{\bm p} - e{\bm A}(\tau),\uparrow} V_{{\bm p},s}(\tau)  \right|^2  \right. \nonumber\\
		&\quad\quad\quad\quad\quad\quad \left. - \left|  u^{\dagger}_{{\bm p} - e{\bm A}(\tau),\downarrow} V_{{\bm p},s}(\tau)  \right|^2  \right], \label{eq37a} \\
	\Delta n^{(+)}_{{\bm p}}
		&= \lim_{\tau \to \infty} \sum_{s} \left[ \left|  v^{\dagger}_{-{\bm p} - e{\bm A}(\tau),\uparrow} U_{-{\bm p},s}(\tau)  \right|^2  \right. \nonumber\\
		&\quad\quad\quad\quad\quad\quad \left. - \left|  v^{\dagger}_{-{\bm p} - e{\bm A}(\tau),\downarrow} U_{-{\bm p},s}(\tau)  \right|^2  \right], \label{eq37b}
\end{align}
\end{subequations}
Notice that the formula is exact since we used no approximations in its derivation, and it can be evaluated on a computer by solving just the Dirac equation (\ref{eq29}).

\subsection{Numerical results} \label{sec3b}

We numerically study the spin-imbalance based on the formalism explained in Sec.~\ref{sec3a}.  We also show that the analytical formula (\ref{eq16}) reproduces the numerical results very well if the perturbation is sufficiently weak and/or is not so slow, for which multi-photon processes only give sub-dominant contributions.

Below, we focus on the spin-imbalance for electrons $\Delta n_{\bm p}^{(-)}$.  With explicit numerical calculations, we checked that the positron's spin-imbalance can be obtained from that of the electron by just flipping the sign of the momentum ${\bm p}$.  This is consistent with the analytical formula (\ref{eq16}) and with the physical intuition that electrons and positrons are created as a pair, whose total momentum is zero.  Numerical results for the total production number $n^{(-)}_{{\bm p},s}$ are discussed in detail in Appendix~\ref{appB}.

Our problem is characterized by eight dimensionless parameters: $e\bar{E}/m^2$, ${\mathcal E}_{\perp}/\bar{E}$, ${\mathcal E}_3/\bar{E}$, $\Omega/m$, $p_{\perp}/m$, $p_3/m$, $\phi$, $\theta_{\bm p}-\theta_{\bm {\mathcal E}}$.  As a demonstration, we consider $e\bar{E}/m^2 = 0.4, {\mathcal E}_3/\bar{E} = 0, \phi = 1$ below, and discuss the other parameters dependence on the spin-imbalance.  Note that the relative size of the spin-imbalance with this parameter choice compared to the production number $n^{(-)}_{{\bm p},s}$ is found to be ${\mathcal O}(1 {\rm -} 10 \;\%)$, as we shall show explicitly below (see also Appendix~\ref{appB}).

We checked that the spin-imbalance is insensitive to the magnitude of the longitudinal perturbation ${\mathcal E}_3$, as expected from the analytical formula (\ref{eq16}).  We also checked that results with different $\phi$ are reproduced very well by shifting the phase $\phi$ as $\phi \to \phi \mp \Omega p_3/e\bar{E}$, which is consistent with the analytical formula (\ref{eq16}).

\subsubsection{Numerical details}\label{sec3b1}

Let us briefly explain some details of our numerical calculations.  As explained in Sec.~\ref{sec3a}, we numerically solved the Dirac equation (\ref{eq29}) in the presence of the gauge field configuration (\ref{eq22}) to obtain the mode functions $U,V$, with which the spin-imbalance $\Delta n_{\bm p}$ can be evaluated through Eqs.~(\ref{eq37a}) and (\ref{eq37b}).  The boundary conditions for the mode functions $U,V$ are fixed by the plane wave boundary conditions (\ref{eq30a}) and (\ref{eq30b}) set at $x^0 = -\tau$, and the Dirac equation was solved forward in time until $x^0 = \tau$ with the Runge-Kutta method.  Notice that it is sufficient to solve the Dirac equation in the region $x^0 \in [-\tau, \tau]$ because the solution outside of that region $|x^0| > \tau$ is trivially given by the plane waves, and the time evolution becomes non-trivial only inside of $x^0 \in [-\tau, \tau]$.  Below, we take $m \tau = 100$.  We carefully checked that this $\tau$ is sufficiently large such that the results presented below are insensitive to the values of $\tau$ within the numerical accuracy.  The numerical accuracy of our results was carefully checked by the normalization condition of $U,V$ (\ref{eq31}), and the error bar was found to be less than $10^{-5}\;\%$.

\subsubsection{$\Omega$-dependence} \label{sec3b2}

\begin{figure*}[!t]
\begin{center}
\hspace*{-10mm}\includegraphics[clip, width=0.365\textwidth]{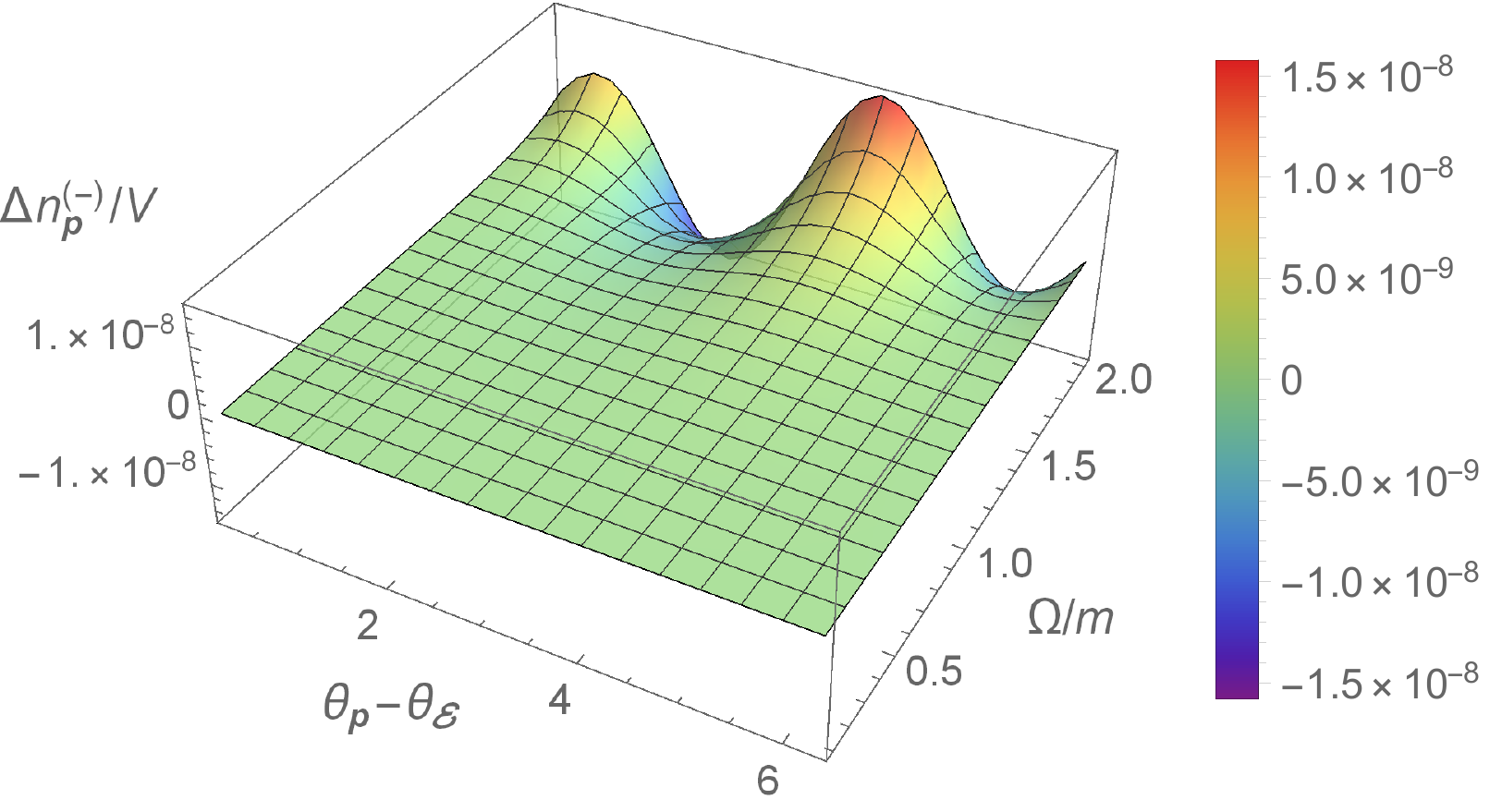}
\hspace*{-1mm}\includegraphics[clip, width=0.345\textwidth]{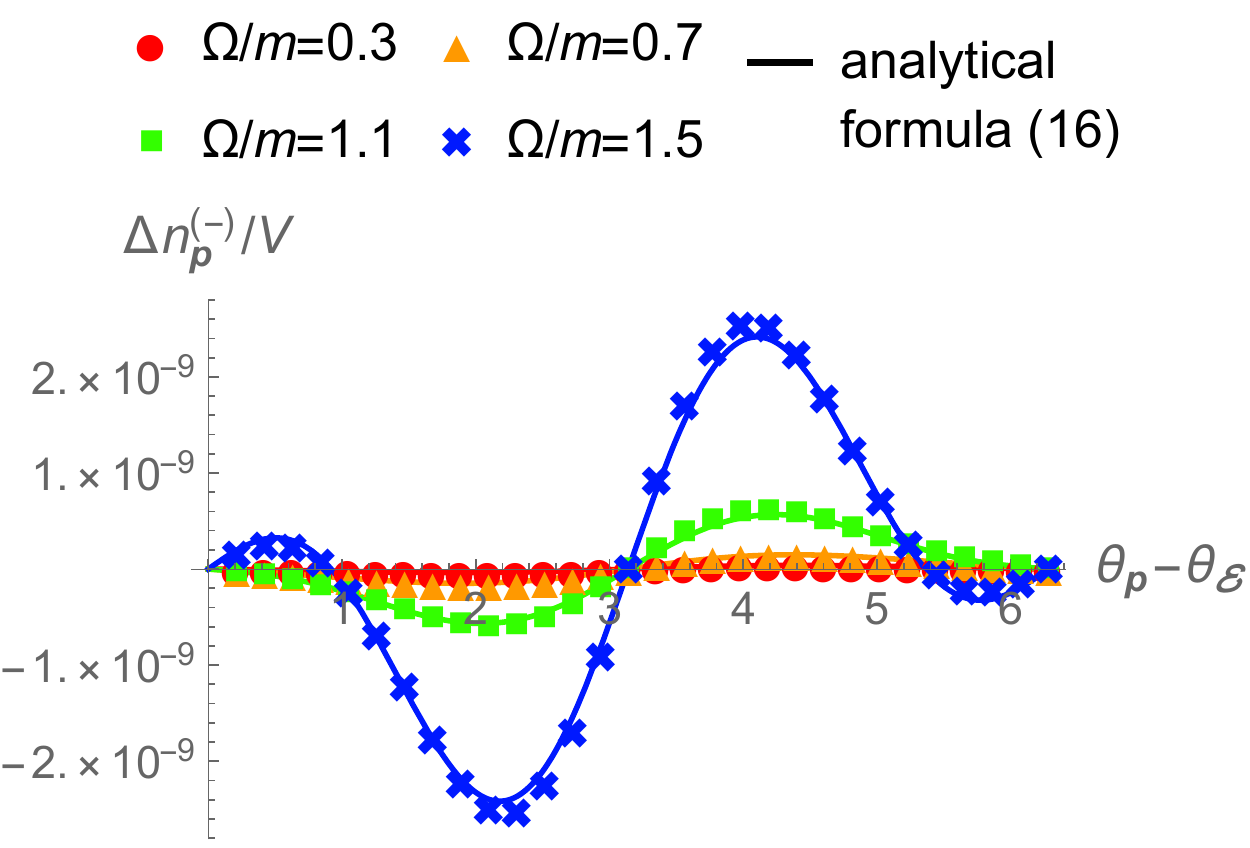}
\hspace*{-1mm}\includegraphics[clip, width=0.345\textwidth]{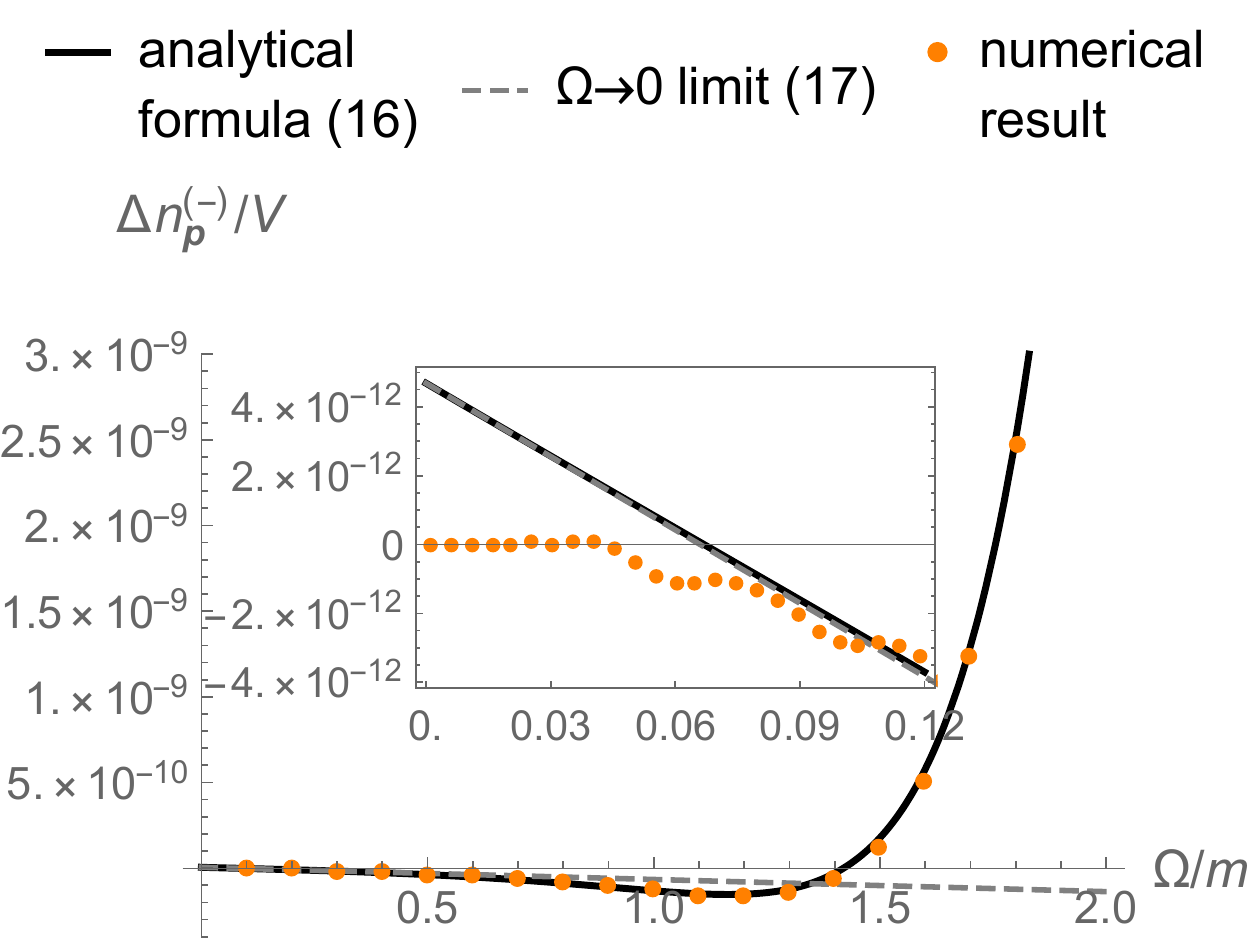}
\vspace*{3mm}\hfill \\
\mbox{(i) Small frequency $\Omega/m \in [0,2]$}\hfill \\
\vspace*{6mm}
\hspace*{-10mm}\includegraphics[clip, width=0.365\textwidth]{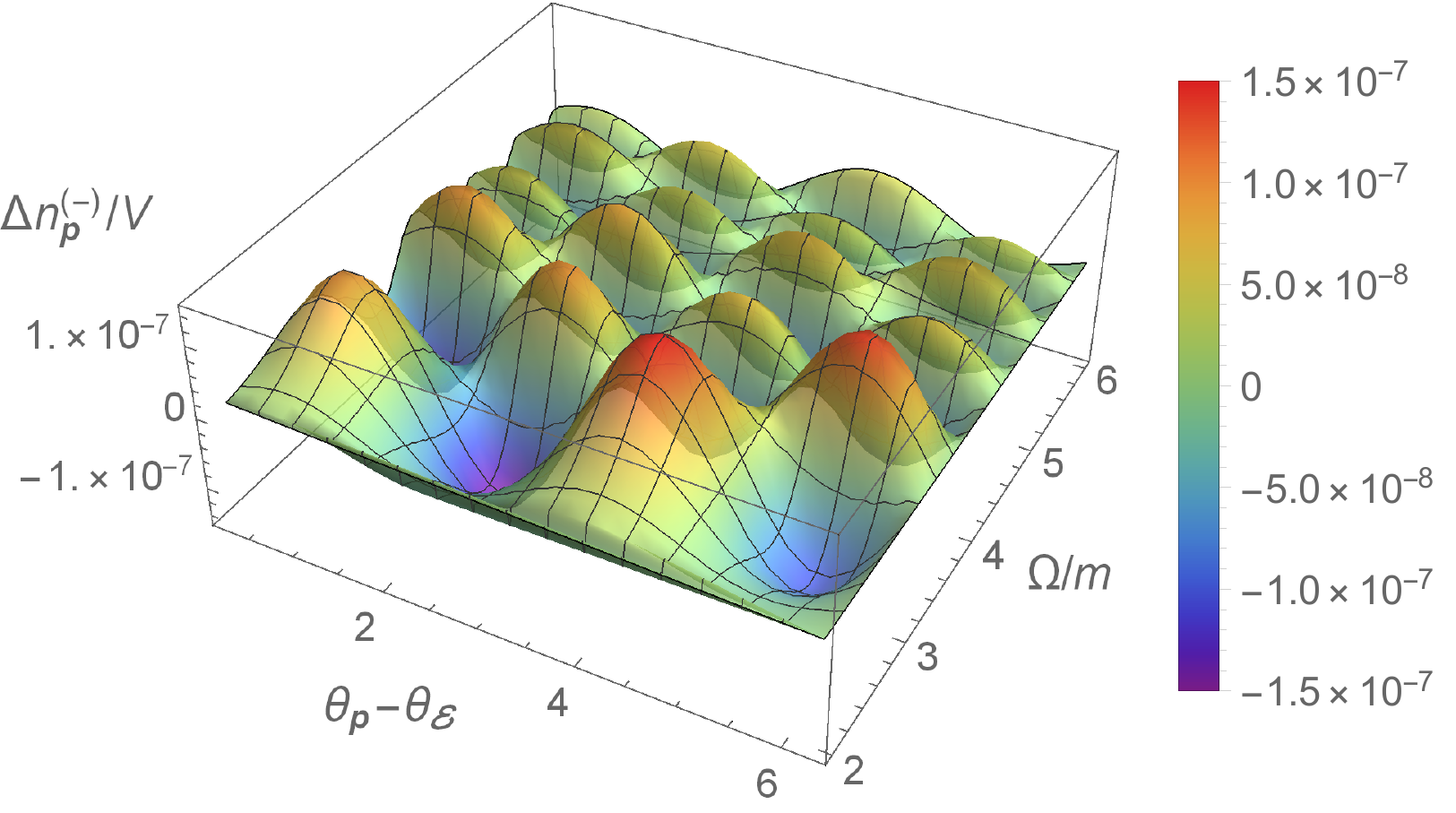}
\hspace*{-1mm}\includegraphics[clip, width=0.345\textwidth]{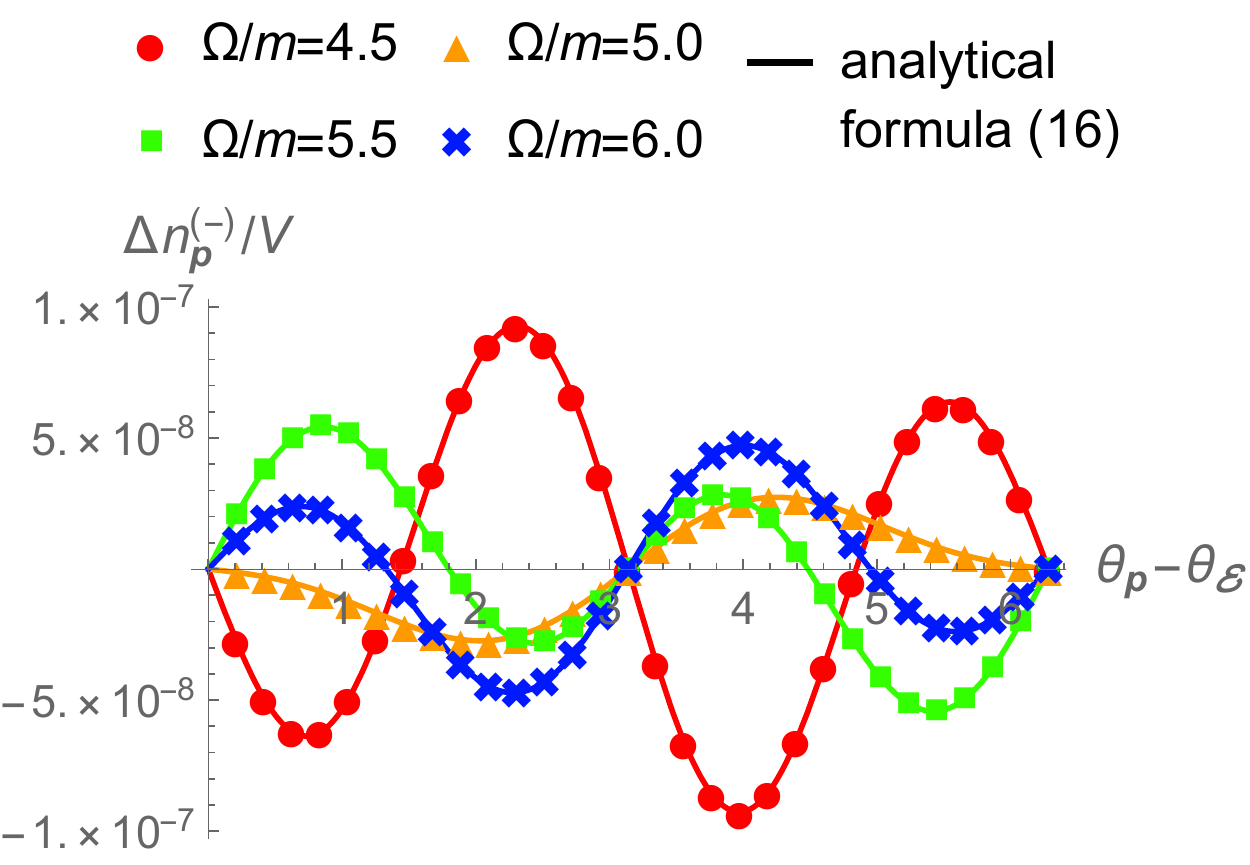}
\hspace*{-1mm}\includegraphics[clip, width=0.345\textwidth]{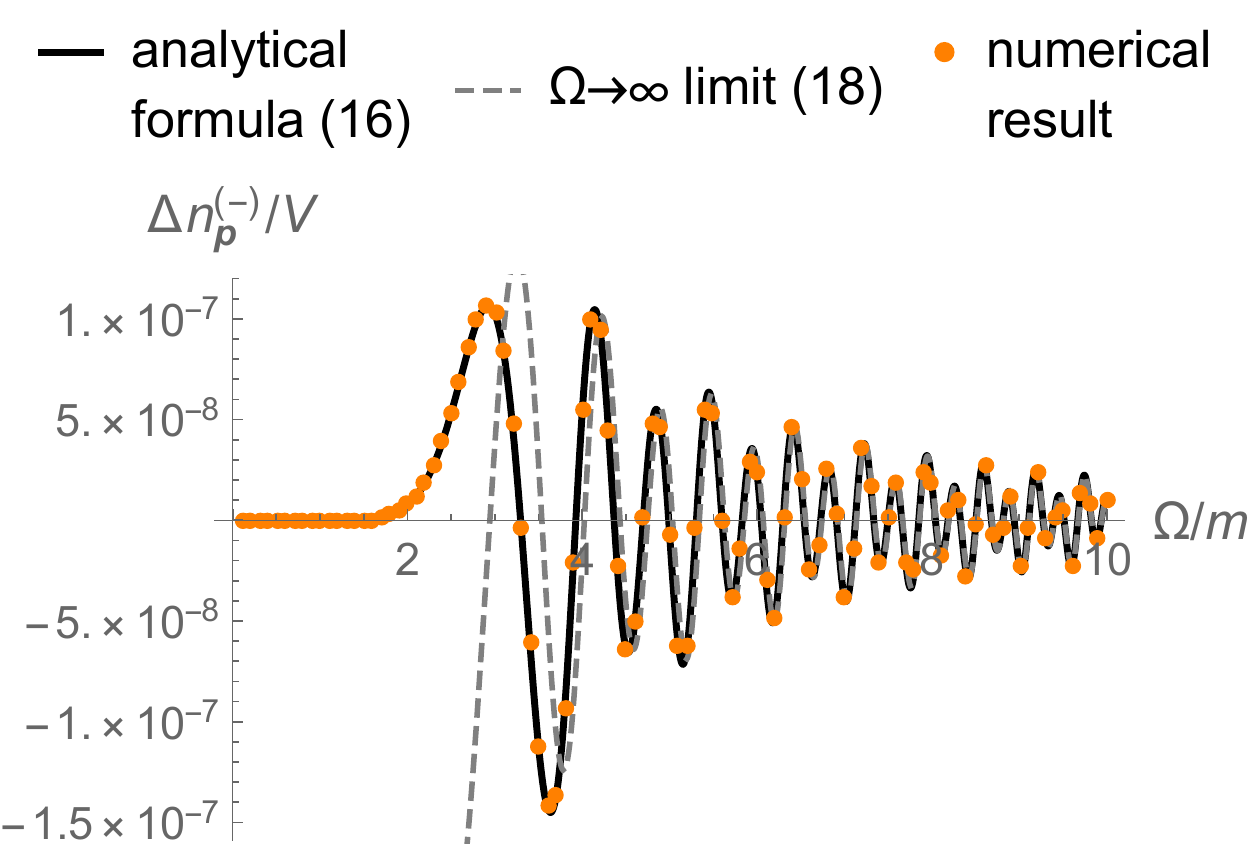}
\vspace*{3mm} \hfill \\
\mbox{(ii) Large frequency $\Omega/m \in [2,10]$}
\caption{\label{fig1} (color online) The numerical results of the spin-imbalance $\Delta n^{(-)}_{\bm p}$ as a function of $(\theta_{\bm p}, \Omega)$ (left); as a function of $\theta_{\bm p}$ for several values of $\Omega$ (center); and as a function of $\Omega$ for fixed $\theta_{\bm p}-\theta_{\bm {\mathcal E}} = \pi/4$ (right).  As a comparison, the analytical results (\ref{eq16}) are plotted as the lines in the center and right panels.  The top (i) and bottom (ii) panels distinguish the size of the frequency $\Omega$.  The parameters are fixed as $e\bar{E}/m^2 = 0.4, {\mathcal E}_{\perp}/\bar{E} = 0.025, {\mathcal E}_3/\bar{E} = 0, p_{\perp}/m = 1, p_3/m = 0, \phi = 1,\ {\rm and\ }m\tau=100$. }
\end{center}
\end{figure*}

The numerical results for the spin-imbalance $\Delta n^{(-)}_{\bm p}$ for a weak perturbation ${\mathcal E}_{\perp}/\bar{E} = 0.025$ as a function of the azimuthal angle $\theta_{\bm p}$ and the frequency $\Omega$ are plotted in Fig.~\ref{fig1}.  It is evident that the spin-imbalance, indeed, becomes non-vanishing for a transverse perturbation.

As a comparison, the analytical formula (\ref{eq16}) is also displayed in Fig.~\ref{fig1}.  We find that the analytical formula (\ref{eq16}) is in excellent agreement with the numerical results except for very small $\Omega/m \lesssim 0.1$.  This disagreement comes from multi-photon effects.  In fact, the formula (\ref{eq16}) only takes into account the one-photon process described by the diagram (\ref{eq7}).  When the frequency becomes small such that the Keldysh parameter for the weak field $\gamma_{\rm K} = m\Omega/e|{\bm {\mathcal E}}|$ becomes small $\gamma_{\rm K} \lesssim 1$, i.e., $\Omega \lesssim e|{\bm {\mathcal E}}|/m$, the weak field ${\bm {\mathcal E}}$ can interact with a vacuum loop many times \cite{bre70, pop72, kel65, tay14}, which cannot be captured by the formula (\ref{eq16}).  Note that we will discuss more about the validity of the formula (\ref{eq16}) in Sec.~\ref{sec3b3}

Figure~\ref{fig1} shows that the multi-photon processes suppress the spin-imbalance.  This is a reasonable result because the weak field ${\bm {\mathcal E}}$ cannot be distinguished by the strong slow field $\bar{{\bm E}}$ in the limit of $\Omega \to 0$.  Then, the superposition of ${\bm {\mathcal E}}$ just amounts to changing the direction of the total electric field ${\bm E} = \bar{\bm E} + {\bm {\mathcal E}}$.  Therefore, the production is dominated by the usual Schwinger mechanism (\ref{eq1}), so that the spin-imbalance vanishes.

The spin-imbalance exhibits an oscillating behavior in terms of $\theta_{\bm p}$ and $\Omega$, but the behavior is quite different depending on the size of the frequency $\Omega$: For small frequency $\Omega \lesssim \sqrt{m^2+p_{\perp}^2}, \sqrt{e\bar{E}}$, the spin-imbalance is just a monotonically increasing function of $\Omega$.  This is because the production numbers, $n^{(-)}_{{\bm p}, \uparrow}$ and $n^{(-)}_{{\bm p}, \downarrow}$, are enhanced by the dynamically assisted Schwinger mechanism (see also Fig.~\ref{fig6} in Appendix~\ref{appB}), so that their absolute difference $| \Delta n^{(-)}_{\bm p} |$ also increases.  The azimuthal angle $\theta_{\bm p}$-dependence is determined by the strength of the effective magnetic field ${\mathcal E}_{\perp}p_{\perp}/\bar{E}\Omega$ (see Eq.~(\ref{eq17})).  Since the spin-imbalance is suppressed by the multi-photon effects for very small $\Omega/m \lesssim 0.1$ for our parameter choices, a not-very-small $\Omega/m \gtrsim 0.1$ is required for the spin-imbalance to be manifest.  Then, ${\mathcal E}_{\perp}p_{\perp}/\bar{E}\Omega = 0.025/\Omega \lesssim 1$ holds for non-vanishing spin-imbalance, so that the $\theta_{\bm p}$-dependence is given by $\Delta n^{(-)}_{\bm p} \propto \sin (\theta_{\bm p} - \theta_{\bm a} )$.  One can change the $\theta_{\bm p}$-dependence by, for example, increasing $p_{\perp}/\sqrt{e\bar{E}}$ (cf., Fig.~\ref{fig4}).  For this case, $\Delta n^{(-)}_{\bm p} \propto \sin (2(\theta_{\bm p} - \theta_{\bm a}) )$ holds, but the spin-imbalance is strongly suppressed by the exponential factor $\exp[ - \pi (m^2 + p_{\perp}^2)/e\bar{E}  ]$.  For large frequency $\Omega \gtrsim \sqrt{m^2+p_{\perp}^2}, \sqrt{e\bar{E}}$, the spin-imbalance oscillates in $\Omega$.  The oscillation becomes faster with increasing $\Omega$, which is reminiscent of the Franz-Keldysh oscillation \cite{tay19}.  The frequency of the oscillation is reproduced very well by the phase factor $\varphi \propto \Omega^2 + {\mathcal O}(\Omega^0)$ (see Eq.~(\ref{eq19})).  Notice that the $\theta_{\bm p}$-dependence is changed from $\Delta n^{(-)}_{\bm p} \propto \sin (\theta_{\bm p} - \theta_{\bm a} )$ to $\Delta n^{(-)}_{\bm p} \propto \sin (2(\theta_{\bm p} - \theta_{\bm a}) )$.  This occurs for subcritical field strength $e\bar{E} \lesssim m^2 + p_{\perp}^2$ because the interference term (i.e., the first term) in Eq.~(\ref{eq18}) becomes exponentially suppressed compared to the perturbative term (i.e., the second term).  For supercritical field strength $e\bar{E} \gtrsim m^2 + p_{\perp}^2$, the spin-imbalance shows up the same $\theta_{\bm p}$-dependence as the low-frequency case $\Delta n^{(-)}_{\bm p} \propto \sin (\theta_{\bm p} - \theta_{\bm a} )$ because the interference term is free from the exponential suppression and the same dimensionless quantity ${\mathcal E}_{\perp}p_{\perp}/\bar{E}\Omega$ determines the $\theta_{\bm p}$-dependence.

\subsubsection{${\mathcal E}_{\perp}$-dependence} \label{sec3b3}

\begin{figure*}[!t]
\begin{center}
\hspace*{-10mm}\includegraphics[clip, width=0.365\textwidth]{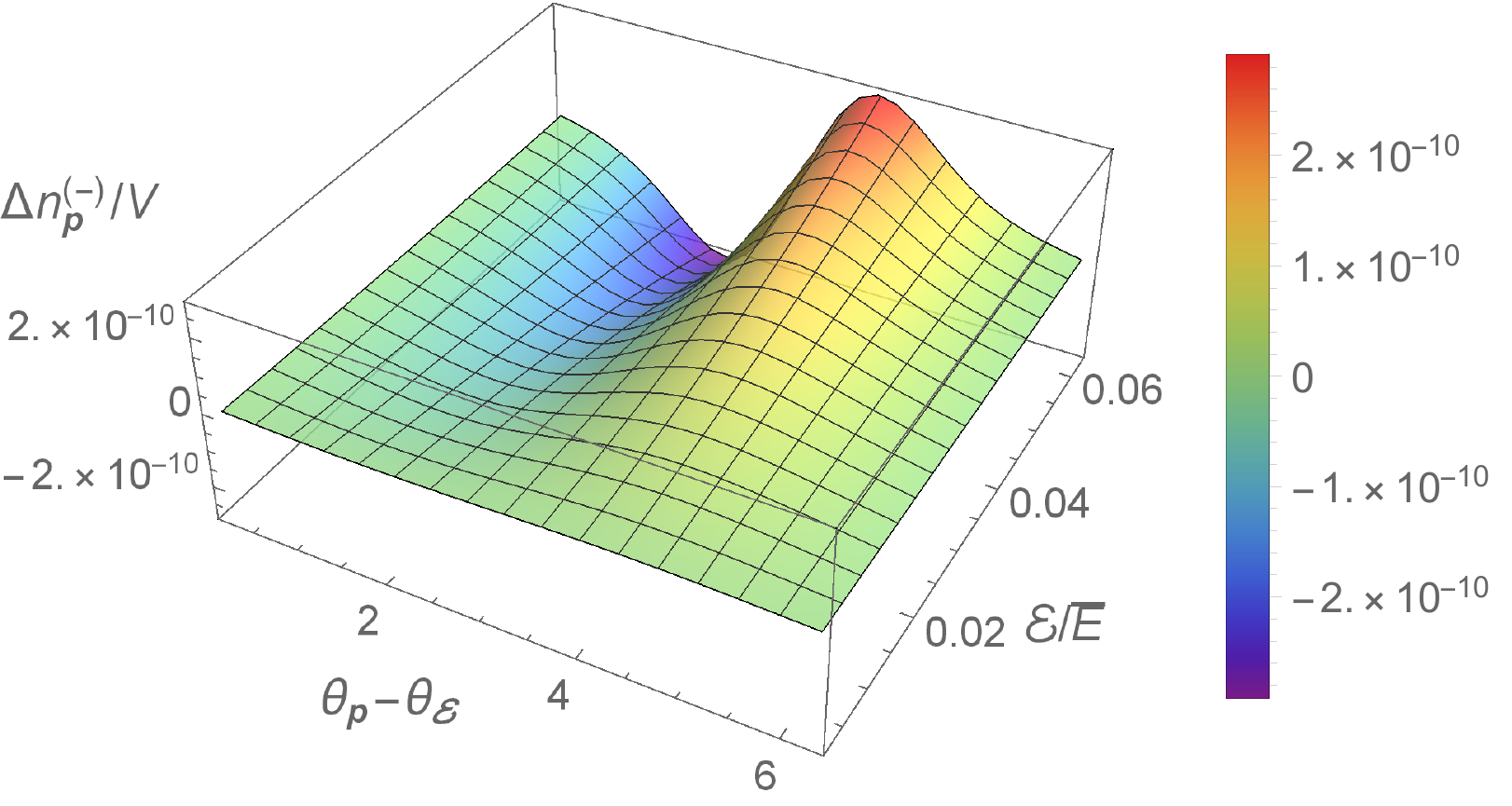}
\hspace*{-1mm}\includegraphics[clip, width=0.345\textwidth]{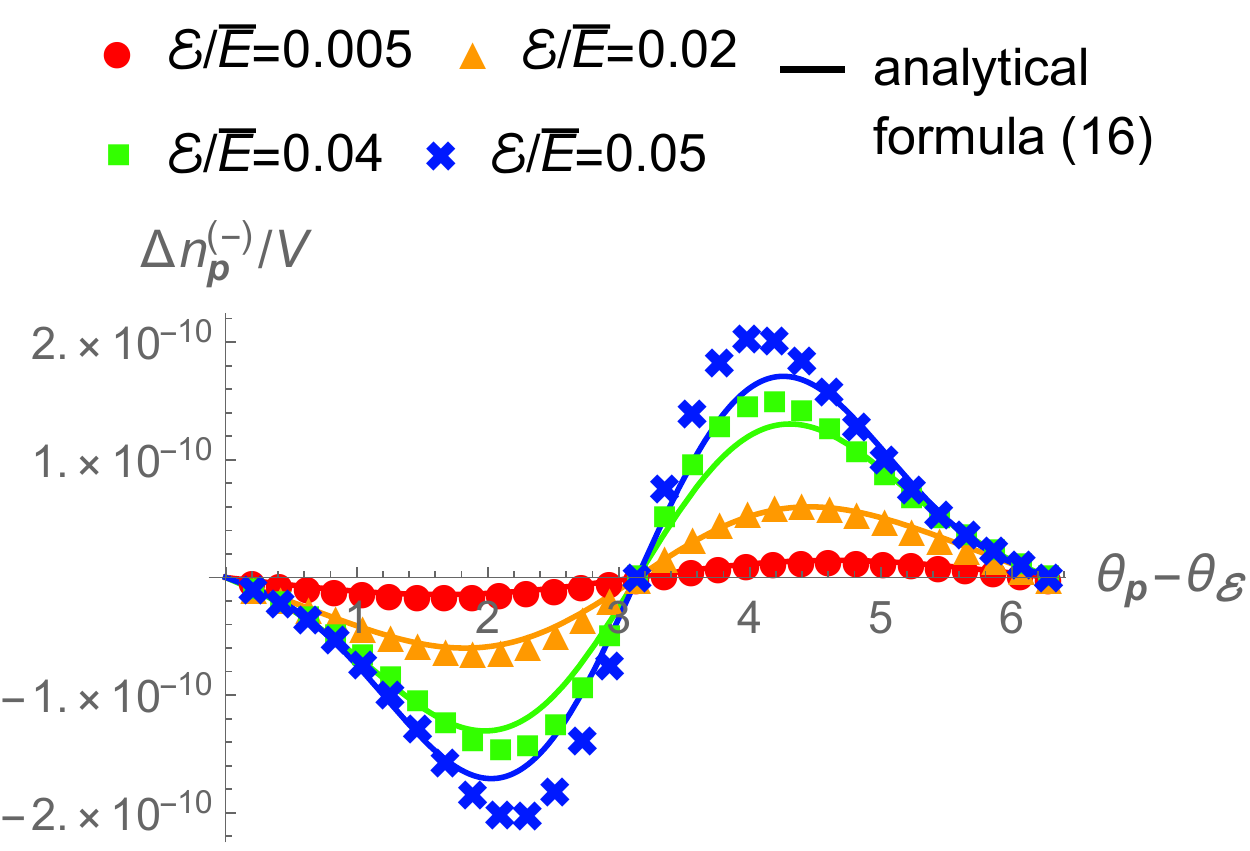}
\hspace*{-1mm}\includegraphics[clip, width=0.345\textwidth]{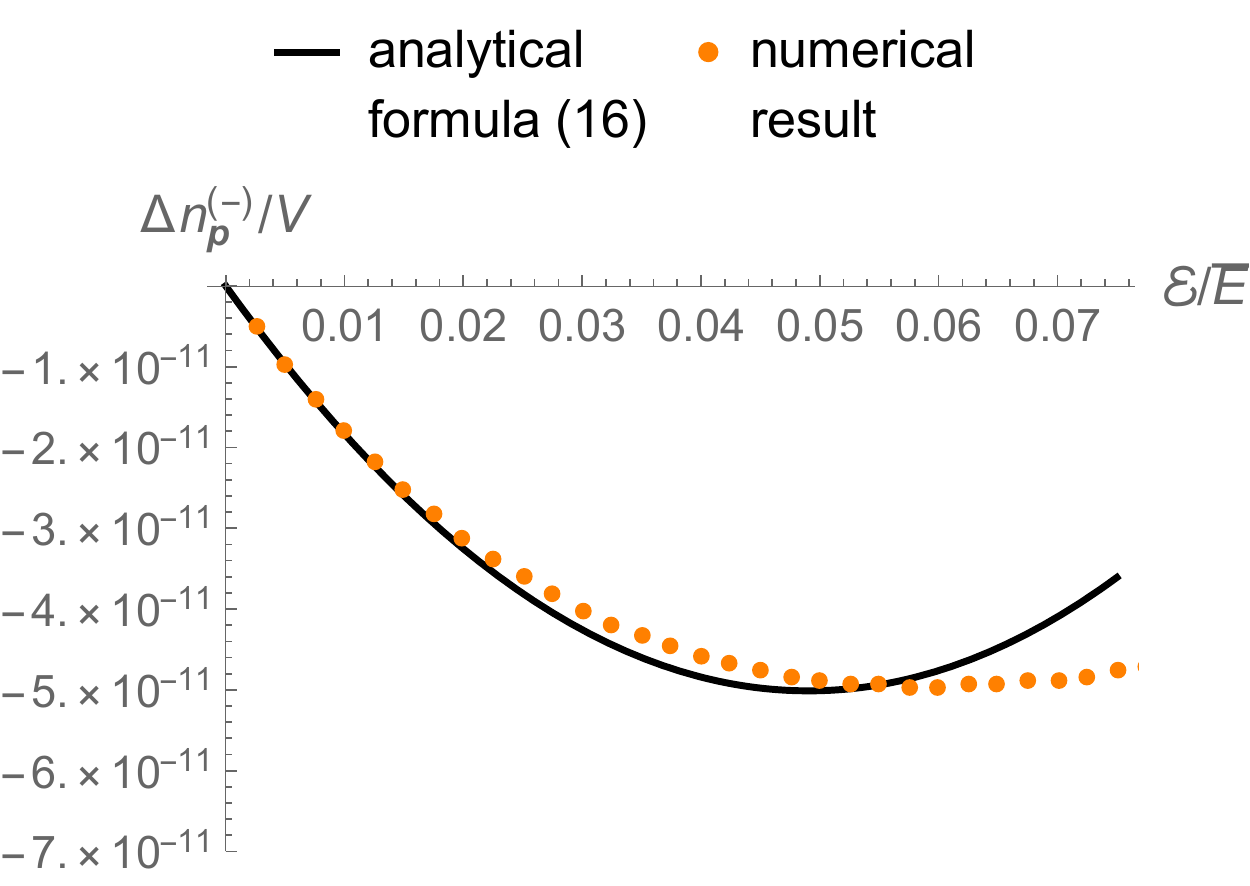}
\vspace*{3mm}\hfill \\
\mbox{(i) Weak perturbation ${\mathcal E}_{\perp}/\bar{E} \in [0,0.05]$}\hfill \\
\vspace*{6mm}
\hspace*{-10mm}\includegraphics[clip, width=0.365\textwidth]{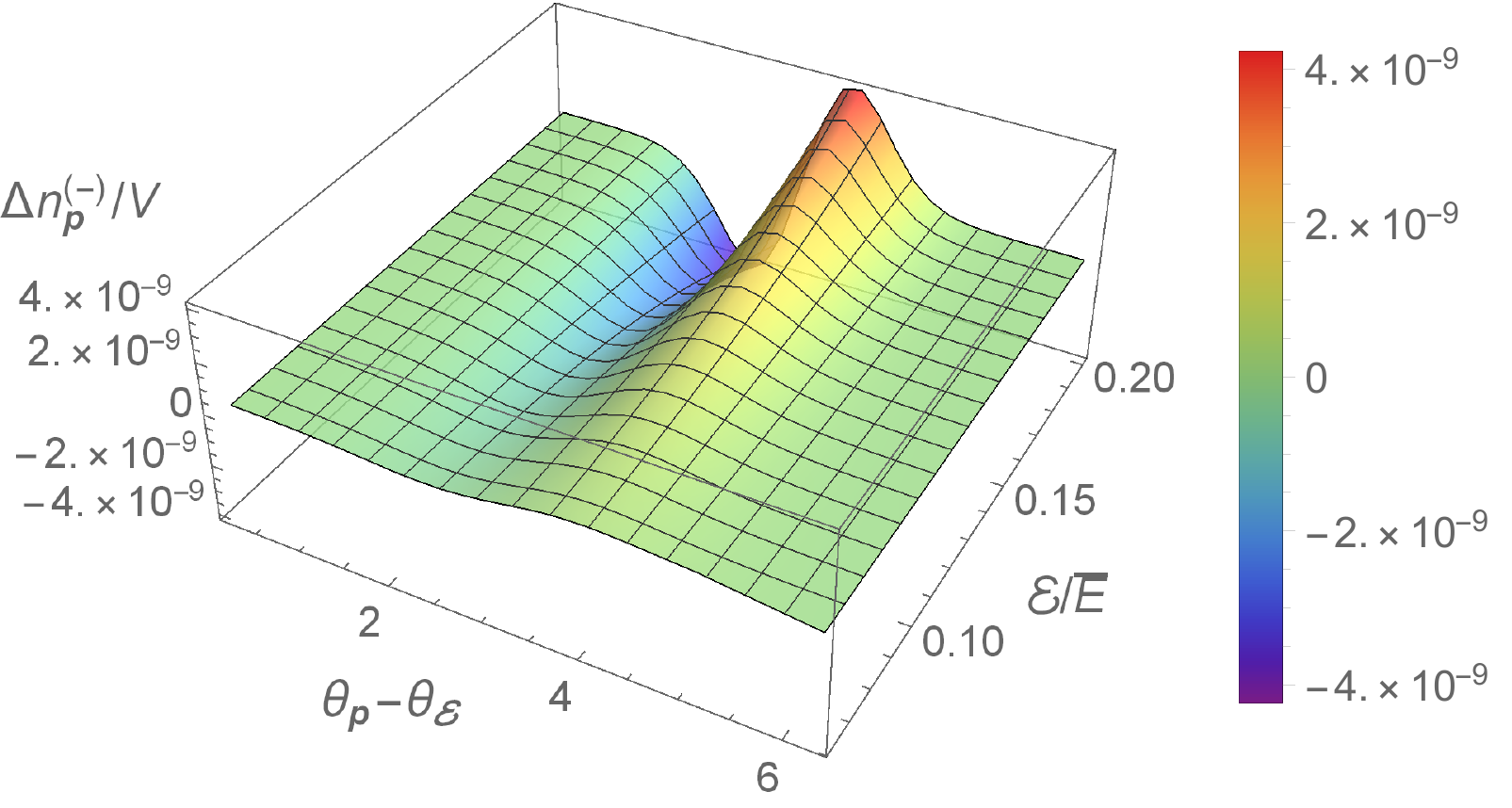}
\hspace*{-1mm}\includegraphics[clip, width=0.345\textwidth]{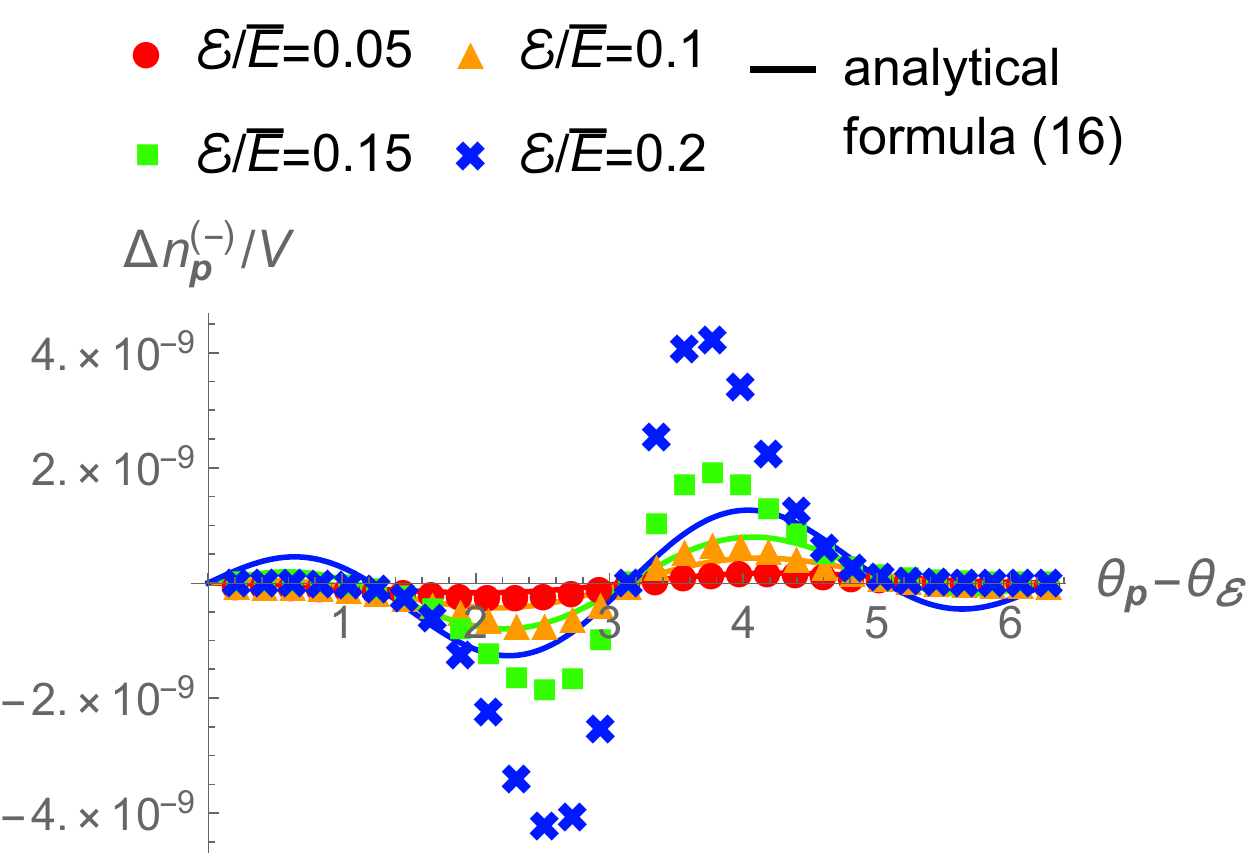}
\hspace*{-1mm}\includegraphics[clip, width=0.345\textwidth]{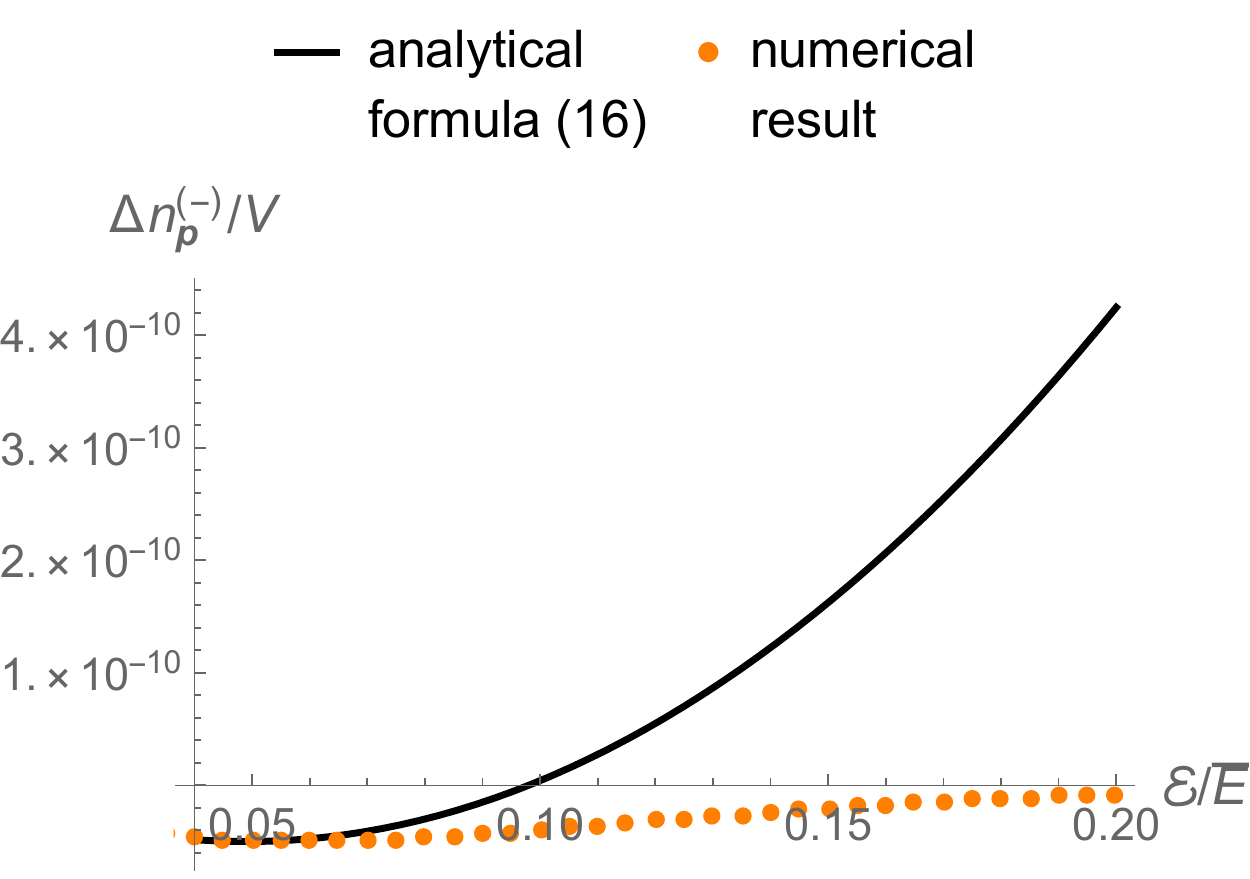}
\vspace*{3mm} \hfill \\
\mbox{(ii) Strong perturbation ${\mathcal E}_{\perp}/\bar{E} \in [0.05,0.2]$}
\caption{\label{fig2} (color online) The numerical results of the spin-imbalance $\Delta n^{(-)}_{{\bm p}}$ as a function of $(\theta_{\bm p}, {\mathcal E}_{\perp})$ (left); as a function of $\theta_{\bm p}$ for several values of ${\mathcal E}_{\perp}$ (center); and as a function of ${\mathcal E}_{\perp}$ for fixed $\theta_{\bm p}-\theta_{\bm {\mathcal E}} = \pi/4$ (right).  As a comparison, the analytical results (\ref{eq16}) are plotted as the lines in the center and right panels.  The top (i) and bottom (ii) panels distinguish the size of the transverse perturbation ${\mathcal E}_{\perp}$.  $\Omega$ is fixed as $\Omega/m = 0.5$.  The other parameters are the same as in Fig.~\ref{fig1}, i.e., $e\bar{E}/m^2 = 0.4, {\mathcal E}_3/\bar{E} = 0, p_{\perp}/m = 1, p_3/m = 0, \phi = 1,\ {\rm and\ }m\tau=100$. 
}
\end{center}
\end{figure*}

Figure~\ref{fig2} shows the numerical results for the spin-imbalance $\Delta n_{\bm p}^{(-)}$ for small $\Omega/m=0.5$ as a function of the azimuthal angle $\theta_{\bm p}$ and the strength of the transverse perturbation ${\mathcal E}_{\perp}$.  The analytical results (\ref{eq16}) are also displayed as a comparison.

\begin{figure}[!t]
\includegraphics[clip, width=0.4\textwidth]{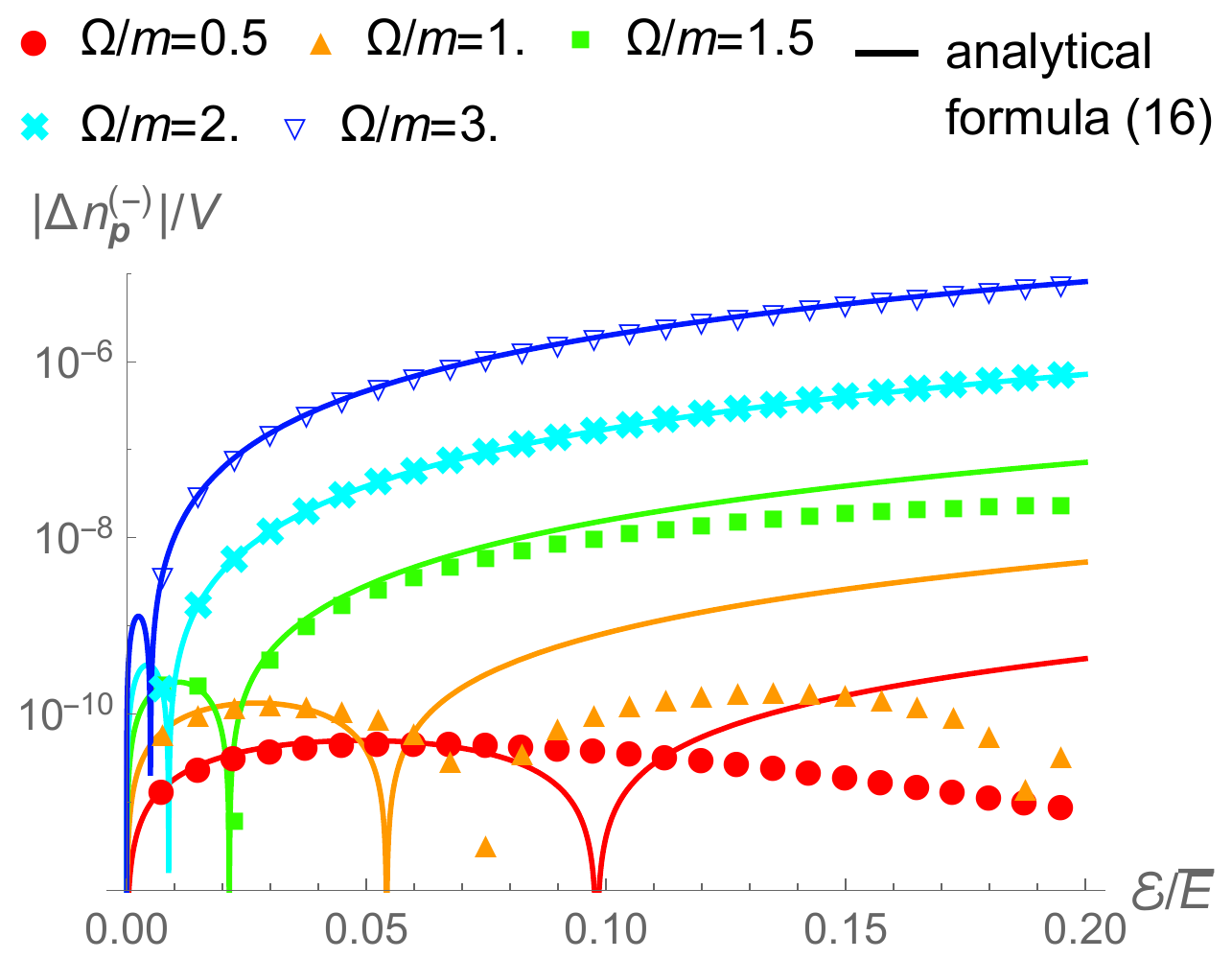}
\caption{\label{fig3} (color online) A comparison between the numerical results (points) and the analytical results (lines) for the spin-imbalance $\Delta n^{(-)}_{\bm p}$ as a function of the strength of the perturbation ${\mathcal E}_{\perp}$ for several values of the frequency $\Omega$.  The parameters are the same as in Fig.~\ref{fig2}, i.e., $e\bar{E}/m^2 = 0.4, {\mathcal E}_3/\bar{E} = 0, p_{\perp}/m = 1, p_3/m = 0, \phi = 1,\ {\rm and\ }m\tau=100$. }
\end{figure}

As explained, the formula (\ref{eq16}) is valid when the production is dominated by the one-photon process.  Multi-photon processes become important for (a) very small $\Omega$ far below the mass gap $\Omega \ll 2m$ because one photon is not energetic enough to excite a pair from the vacuum; and (b) large $|{\bm {\mathcal E}}|$ because the scattering amplitude involving $n$-photons is basically proportional to $|{\bm {\mathcal E}}|^n$.  Therefore, the formula (\ref{eq16}) starts to deviate from the numerical results not only for very small $\Omega$ (see Fig.~\ref{fig1}), but also for large $|{\bm {\mathcal E}}|$ (see Fig.~\ref{fig2}).  However, the disagreement for large $|{\bm {\mathcal E}}|$ occurs only for small $\Omega$ below the mass gap $\Omega \lesssim 2m$ provided that $|{\bm {\mathcal E}}|$ is subcritical $e|{\bm {\mathcal E}}|/m^2 \ll 1$.  This is clearly demonstrated in Fig.~\ref{fig3}, in which comparisons between the analytical and numerical results are made for several different values of $|{\bm {\mathcal E}}|$ and $\Omega$.  This is because one photon is enough to excite a pair from the vacuum for large $\Omega$ above the mass gap $\Omega \gtrsim 2m$, and multi-photon processes are strongly suppressed by $(e|{\bm {\mathcal E}}|/m^2)^n$.

The azimuthal angle $\theta_{\bm p}$-dependence is also affected by the size of ${\mathcal E}_{\perp}$ (see Fig.~\ref{fig2}).  For sufficiently small ${\mathcal E}_{\perp}$, the production is dominated by the one-photon process and is consistent with the formula (\ref{eq16}).  Therefore, the $\theta_{\bm p}$-dependence is controlled by the size of ${\mathcal E}_{\perp}p_{\perp}/\bar{E}\Omega$ as explained in Sec.~\ref{sec3b2}.   For large ${\mathcal E}_{\perp}$, multi-photons processes become important, and the $\theta_{\bm p}$-dependence deviates from the formula (\ref{eq16}).  The numerical results show that the multi-photon processes make the $\theta_{\bm p}$-dependence sharper, and strongly suppress collinear production around $\theta_{\bm p} - \theta_{\bm {\mathcal E}} \sim 0$ (see also the total production number in Fig.~\ref{fig7} in Appendix~\ref{appB}).  It is interesting to study these multi-photon effects in the azimuthal angle $\theta_{\bm p}$-dependence by extending our perturbative method.  We leave this for future work.

\subsubsection{$p_{\perp}$-dependence} \label{sec3b4}

\begin{figure*}[t!]
\begin{center}
\hspace*{-10mm}\includegraphics[clip, width=0.365\textwidth]{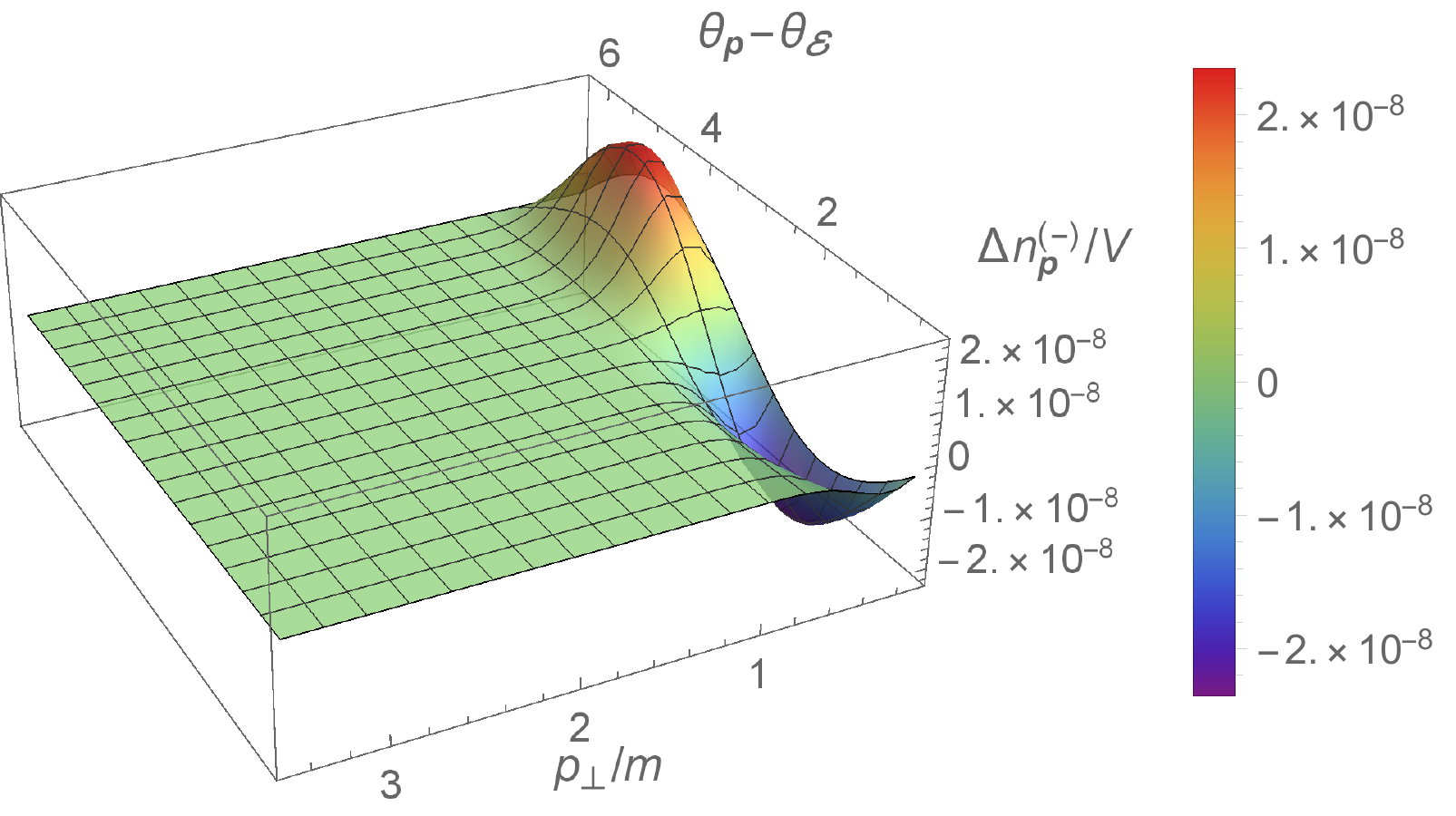}
\hspace*{-1mm}\includegraphics[clip, width=0.345\textwidth]{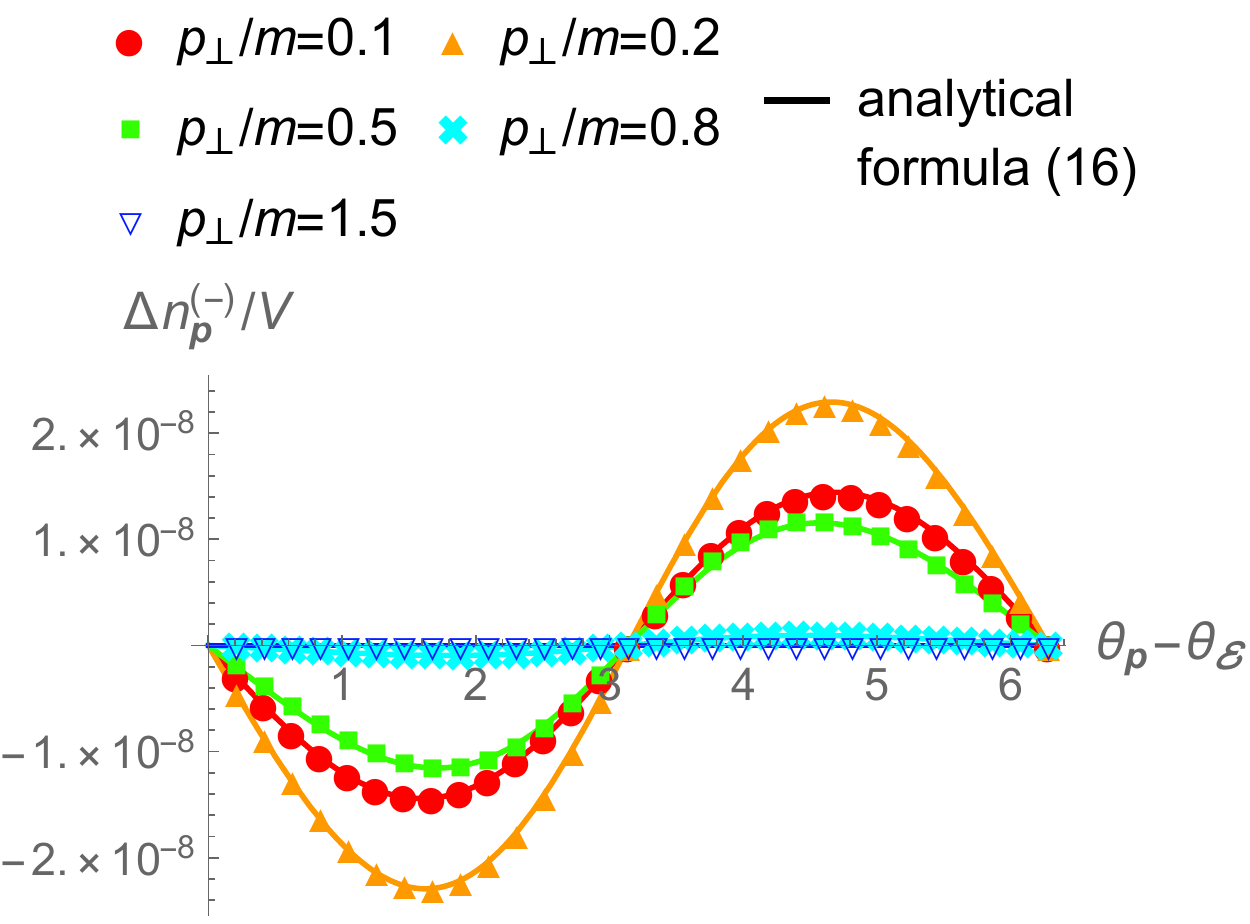}
\hspace*{-1mm}\includegraphics[clip, width=0.345\textwidth]{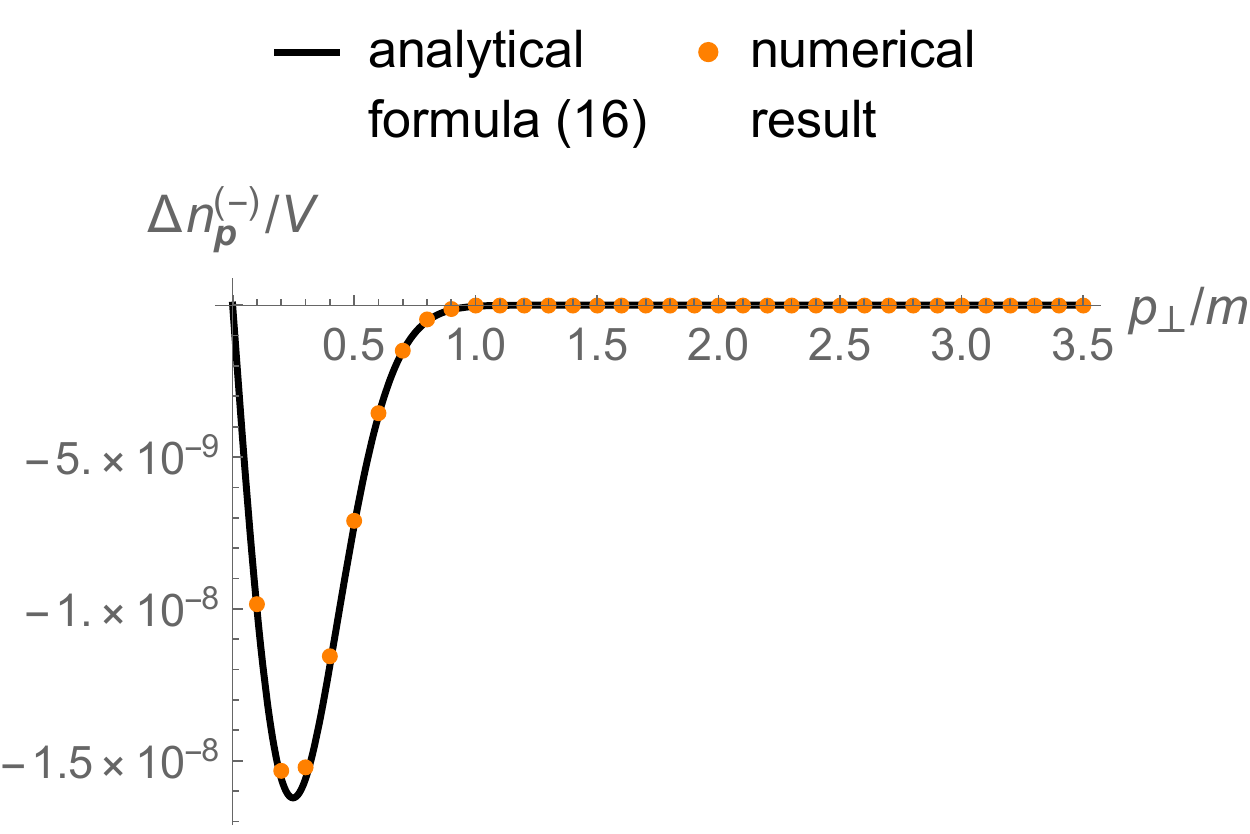}
\vspace*{-1mm}\hfill \\
\mbox{(i) Small frequency $\Omega/m =0.5 $ and weak perturbation ${\mathcal E}_{\perp}/\bar{E} = 0.025 $}\hfill \\
\vspace*{1mm}
\hspace*{-10mm}\includegraphics[clip, width=0.365\textwidth]{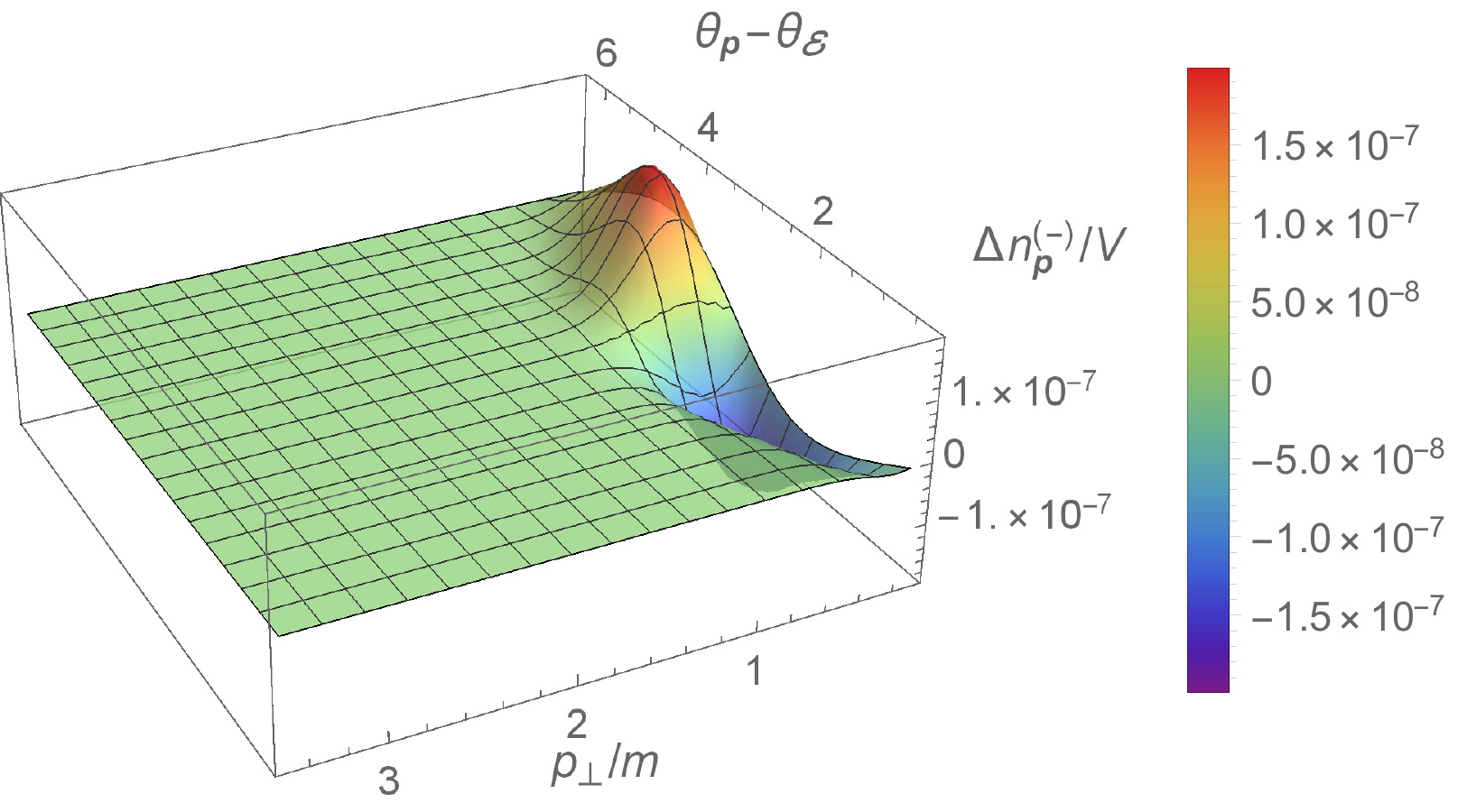}
\hspace*{-1mm}\includegraphics[clip, width=0.345\textwidth]{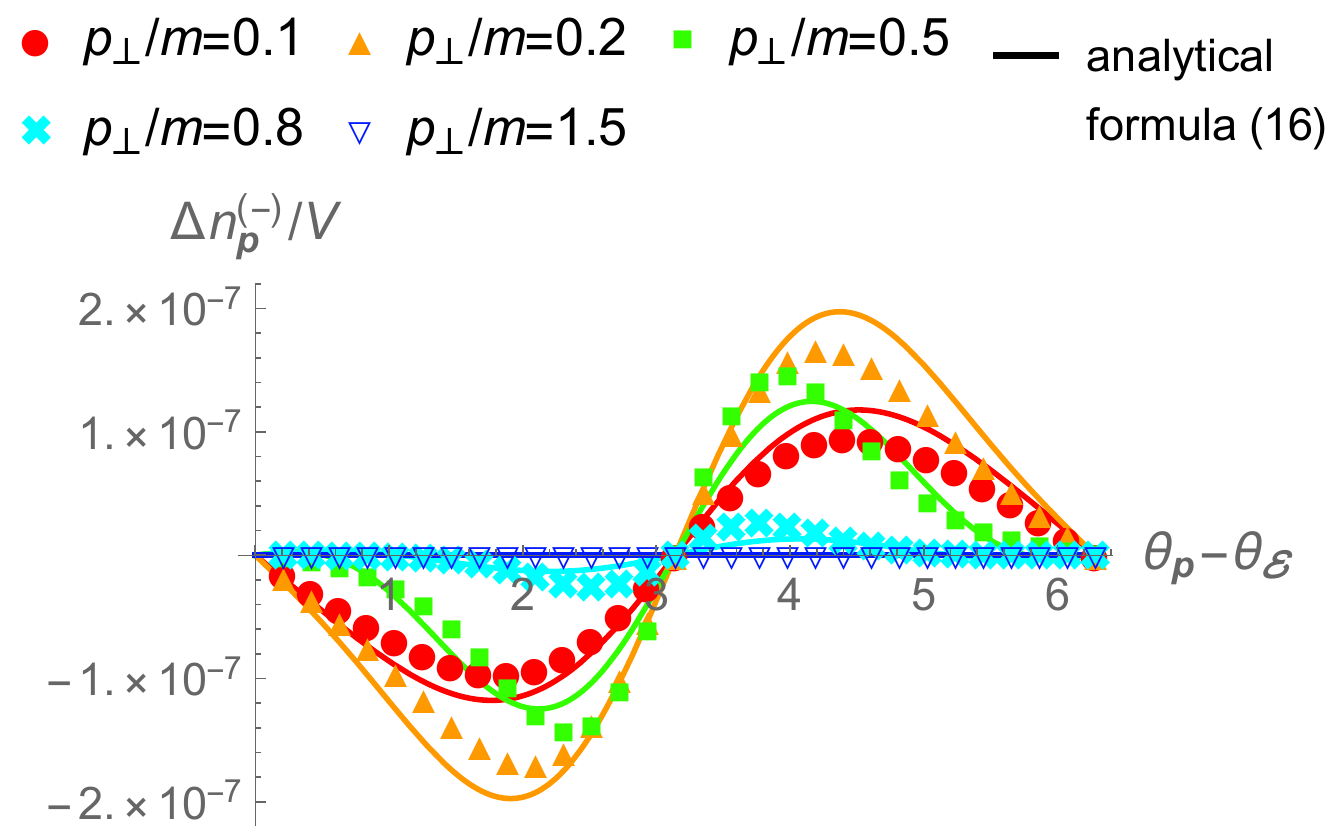}
\hspace*{-1mm}\includegraphics[clip, width=0.345\textwidth]{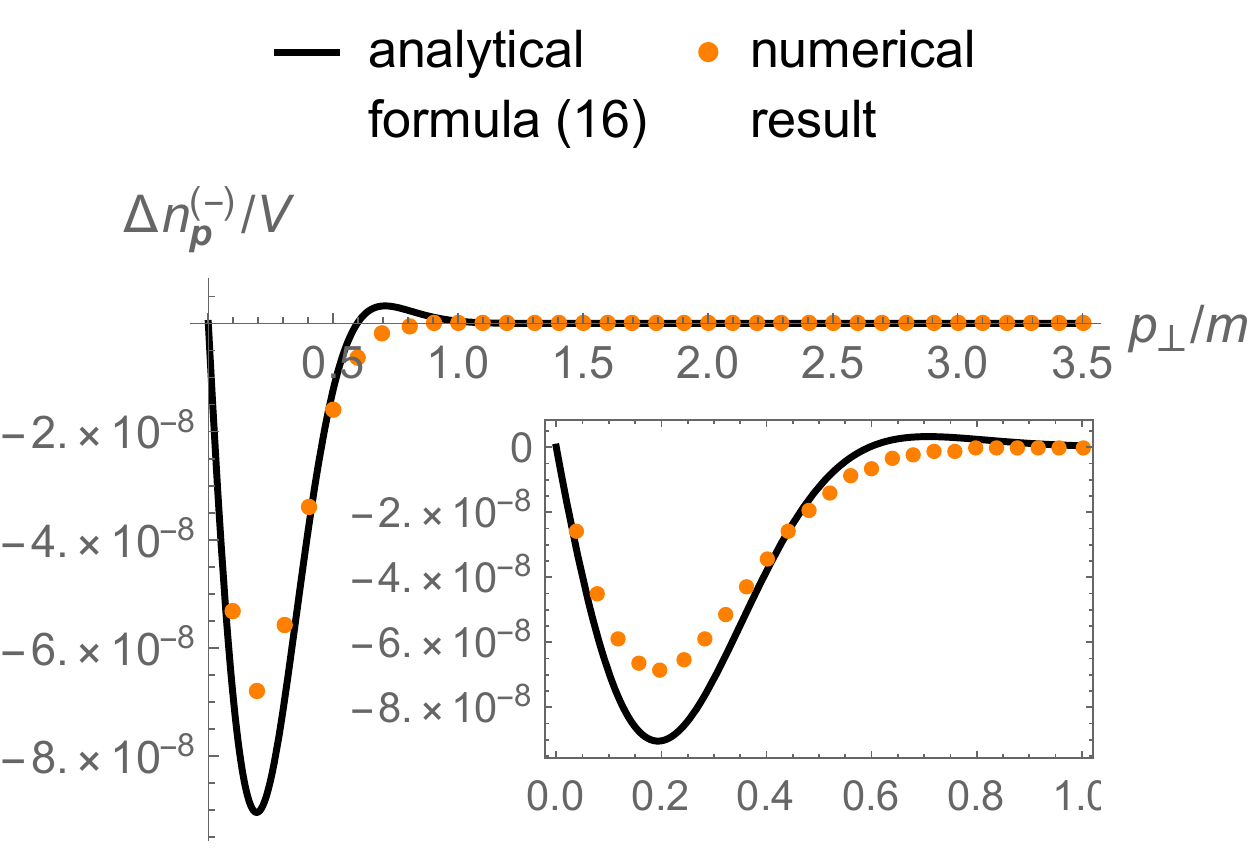}
\vspace*{-1mm} \hfill \\
\mbox{(ii) Small frequency $\Omega/m =0.5 $ and strong perturbation ${\mathcal E}_{\perp}/\bar{E} = 0.2 $}\hfill \\
\vspace*{1mm}
\hspace*{-10mm}\includegraphics[clip, width=0.365\textwidth]{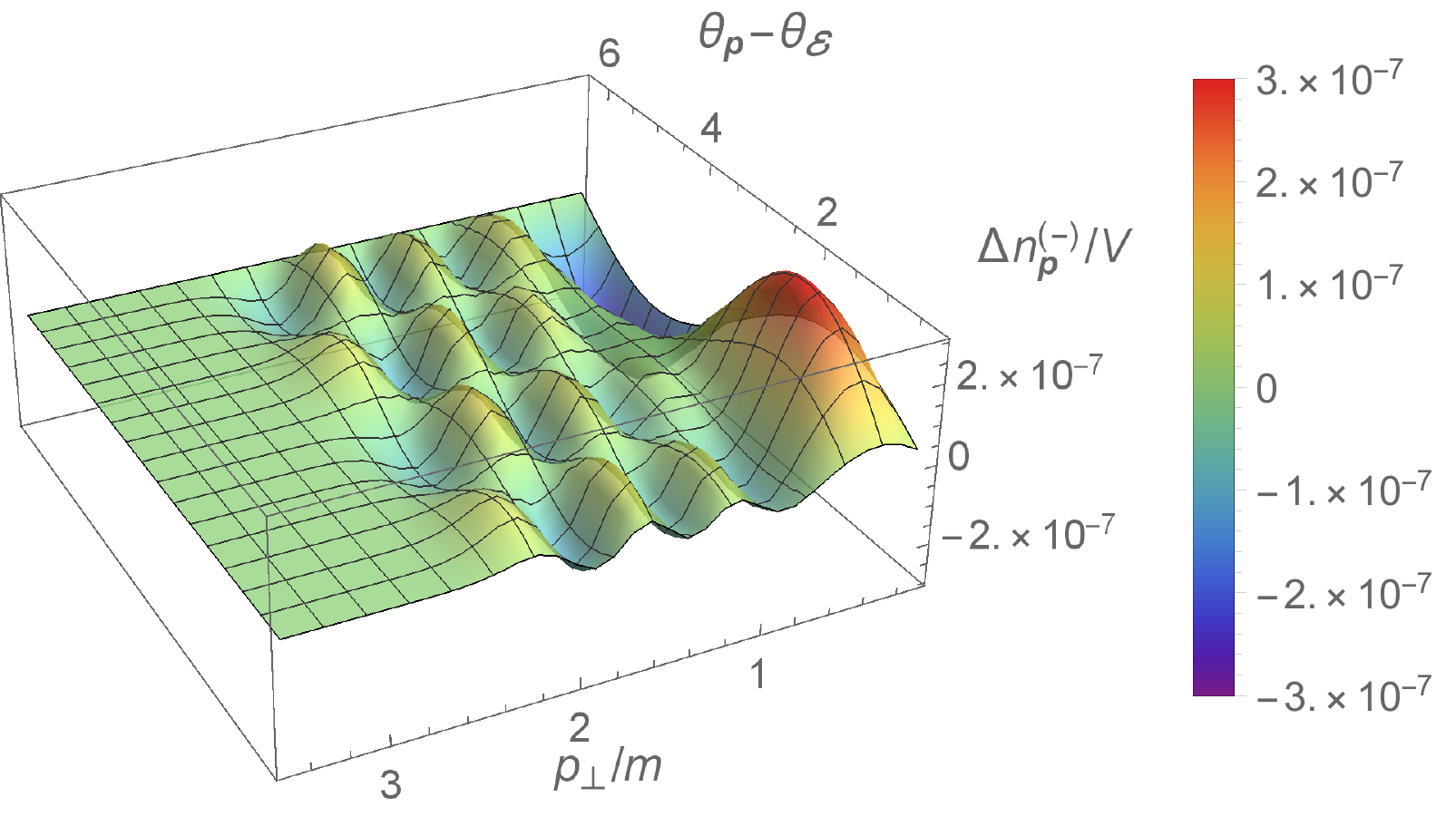}
\hspace*{-1mm}\includegraphics[clip, width=0.345\textwidth]{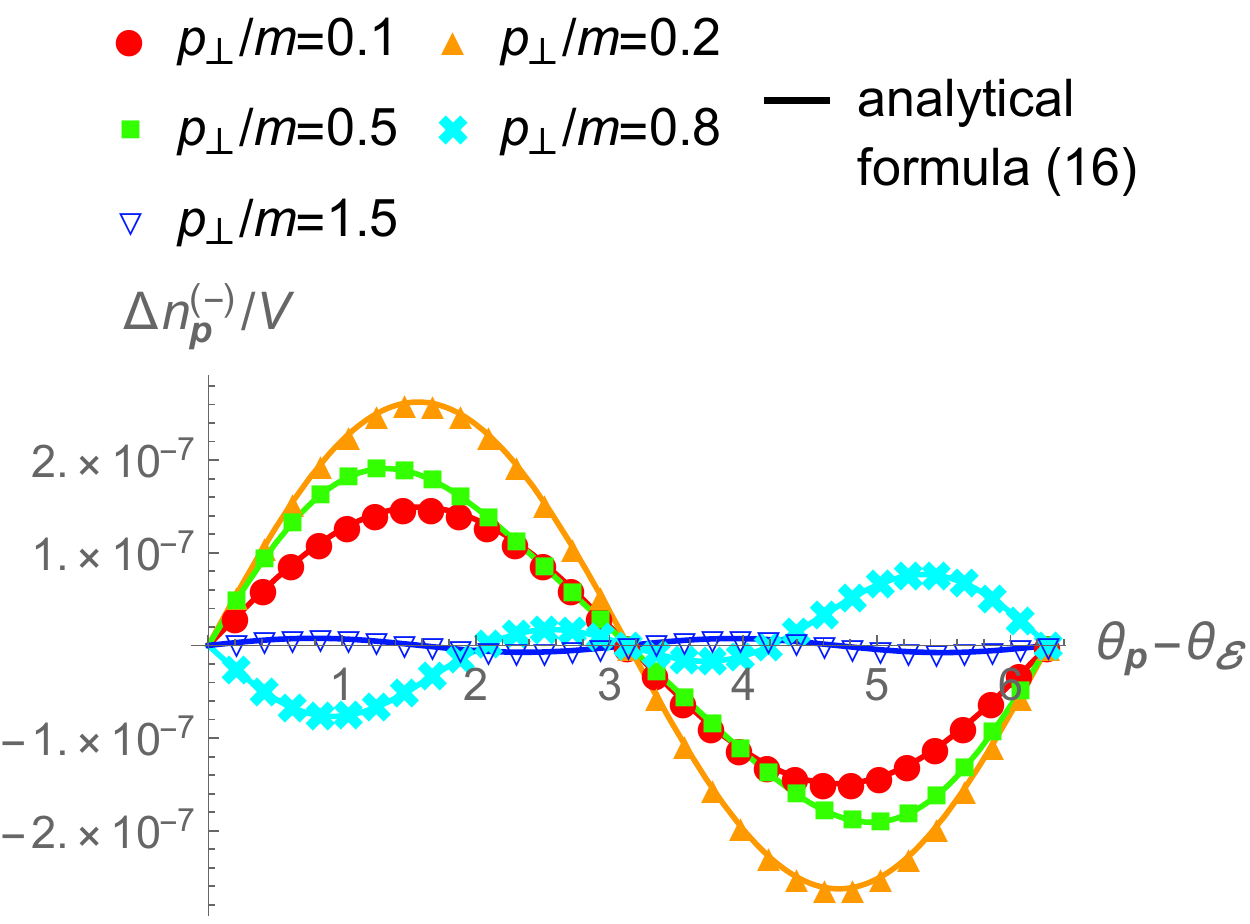}
\hspace*{-1mm}\includegraphics[clip, width=0.345\textwidth]{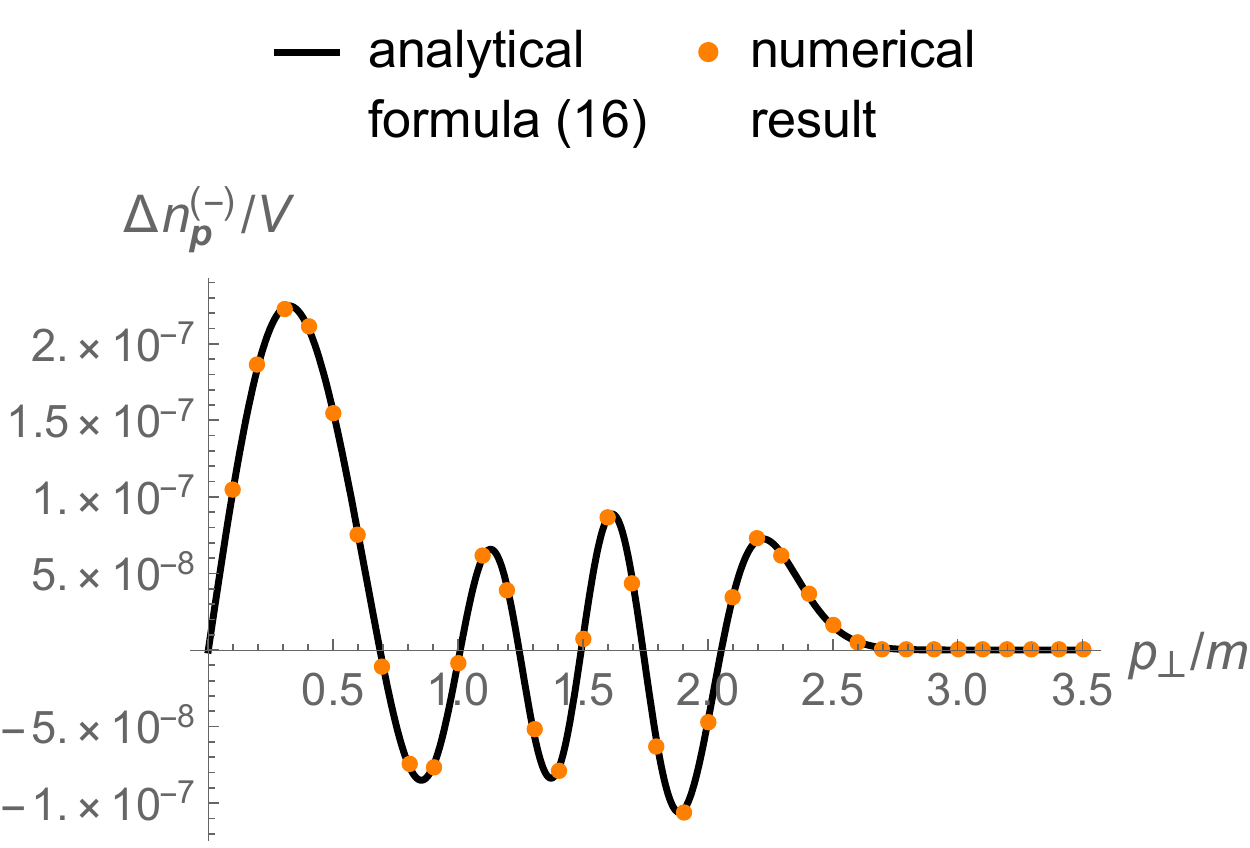}
\vspace*{-1mm} \hfill \\
\mbox{(iii) Large frequency $\Omega/m =5.0 $ and weak perturbation ${\mathcal E}_{\perp}/\bar{E} = 0.025 $}\hfill \\
\vspace*{1mm}
\hspace*{-10mm}\includegraphics[clip, width=0.365\textwidth]{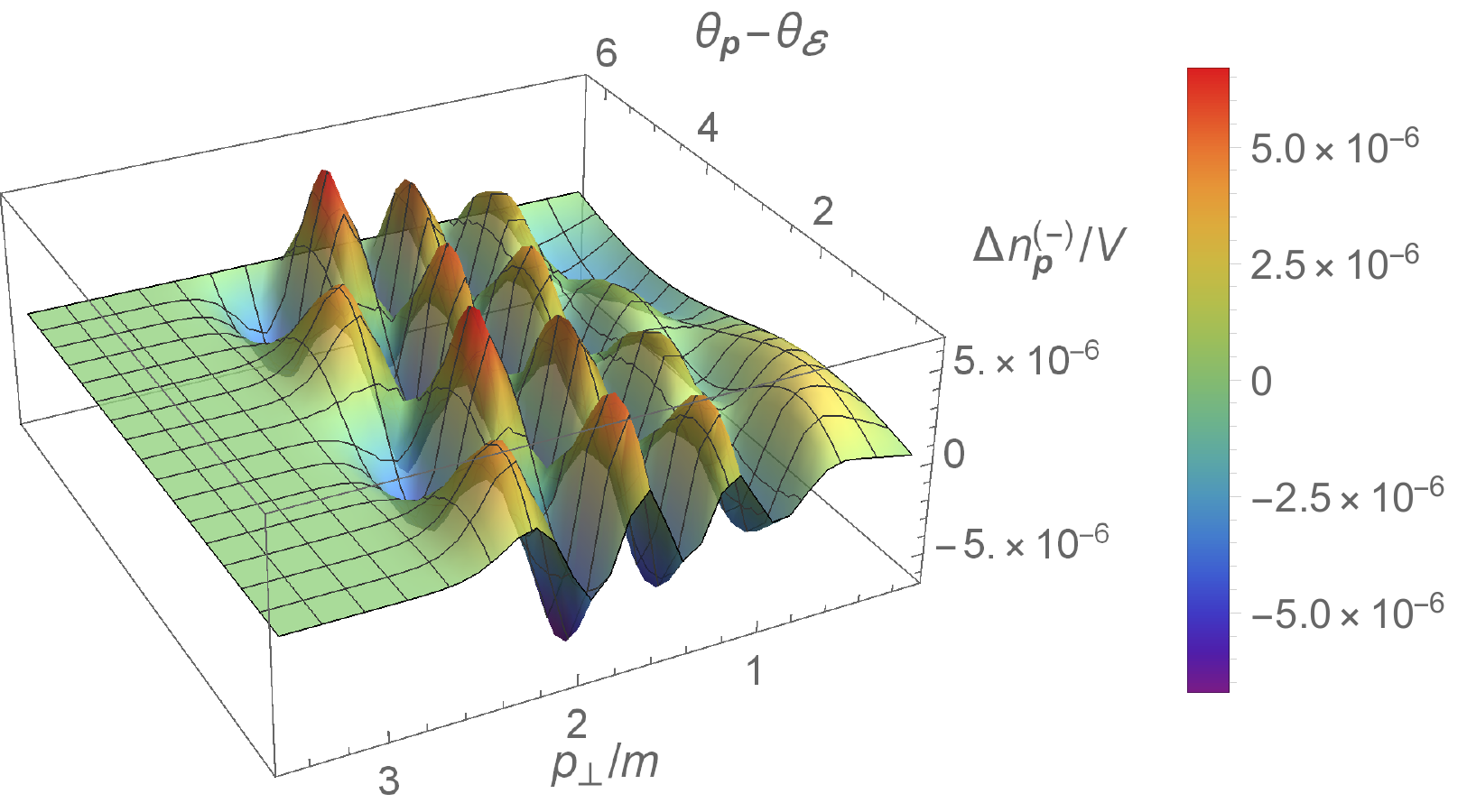}
\hspace*{-1mm}\includegraphics[clip, width=0.345\textwidth]{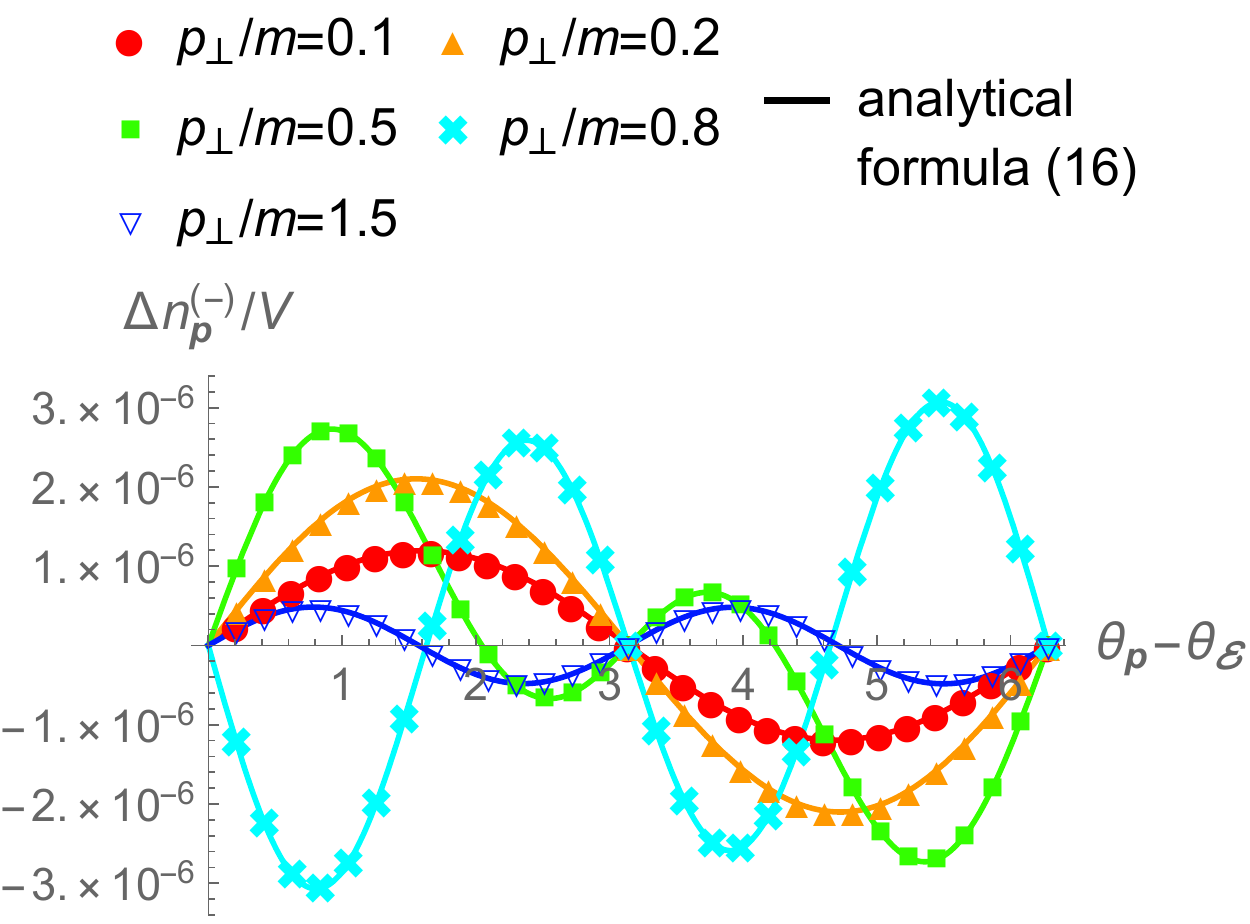}
\hspace*{-1mm}\includegraphics[clip, width=0.345\textwidth]{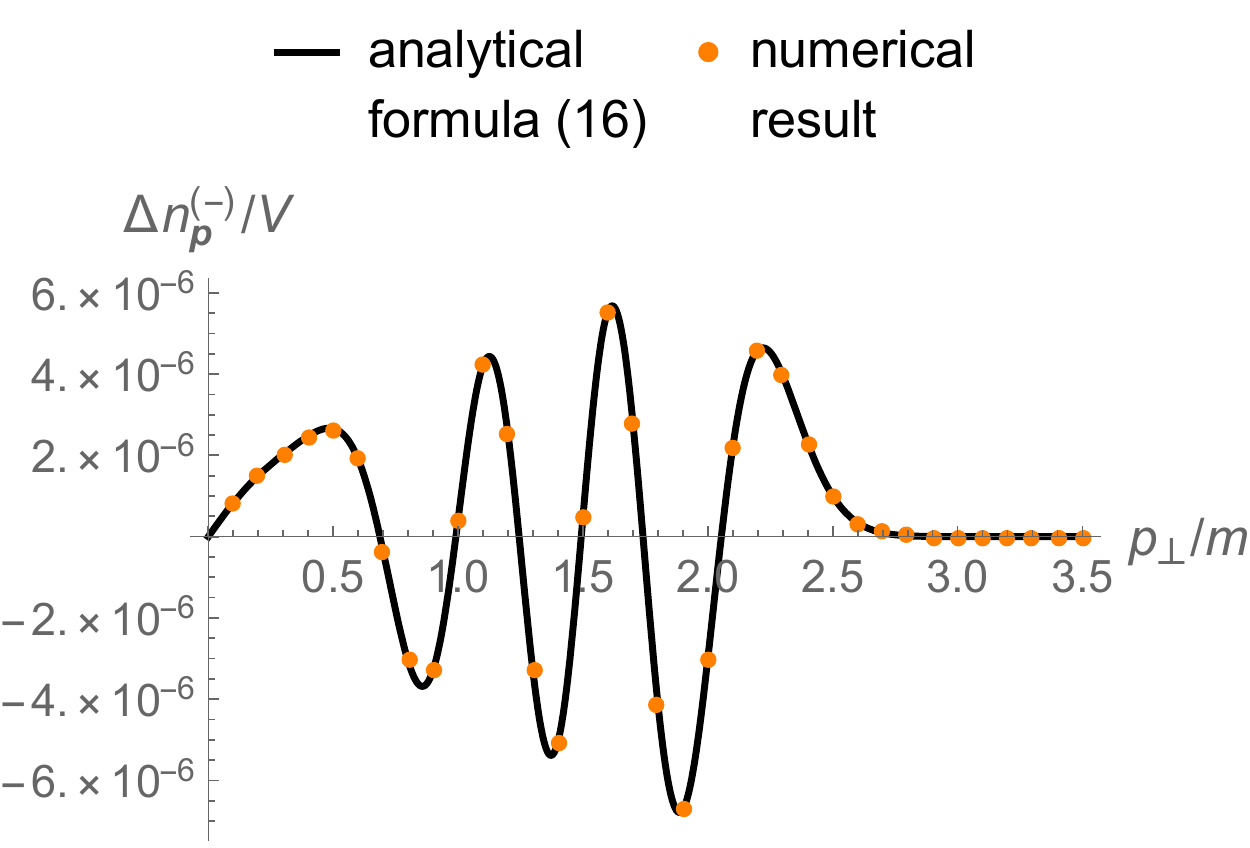}
\vspace*{-1mm} \hfill \\
\mbox{(iv) Large frequency $\Omega/m =5.0 $ and strong perturbation ${\mathcal E}_{\perp}/\bar{E} = 0.2 $}
\caption{\label{fig4} (color online) The numerical results of the spin-imbalance $\Delta n^{(-)}_{\bm p}$ as a function of $(\theta_{\bm p}, p_{\perp})$ (left); as a function of $\theta_{\bm p}$ for several values of $p_{\perp}$ (center); and as a function of $p_{\perp}$ for fixed $\theta_{\bm p}-\theta_{\bm {\mathcal E}} = \pi/4$ (right).  As a comparison, the analytical results (\ref{eq16}) are plotted as the lines in the center and right panels.  The upper (i) and lower (ii) panels distinguish the size of the frequency $\Omega$ and the perturbation ${\mathcal E}_{\perp}$.  The other parameters are the same as in Fig.~\ref{fig1}, i.e., $e\bar{E}/m^2 = 0.4, {\mathcal E}_3/\bar{E} = 0, p_3/m = 0, \phi = 1,\ {\rm and\ }m\tau=100$.   }
\end{center}
\end{figure*}

Figure~\ref{fig4} shows the spin-imbalance $\Delta n^{(-)}_{\bm p}$ as a function of the azimuthal angle $\theta_{\bm p}$ and the transverse momentum $p_{\perp}$ together with the analytical formula (\ref{eq16}) as a comparison.  Several different values of the frequency $\Omega$ and the field strength ${\mathcal E}_{\perp}$ are considered.  It is found that the formula (\ref{eq16}) gives an accurate description of the numerical results except for the case with small frequency and strong perturbation (i.e., panel (ii) of Fig.~\ref{fig4}), which is consistent with the previous discussions in Secs.~\ref{sec3b2} and \ref{sec3b3}.

For small $\Omega \lesssim \sqrt{e\bar{E}}, \sqrt{m^2 + p_{\perp}^2}$, the spin-imbalance is suppressed exponentially as $\Delta n^{(-)}_{\bm p} \propto \exp[-\pi (m^2+p_{\perp}^2)/e\bar{E}]$ (see Eq.~(\ref{eq17})).  Therefore, the spin-imbalance can be manifest only for small values of $p_{\perp} \lesssim \sqrt{e\bar{E}}$.  The $\theta_{\bm p}$-dependence  is determined by the strength of the effective magnetic field ${\mathcal E}_{\perp}p_{\perp}/\bar{E} \Omega$ (see Eq.~(\ref{eq17})).  Thus, the $\theta_{\bm p}$-dependence of the spin-imbalance changes by increasing $p_{\perp}$ from $\Delta n^{(-)}_{\bm p} \propto \sin (\theta_{\bm p} - \theta_{\bm a} )$ to $ \propto \sin (2(\theta_{\bm p} - \theta_{\bm a}) )$, but the latter is not manifest in Fig.~\ref{fig4} because of the exponential suppression.

For large $\Omega \gtrsim \sqrt{e\bar{E}}, \sqrt{m^2+p_{\perp}^2}$, the spin-imbalance is not suppressed exponentially until $p_{\perp} \lesssim \Omega$.  The spin-imbalance is free from the exponential suppression because the perturbative term (i.e., the second term) in Eq.~(\ref{eq18}) dominates the production.  The dominance of the perturbative term results in that the $\theta_{\bm p}$-dependence $\Delta n^{(-)}_{\bm p} \propto \sin (2(\theta_{\bm p} - \theta_{\bm a}) )$ at large $p_{\perp}$.  Also, the spin-imbalance exhibits an oscillating behavior in $p_{\perp}$, which is reminiscent of the Franz-Keldysh oscillation \cite{tay19}.  The frequency of the oscillation is determined by the phase factor $\varphi$ (\ref{eq19}).

\subsubsection{$p_{3}$-dependence} \label{sec3b5}

\begin{figure*}[t!]
\begin{center}
\hspace*{-10mm}\includegraphics[clip, width=0.365\textwidth]{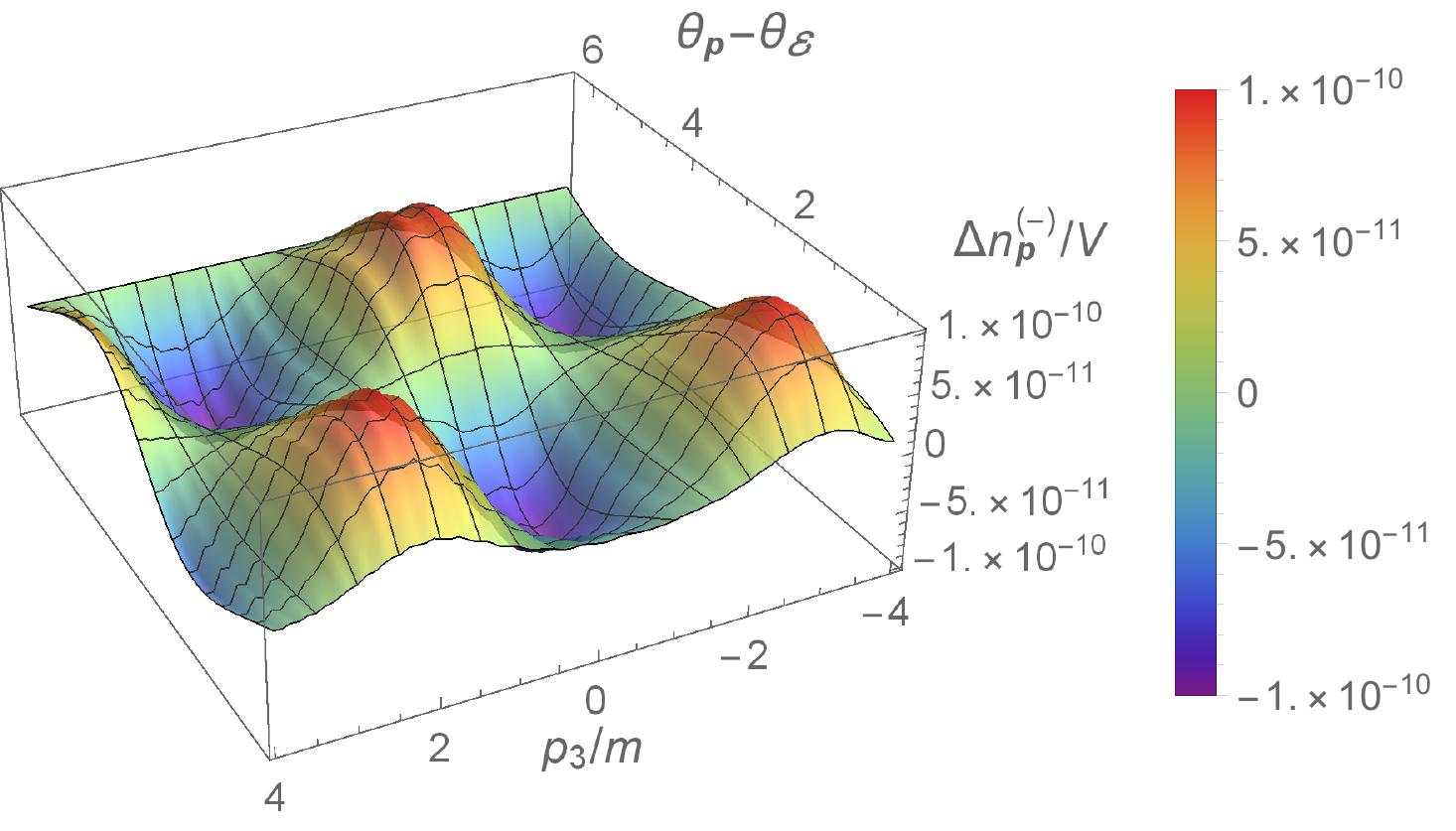}
\hspace*{-1mm}\includegraphics[clip, width=0.345\textwidth]{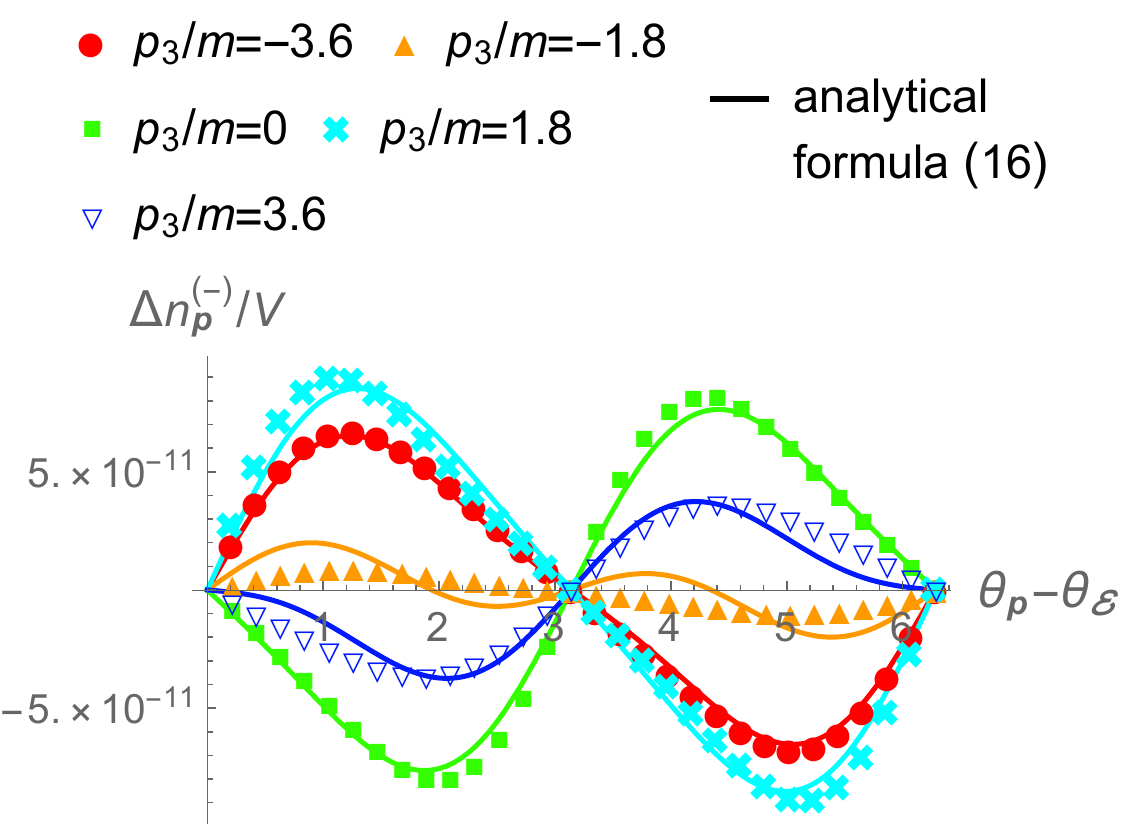}
\hspace*{-1mm}\includegraphics[clip, width=0.34\textwidth]{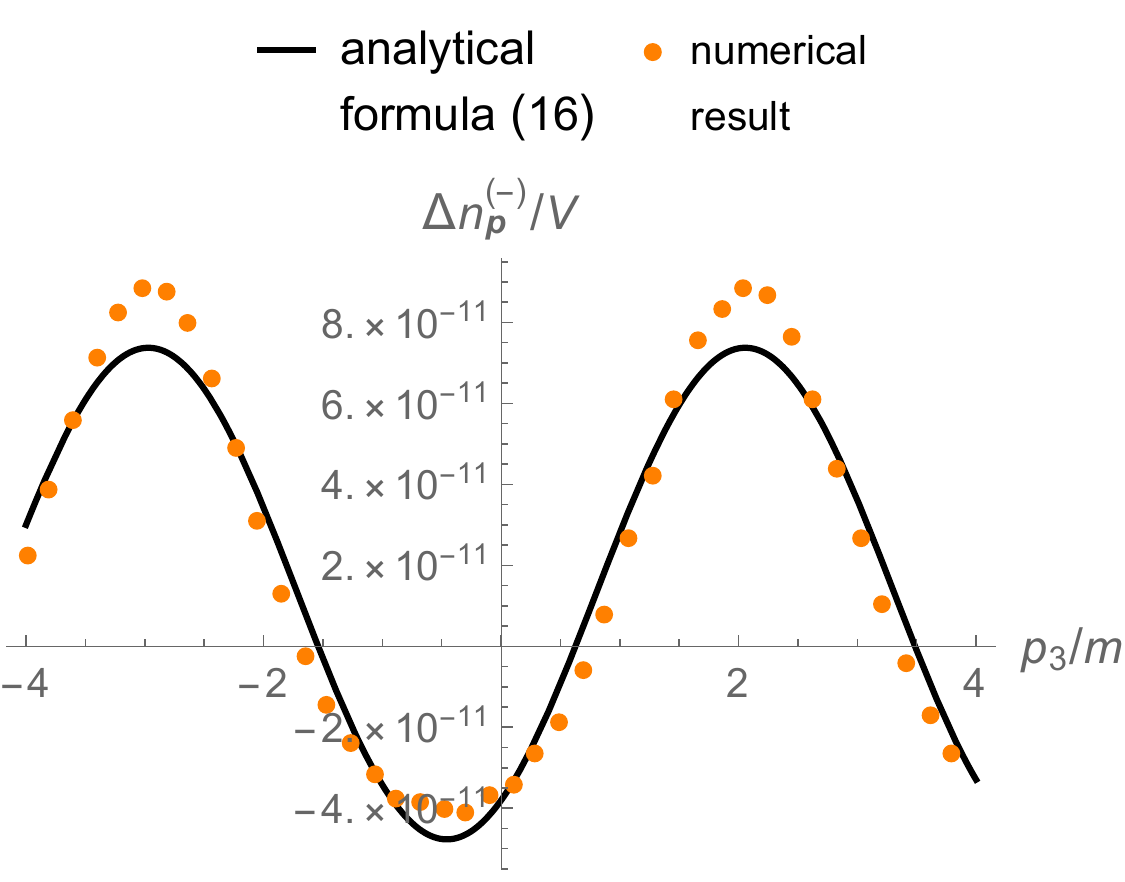}
\vspace*{-1mm}\hfill \\
\mbox{(i) Small frequency $\Omega/m =0.5 $ and weak perturbation ${\mathcal E}_{\perp}/\bar{E} = 0.025 $}\hfill \\
\vspace*{1mm}
\hspace*{-10mm}\includegraphics[clip, width=0.365\textwidth]{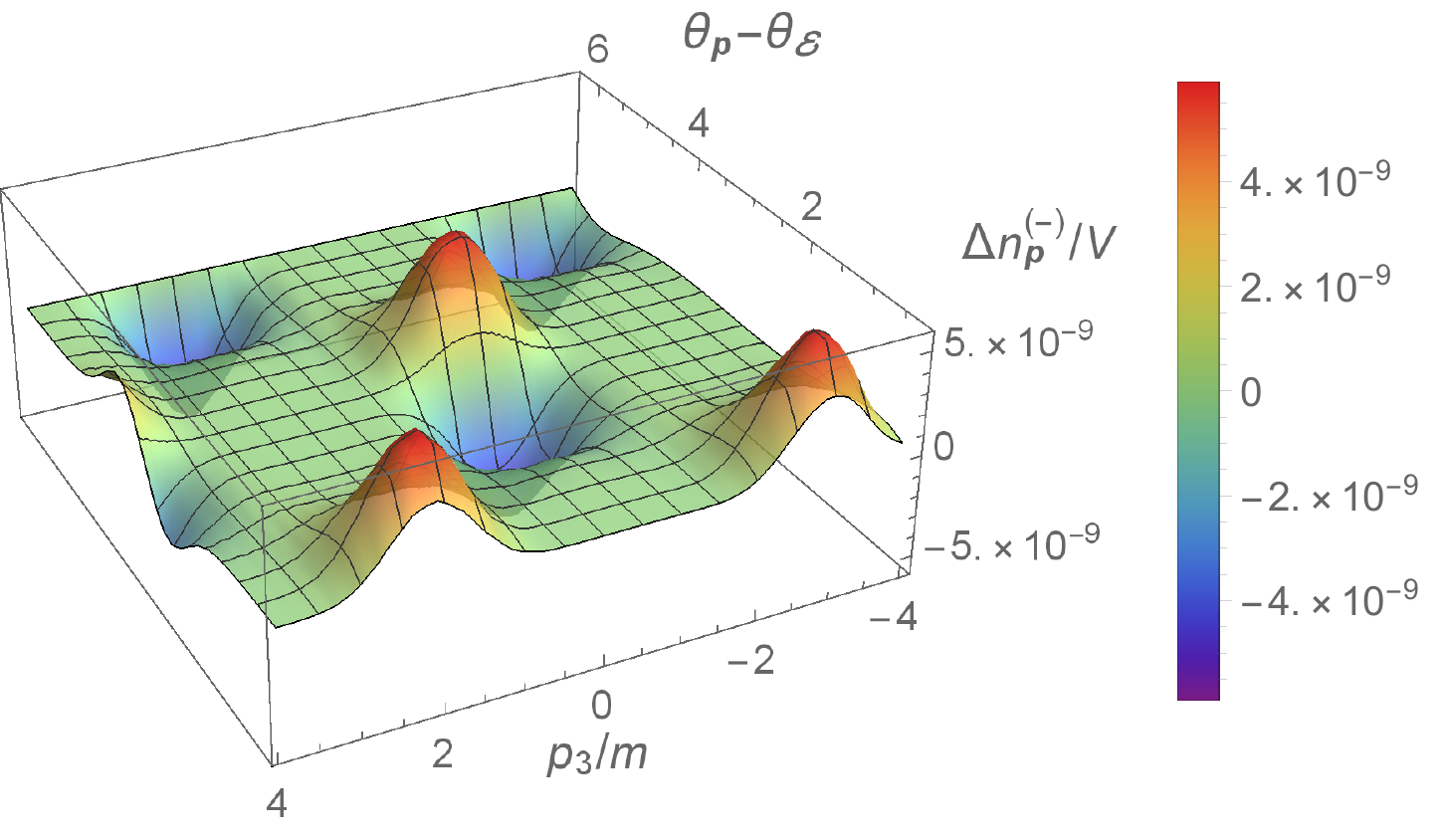}
\hspace*{-1mm}\includegraphics[clip, width=0.345\textwidth]{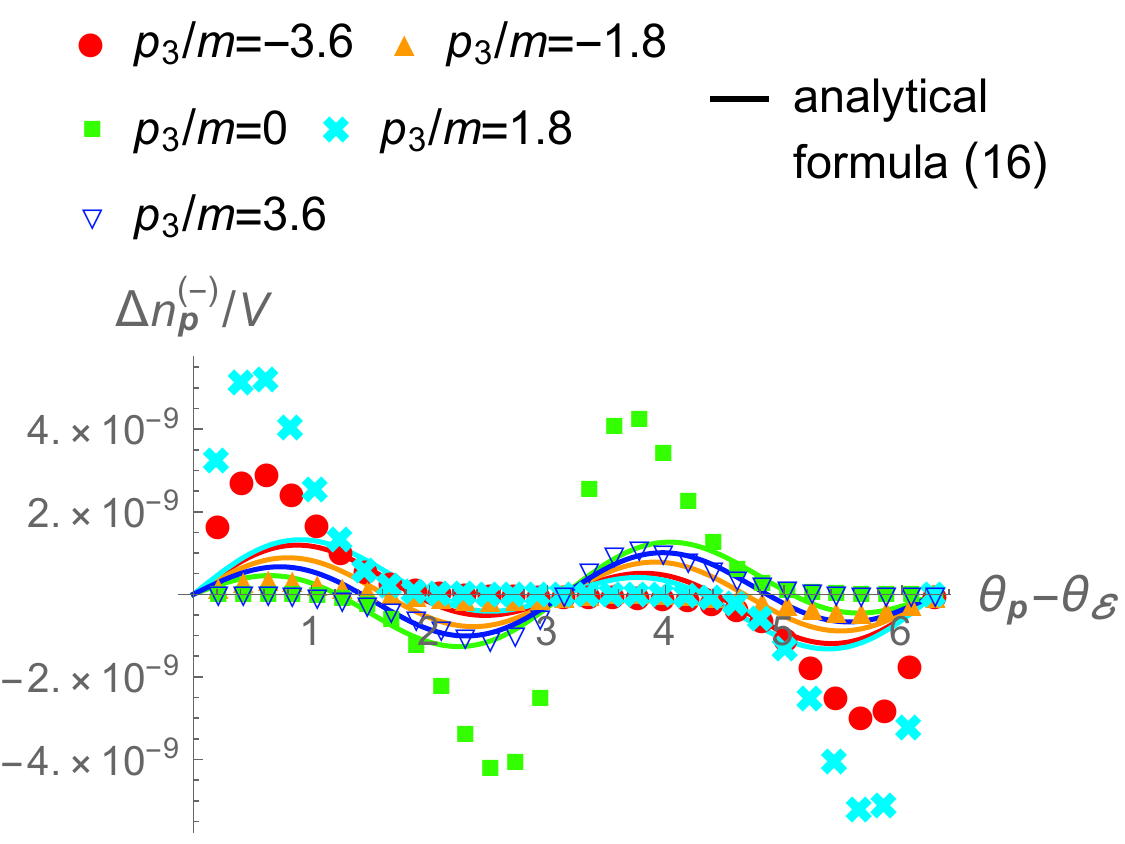}
\hspace*{-1mm}\includegraphics[clip, width=0.34\textwidth]{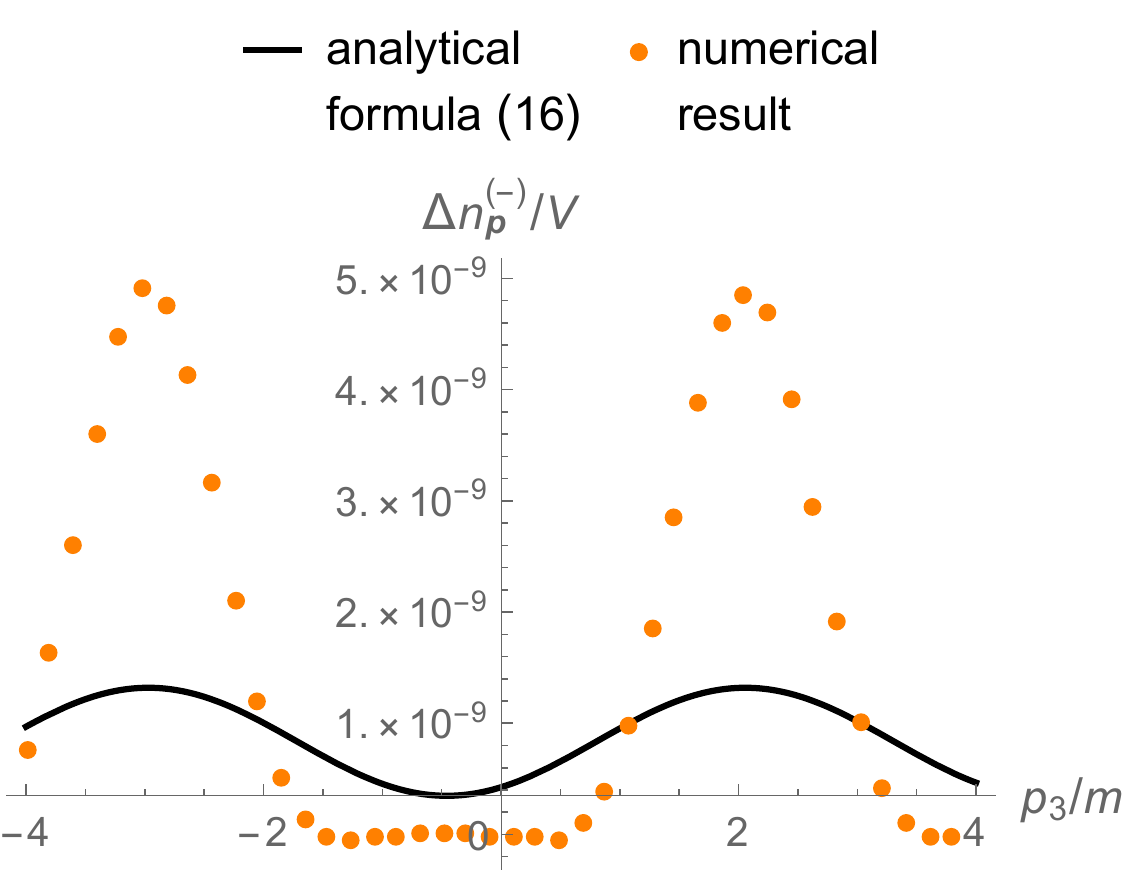}
\vspace*{-1mm} \hfill \\
\mbox{(ii) Small frequency $\Omega/m =0.5 $ and strong perturbation ${\mathcal E}_{\perp}/\bar{E} = 0.2 $}\hfill \\
\vspace*{1mm}
\hspace*{-10mm}\includegraphics[clip, width=0.365\textwidth]{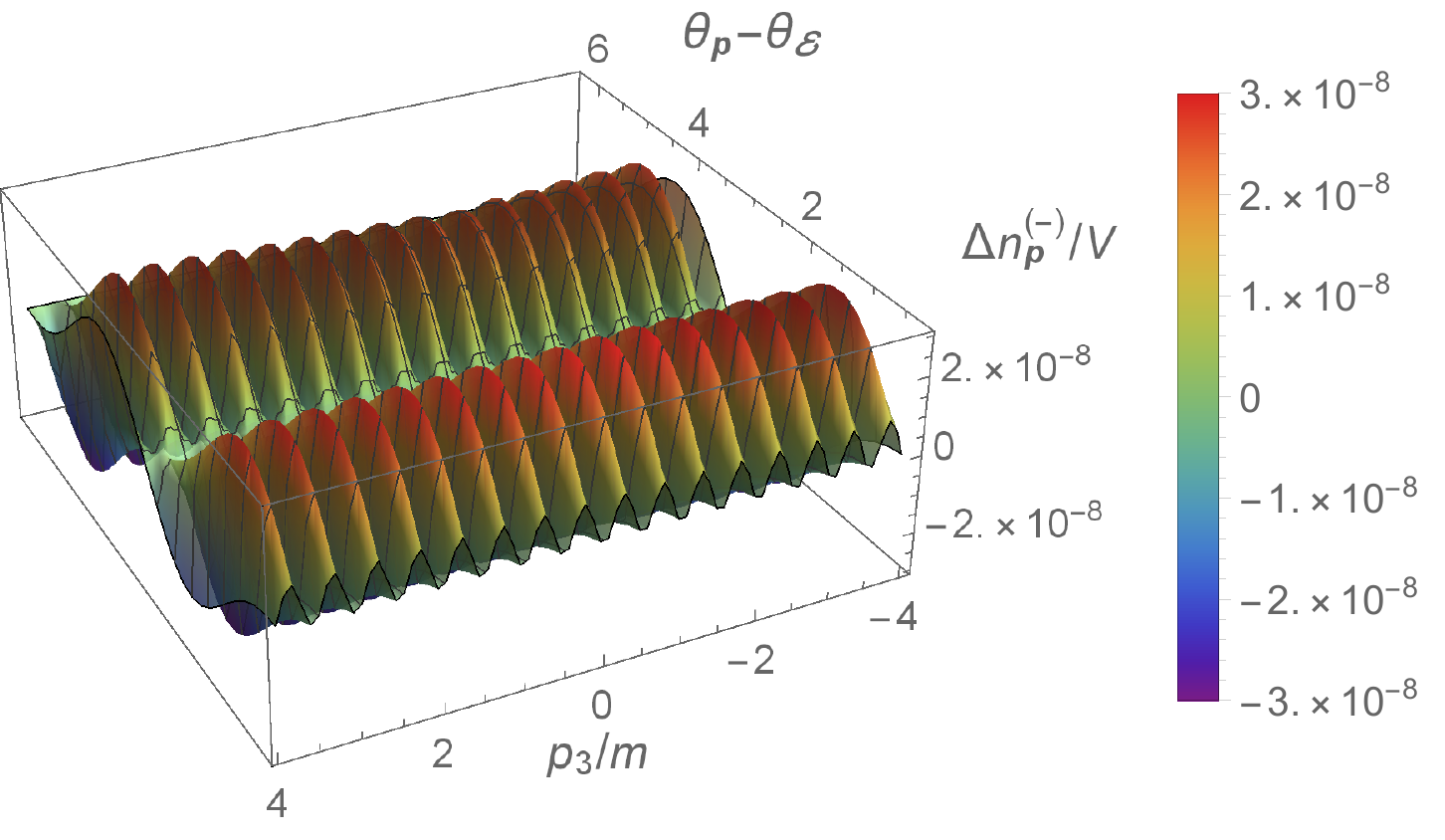}
\hspace*{-1mm}\includegraphics[clip, width=0.345\textwidth]{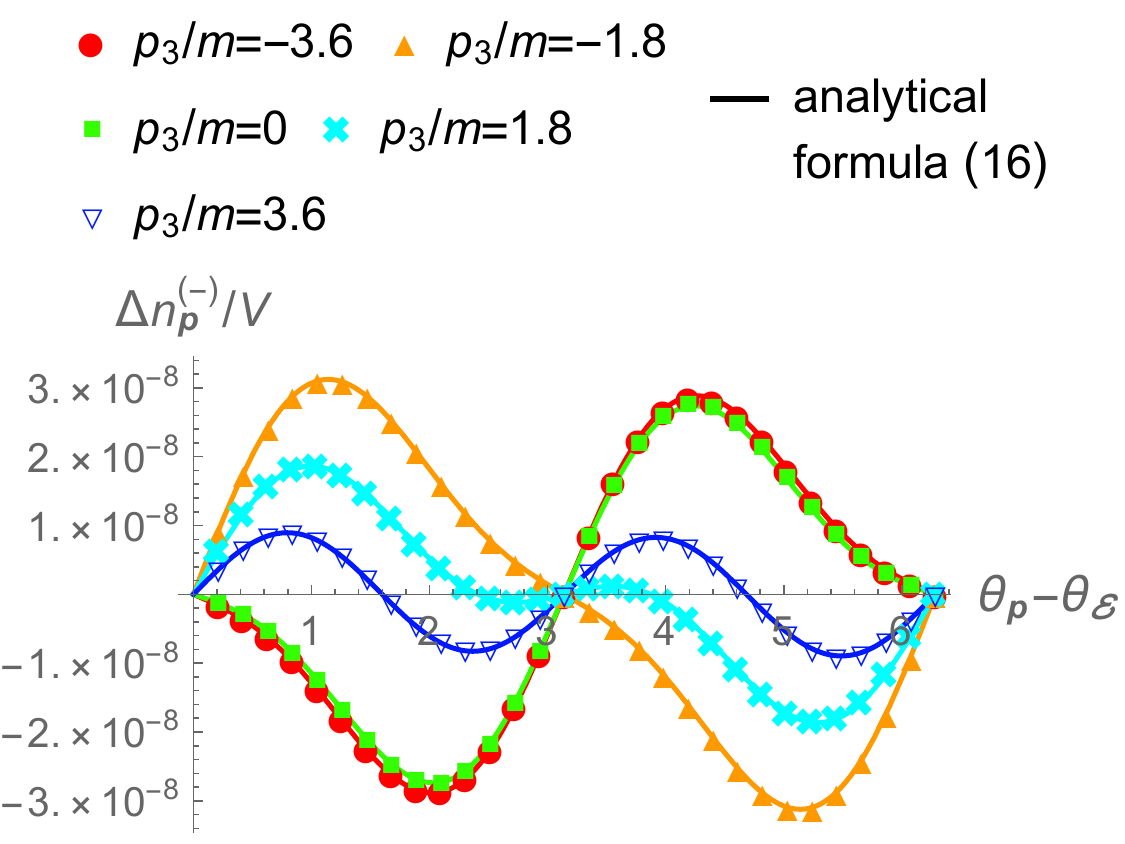}
\hspace*{-1mm}\includegraphics[clip, width=0.34\textwidth]{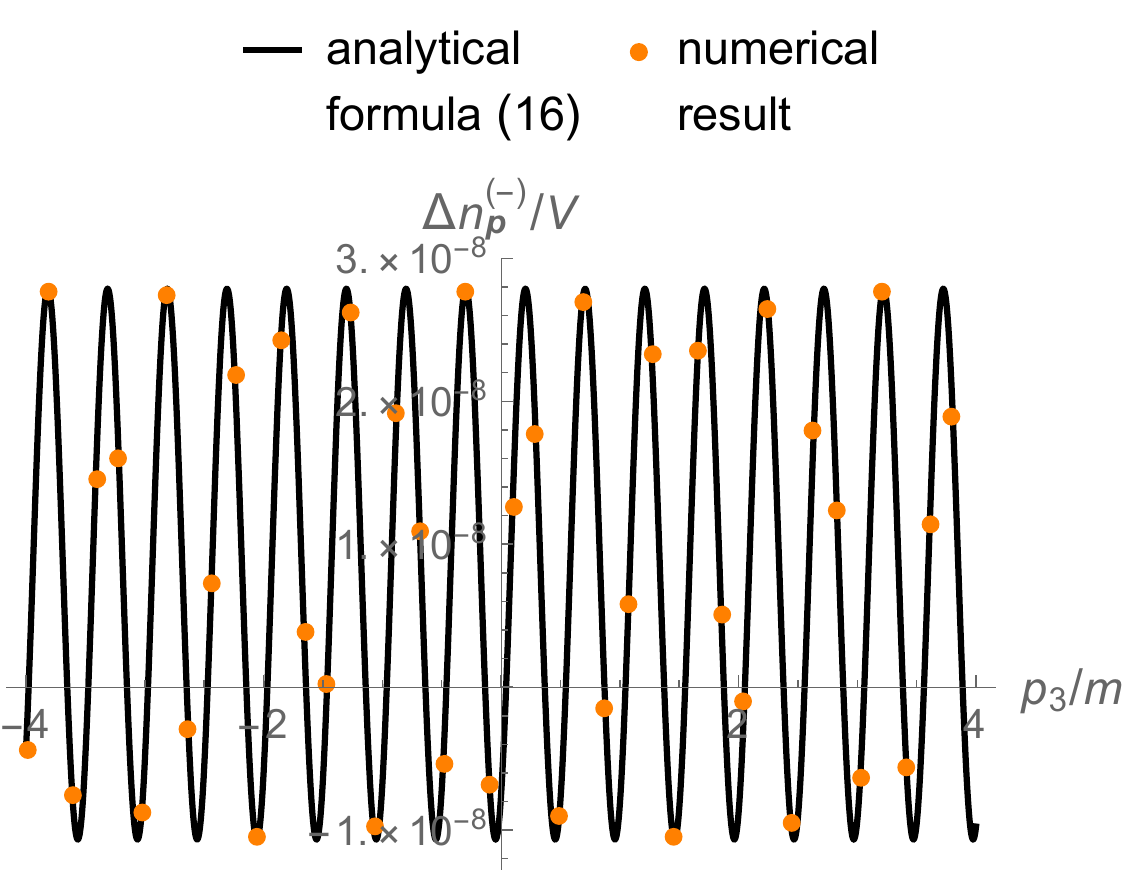}
\vspace*{-1mm} \hfill \\
\mbox{(iii) Large frequency $\Omega/m =5.0 $ and weak perturbation ${\mathcal E}_{\perp}/\bar{E} = 0.025 $}\hfill \\
\vspace*{1mm}
\hspace*{-10mm}\includegraphics[clip, width=0.365\textwidth]{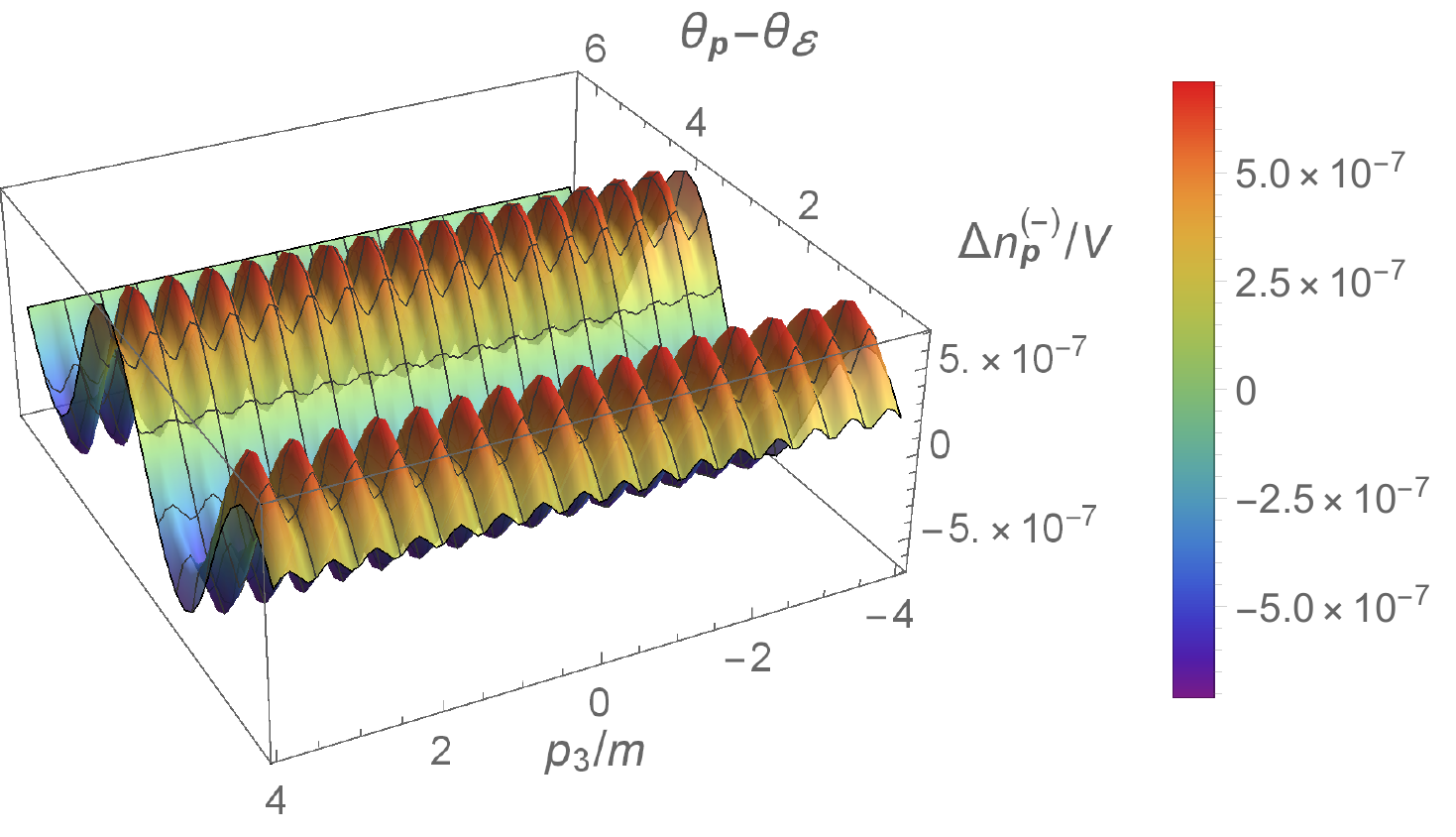}
\hspace*{-1mm}\includegraphics[clip, width=0.345\textwidth]{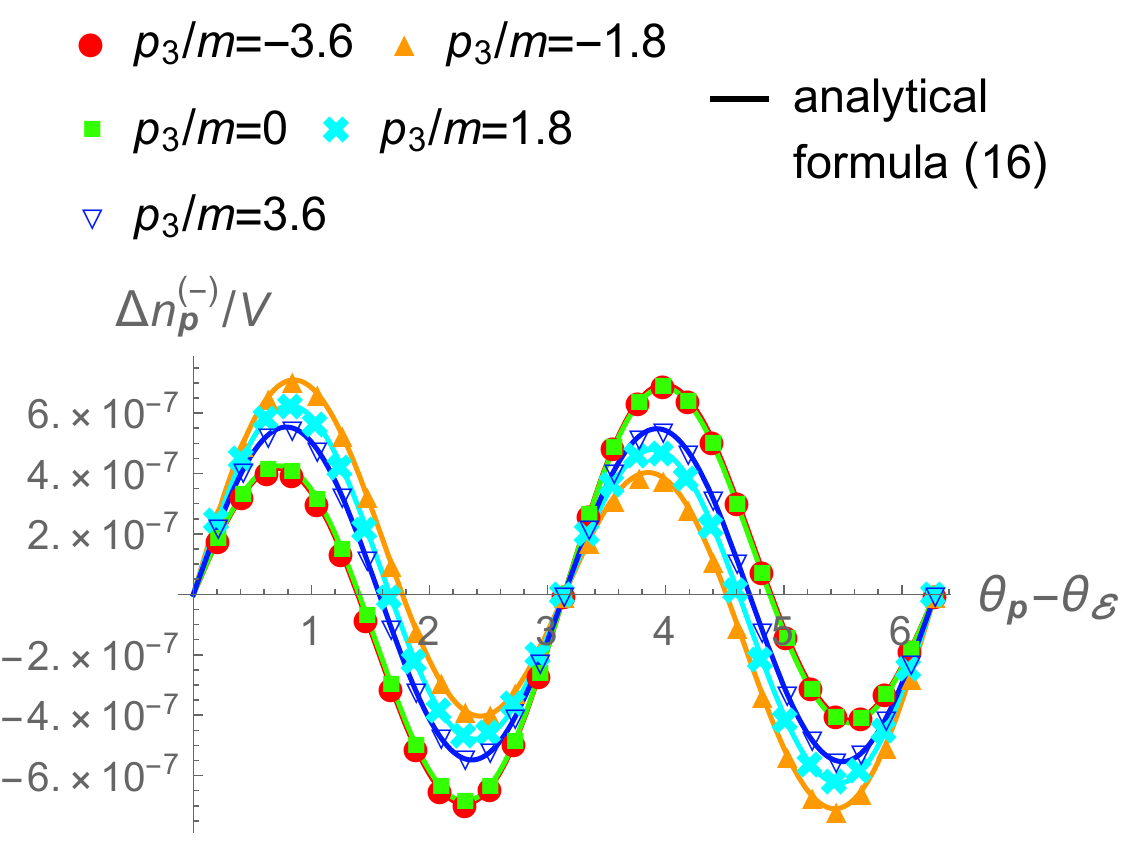}
\hspace*{-1mm}\includegraphics[clip, width=0.34\textwidth]{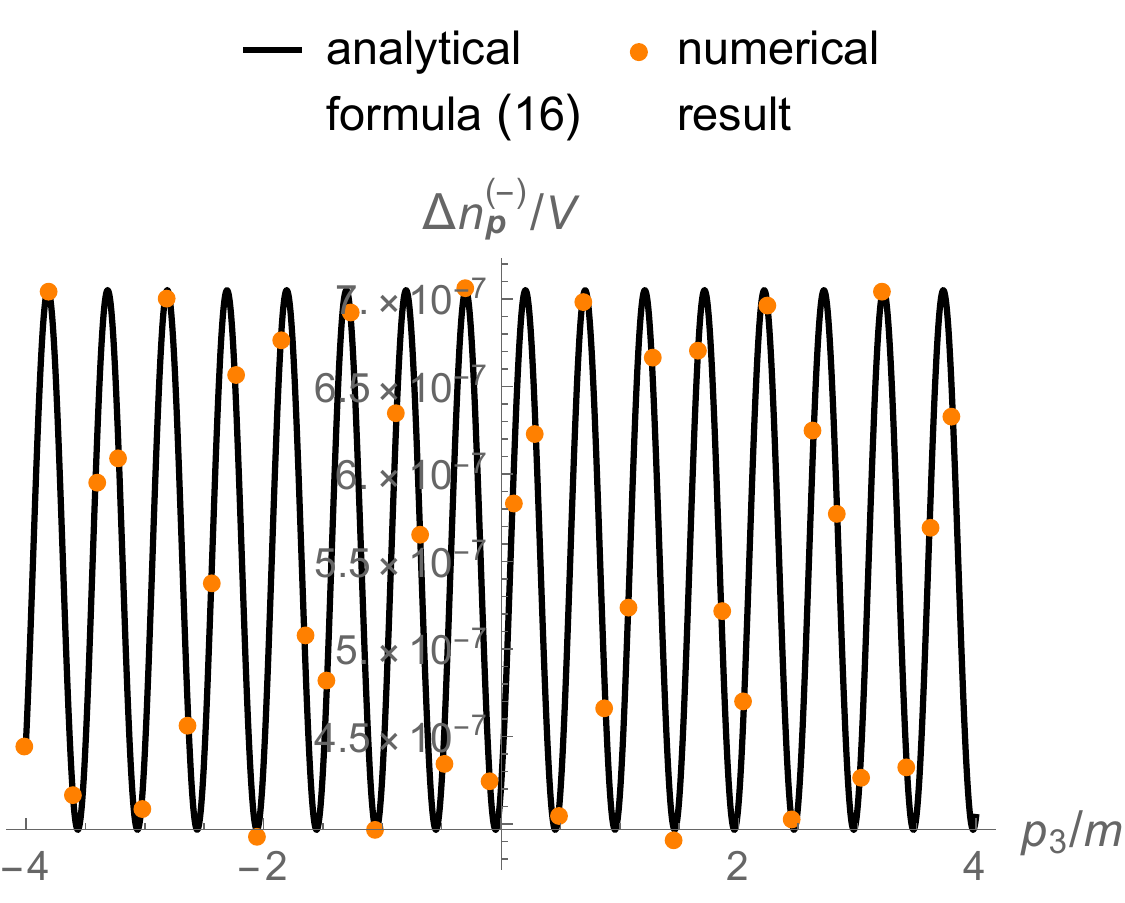}
\vspace*{-1mm} \hfill \\
\mbox{(iv) Large frequency $\Omega/m =5.0 $ and strong perturbation ${\mathcal E}_{\perp}/\bar{E} = 0.2 $}
\caption{\label{fig5} (color online) The numerical results of the spin-imbalance $\Delta n^{(-)}_{\bm p}$ as a function of $(\theta_{\bm p}, p_{3})$ (left); as a function of $\theta_{\bm p}$ for several values of $p_{3}$ (center); and as a function of $p_{3}$ for fixed $\theta_{\bm p}-\theta_{\bm {\mathcal E}} = \pi/4$ (right).  As a comparison, the analytical results (\ref{eq16}) are plotted as the lines in the center and right panels.  The upper (i) and lower (ii) panels distinguish the size of the frequency $\Omega$ and the perturbation ${\mathcal E}_{\perp}$.  The other parameters are the same as in Fig.~\ref{fig1}, i.e., $e\bar{E}/m^2 = 0.4, {\mathcal E}_3/\bar{E} = 0, p_{\perp}/m = 1, \phi = 1,\ {\rm and\ }m\tau=100$.  }
\end{center}
\end{figure*}

The canonical longitudinal momentum $p_3$-dependence is investigated in Fig.~\ref{fig5}.  It is found that the spin-imbalance $\Delta n_{\bm p}$ shows an oscillation in $p_3$, whose frequency is solely determined by $\Omega$ and is irrelevant to the other parameters.

This is consistent with the analytical formula (\ref{eq16}).  Intuitively, as explained in Sec.~\ref{sec2b}, in the presence of the strong slow field $\bar{E}$, $p_3$ is related to the production time  $t_{\rm prod} $ at which a pair is created as $t_{\rm prod} = - p_{3}/e\bar{E}$.  Therefore, reflecting the oscillating weak field configuration ${\mathcal A}$, the spin-imbalance $\Delta n_{{\bm p}}$ also exhibits an oscillating behavior in $p_3$ as ${\mathcal A}(t_{\rm prod}) \propto \sin( \phi - \Omega p_3 / e\bar{E}  )$.  Notice that the presence of the strong slow field $\bar{E}$ is essential for the oscillation.  Indeed, the purely perturbative formula without $\bar{E}$ (\ref{eq14}) is not oscillating in $p_3$ because there is no preferred time for the perturbative particle production and $t_{\rm prod}$ is independent of $p_3$.

\section{Summary and discussion} \label{sec4}

We discussed electron and positron pair production from the vacuum (the Schwinger mechanism) in the presence of a strong slow electric field superimposed by a weak monochromatic electric wave as a perturbation.  We argued, both analytically and numerically, that the Schwinger mechanism becomes spin-dependent if the perturbation is transverse with respect to the strong electric field.  Namely, in Sec.~\ref{sec2}, we derived an analytical formula for the production number and the spin-imbalance from the Schwinger mechanism with a weak perturbation pointing in an arbitrary direction based on a perturbation theory in the Furry picture.  In Sec.~\ref{sec3}, we solved the Dirac equation numerically to evaluate the spin-imbalance exactly.  We found excellent agreement between the numerical results and the analytical formula if the perturbation is sufficiently weak and/or is not very slow.  We also showed that the spin-imbalance (i) is strongly suppressed by an exponential of the critical field strength if the frequency of the perturbation is small, while it is suppressed only weakly by powers of the critical field strength if the frequency is large enough; and that it (ii) exhibits non-trivial oscillating behaviors in terms of the frequency, the azimuthal angle, and the momentum of produced particles.

Let us briefly discuss the implications of our results for laser experiments.  With current laser technology, it is difficult to create not only a very strong electric field but also a very high frequency field exceeding the electron mass scale \cite{pia12}.  Thus, Eq.~(\ref{eq17}) is appropriate for current lasers, and the spin-imbalance would be suppressed exponentially by the critical field strength.  Unfortunately, this implies that it is still difficult to observe the spin-imbalance with current laser technology.  However, it is possible to optimize the field configuration so as to maximize the production number \cite{koh13, heb14, lin15}.  In fact, our spin-dependent Schwinger mechanism is not limited to the monochromatic perturbation, but occurs as long as the perturbation has a transverse electric field, which is not a strong constraint for the optimization procedure.  Therefore, it is important and interesting for current laser experiments to explore the optimization problem of the spin-imbalance.  On the other hand, if a high frequency laser is realized in the future, the spin-imbalance is suppressed only weakly by powers of the critical field strength (see Eq.~(\ref{eq18})), so that it should be testable in experiments.  In particular, the non-trivial oscillating patterns in terms of the frequency of the perturbation, and the azimuthal angle and the transverse momentum of produced particles should serve as a fingerprint of our spin-dependent Schwinger mechanism.

Finally, we note that laser experiments are not the only place where strong fields can appear.  Strong fields naturally appear in many physical systems under extreme conditions (e.g., neutron stars, heavy-ion collisions, the early Universe).  For example, just after a collision of heavy ions at RHIC and the LHC, there appears a very strong (chromo-)electromagnetic field (sometimes called ``glasma''), which is as strong as ${\mathcal O}(1\;{\rm GeV})$ \cite{low75, nus75, kov95a, kov95b, lap06}.  On top of the glasma, there also exist energetic jets, whose momentum scale is ${\mathcal O}(1 {\rm -} 100\;{\rm GeV})$, coming from initial hard collisions.  It is widely recognized that the Schwinger mechanism due to glasma plays an essential role in the formation of quark-gluon plasma in heavy-ion collisions \cite{low75, nus75, gle83, kaj85, gat87, tan09, tay17}, but interaction effects due to the presence of jets are not well understood.  Our results suggest that the interaction effects have a significant impact on the Schwinger mechanism, and may induce longitudinal spin polarization of quarks. This might contribute to the longitudinal spin-polarization observed in heavy-ion collision experiments~\cite{bec2018,nii2019}.  Another example are the asymmetric heavy-ion collisions, e.g., Cu + Au collisions, in which a strong electric field can be generated with a strength much larger than the critical field strength~\cite{deng2015,hiro2014}. This strong electric field points from the Au nucleus to the Cu nucleus, and thus if it is perturbed by charged jets it will emit electron and positron pairs with spin-imbalance in the out-of-plane direction, which may be detected by measuring the spin polarization (along the impact-parameter direction) of the electron and positron moving in the direction perpendicular to the reaction plane.  We will examine these possibilities in the future.

\section*{Acknowledgments}

H.~T. would like to thank Koichi~Hattori, Mamoru~Matsuo, and the RIKEN iTHEMS STAMP working group for useful discussions.  H.~T. is supported by the National Natural Science Foundation of China (NSFC) under Grant No.~11847206. X.-G.~H. is supported by NSFC under Grants No.~11535012 and No.~11675041.

\appendix
\section{Evaluation of Eq.~(\ref{eq7})} \label{appA}

In order to evaluate the integrals in Eq.~(\ref{eq7}), we first note that the analytical expression for the mode function ${}_{\pm}\psi_{{\bm p},s}^{\rm as}$ for the constant electric field configuration (\ref{eq5}) is given by \cite{nik70, tan09}
\begin{subequations}
\begin{align}
	{}_{+} \psi_{{\bm p},s}^{\rm as}(x) \!&=\! \left[ A_{{\bm p}}^{\rm as}(x^0) + B_{{\bm p}}^{\rm as}(x^0) \gamma^0 \frac{ m + {\bm \gamma}_{\perp} \cdot {\bm p}_{\perp}}{\sqrt{m^2 + {\bm p}_{\perp}^2} } \right] \Gamma_s \frac{{\rm e}^{i{\bm p}\cdot {\bm x}}}{(2\pi)^{3/2}} , \label{eqa1a}\\
	{}_{-} \psi_{{\bm p},s}^{\rm as}(x) \!&=\!  \left[ B_{{\bm p}}^{{\rm as}*}(x^0) - A_{{\bm p}}^{{\rm as}*}(x^0) \gamma^0 \frac{ m + {\bm \gamma}_{\perp} \cdot {\bm p}_{\perp}}{\sqrt{m^2 + {\bm p}_{\perp}^2} } \right] \Gamma_s \frac{{\rm e}^{i{\bm p}\cdot {\bm x}}}{(2\pi)^{3/2}} ,  \label{eqa1b}
\end{align}
\end{subequations}
where ${\bm \gamma}_{\perp} = (\gamma^1, \gamma^2)$, ${\bm p}_{\perp} = (p_1,p_2)$ is the transverse momentum with respect to the direction of the electric field, and the scalar functions $ A_{{\bm p}}^{\rm as},  B_{{\bm p}}^{\rm as}$ are
\begin{subequations}
\begin{align}
	&\left\{\begin{array}{l}
		A^{\rm in}_{{\bm p}} =  {\rm e}^{-\frac{i\pi}{8}} {\rm e}^{- \frac{\pi a_{\bm p}}{4}} \sqrt{a_{\bm p}} D_{i a_{\bm p} -1} \left(-{\rm e}^{-\frac{i\pi}{4}} \xi_{\bm p}(x^0) \right) \\
		B^{\rm in}_{{\bm p}} =  {\rm e}^{+\frac{i\pi}{8}} {\rm e}^{-\frac{\pi a_{\bm p}}{4}} D_{i a_{\bm p} } \left(-{\rm e}^{-\frac{i\pi}{4}}\xi_{\bm p}(x^0) \right)
	\end{array}\right., \label{eqa2a}\\
	&\left\{\begin{array}{l}
		A^{\rm out}_{{\bm p}} =  {\rm e}^{-\frac{i\pi}{8}} {\rm e}^{-\frac{\pi a_{\bm p}}{4}}  D_{-i a_{\bm p} } \left({\rm e}^{\frac{i\pi}{4}} \xi_{\bm p}(x^0) \right) \\
		B^{\rm out}_{{\bm p}} =  {\rm e}^{+\frac{i\pi}{8}} {\rm e}^{-\frac{\pi a_{\bm p}}{4}} \sqrt{a_{\bm p}}  D_{-i a_{\bm p} -1} \left({\rm e}^{\frac{i\pi}{4}}\xi_{\bm p}(x^0) \right)
	\end{array}\right. ,  \label{eqa2b}
\end{align}
\end{subequations}
with $D_{\nu}(z)$ being the parabolic cylinder function and
\begin{align}
	a_{\bm p} \equiv \frac{m^2 + {\bm p}_{\perp}^2}{2e\bar{E}},\ \
	\xi_{\bm p} \equiv \sqrt{\frac{2}{e\bar{E}}}(e\bar{E}x^0 + p_{3}).
\end{align}
$\Gamma_s$ are two eigenvectors of $\gamma^0 \gamma^3$ with eigenvalue $+1$, and $s = \uparrow, \downarrow$ specifies the spin direction with respect to the $x^3$-axis.  In the Dirac representation, $\Gamma_s$ can be expressed as
\begin{align}
	\Gamma_{\uparrow} = \frac{1}{\sqrt{2}} \begin{pmatrix} 1 \\ 0 \\ 1 \\ 0 \end{pmatrix}, \
	\Gamma_{\downarrow} = \frac{1}{\sqrt{2}} \begin{pmatrix} 0 \\ 1 \\ 0 \\ -1 \end{pmatrix}.  \label{eqa4}
\end{align}

By plugging Eqs.~(\ref{eqa1a}) and (\ref{eqa1b}) into Eq.~(\ref{eq7}), one obtains
\begin{widetext}
\begin{align}
	n^{(\mp)}_{\pm{\bm p},s}
		&=	\frac{V}{(2\pi)^3}
			\left[
					\left|
							{\rm e}^{-\pi a_{\bm p}}
							- ie \int^{+\infty}_{-\infty} \frac{dk}{2\pi}
								\Biggl\{
										\frac{ p_1 \tilde{\mathcal A}_1 + p_2 \tilde{\mathcal A}_2 }{\sqrt{m^2 + {\bm p}_{\perp}^2}} \times  \int^{+\infty}_{-\infty} dx^0 {\rm e}^{+ikx^0} \left[ - A_{\bm p}^{{\rm out}*} A_{\bm p}^{{\rm in}*} + B_{\bm p}^{{\rm out}*} B_{\bm p}^{{\rm in}*} \right] \right. \right. \nonumber\\
										&\quad\quad\quad\quad\quad\quad\quad\quad\quad\quad\quad\quad\quad\quad + i\sigma \frac{ - p_2 \tilde{\mathcal A}_1 + p_1 \tilde{\mathcal A}_2 }{\sqrt{m^2 + {\bm p}_{\perp}^2}} \times  \int^{+\infty}_{-\infty} dx^0 {\rm e}^{+ikx^0} \left[ A_{\bm p}^{{\rm out}*} A_{\bm p}^{{\rm in}*} + B_{\bm p}^{{\rm out}*} B_{\bm p}^{{\rm in}*} \right] \nonumber\\
										&\quad\quad\quad\quad\quad\quad\quad\quad\quad\quad\quad\quad\quad\quad + \tilde{\mathcal A}_3 \times \int^{+\infty}_{-\infty} dx^0 {\rm e}^{+ikx^0} \left[ A_{\bm p}^{{\rm out}*} B_{\bm p}^{{\rm in}*} + B_{\bm p}^{{\rm out}*} A_{\bm p}^{{\rm in}*} \right] \left.
								\Biggl\}
					\right|^2 \nonumber\\
					&\quad\quad\quad\quad\quad + \left. \left| -ie \int^{+\infty}_{-\infty} \frac{dk}{2\pi} \left\{ \frac{m}{\sqrt{m^2+{\bm p}_{\perp}^2}} (\tilde{\mathcal A}_1 - i\sigma\tilde{\mathcal A}_2)  \times \int^{+\infty}_{-\infty} dx^0 {\rm e}^{+ikx^0} \left[ A_{\bm p}^{{\rm out}*} A_{\bm p}^{{\rm in}*} + B_{\bm p}^{{\rm out}*} B_{\bm p}^{{\rm in}*} \right] \right\} \right|^2
			\right], \label{eqa5}
\end{align}
where $\sigma = +1$ for $s = \uparrow$ and $-1$ for $s=\downarrow$ and $\tilde{\mathcal A}_{\mu}$ represents the Fourier transformation of ${\mathcal A}_{\mu}$,
\begin{align}
	\tilde{\mathcal A}_{\mu}(k) \equiv \int^{+\infty}_{-\infty} dx^0 {\rm e}^{-ikx^0} {\mathcal A}_{\mu}(x^0).
\end{align}
By using Eqs.~(\ref{eqa2a}) and (\ref{eqa2b}), one can rewrite the $x^0$-integrations in Eq.~(\ref{eqa5}) as
\begin{subequations}
\begin{align}
	 \int^{+\infty}_{-\infty} dx^0 {\rm e}^{+ikx^0} A_{\bm p}^{{\rm out}*} A_{\bm p}^{{\rm in}*}
		&= {\rm e}^{+i\pi/4} {\rm e}^{-\pi a_{\bm p}/2} \sqrt{a_{\bm p}} \times I_{a_{\bm p}-i, a_{\bm p}}  , \label{eqa7a} \\
	 \int^{+\infty}_{-\infty} dx^0 {\rm e}^{+ikx^0} A_{\bm p}^{{\rm out}*} B_{\bm p}^{{\rm in}*}
		&= {\rm e}^{-\pi a_{\bm p}/2} \times I_{a_{\bm p}, a_{\bm p}},\\
	 \int^{+\infty}_{-\infty} dx^0 {\rm e}^{+ikx^0} B_{\bm p}^{{\rm out}*} A_{\bm p}^{{\rm in}*}
		&= {\rm e}^{-\pi a_{\bm p}/2} a_{\bm p} \times I_{a_{\bm p}-i, a_{\bm p}-i},\\
	 \int^{+\infty}_{-\infty} dx^0 {\rm e}^{+ikx^0} B_{\bm p}^{{\rm out}*} B_{\bm p}^{{\rm in}*}
		&= {\rm e}^{-i\pi/4} {\rm e}^{-\pi a_{\bm p}/2} \sqrt{a_{\bm p}} \times I_{a_{\bm p}, a_{\bm p}-i}, \label{eqa7d}
\end{align}
\end{subequations}
where
\begin{align}
	I_{\lambda, \lambda'} \equiv \int^{+\infty}_{-\infty} dx^0 {\rm e}^{+ikx^0} D_{-i\lambda} \left( -{\rm e}^{+i\pi/4} \xi_{\bm p}  \right) \left[ D_{-i\lambda'} \left( {\rm e}^{+i\pi/4} \xi_{\bm p}  \right)  \right]^*.
\end{align}
Noting that the parabolic cylinder function $D_{\nu}$ has the integral expression \cite{gra15}
\begin{align}
	D_{\nu}( \pm {\rm e}^{i\pi/4} \xi ) = \frac{{\rm e}^{-i\xi^2/4}}{{\rm e}^{i\pi\nu}\Gamma(-\nu)} \int^{\infty}_0 dy y^{-\nu-1}{\rm e}^{\mp i\xi y - iy^2/2},
\end{align}
one can evaluate $I_{\lambda, \lambda'}$ analytically as
\begin{align}
	I_{\lambda, \lambda'}
		&= \pi \sqrt{\frac{2}{e\bar{E}}}{\rm e}^{-\pi (\lambda + \lambda^{'*})/4} \Theta(-k) \left( \frac{-k}{\sqrt{2e\bar{E}}}  \right)^{i(\lambda-\lambda^{'*})-1} \exp\left[ -i\left( \frac{k^2}{4e\bar{E}} + \frac{kp_3}{e\bar{E}}  \right) \right] {}_1\tilde{F}_1\left( -i\lambda^{'*}; -i(\lambda^{'*}-\lambda); i \frac{k^2}{2e\bar{E}}   \right), \label{eqa10}
\end{align}
where ${}_1\tilde{F}_1(a;b;z) \equiv {}_1 F_1(a;b;z)/\Gamma(b)$ is the regularized hypergeometric function and $\Theta$ is the step function.  By substituting Eq.~(\ref{eqa10}) into Eqs.~(\ref{eqa7a})-(\ref{eqa7d}) and simplifying Eq.~(\ref{eqa5}), one obtains
\begin{align}
	n^{(\mp)}_{\pm{\bm p},s}
		&=	\frac{V}{(2\pi)^3}  {\rm e}^{- 2 \pi a_{\bm p}}
			\left[
					\left|
							1
							- e \int^{\infty}_0 dk {\rm e}^{-i \frac{kp_3}{e\bar{E}}}
								\Biggl\{
										\frac{ p_1 \tilde{\mathcal A}_1 + p_2 \tilde{\mathcal A}_2 }{e\bar{E}} {\rm Re}\left[ {\rm e}^{-i\frac{k^2}{4e\bar{E}}} {}_1\tilde{F}_1\left( 1 - ia_{\bm p}; 1; i \frac{k^2}{2e\bar{E}} \right)  \right] \right. \right. \nonumber\\
										&\quad\quad\quad\quad\quad\quad\quad\quad\quad\quad\quad\quad\quad\quad\quad\quad\quad - \sigma \frac{ - p_2 \tilde{\mathcal A}_1 + p_1 \tilde{\mathcal A}_2 }{e\bar{E}} {\rm Im}\left[ {\rm e}^{-i\frac{k^2}{4e\bar{E}}} {}_1\tilde{F}_1\left( 1 - ia_{\bm p}; 1; i \frac{k^2}{2e\bar{E}} \right)  \right] \nonumber\\
										&\quad\quad\quad\quad\quad\quad\quad\quad\quad\quad\quad\quad\quad\quad\quad\quad\quad - i \frac{k \tilde{\mathcal A}_3}{e\bar{E}}  a_{\bm p} {\rm Re}\left[ {\rm e}^{-i\frac{k^2}{4e\bar{E}}} {}_1\tilde{F}_1\left( 1 - ia_{\bm p}; 2; i \frac{k^2}{2e\bar{E}} \right)  \right] \left.
								\Biggl\}
					\right|^2 \nonumber\\
					&\quad\quad\quad\quad\quad\quad\quad\quad + \left. \left| e \int^{\infty}_0 dk {\rm e}^{-i \frac{kp_3}{e\bar{E}}} \frac{m}{\sqrt{e\bar{E}}} \frac{ \tilde{\mathcal A}_1 + i \sigma \tilde{\mathcal A}_2 }{\sqrt{e\bar{E}}} {\rm Im}\left[ {\rm e}^{-i\frac{k^2}{4e\bar{E}}} {}_1\tilde{F}_1\left( 1 - ia_{\bm p}; 1; i \frac{k^2}{2e\bar{E}} \right)  \right] \right|^2
			\right].  \label{eqa11}
\end{align}

For the monochromatic perturbation (\ref{eq6}), $\tilde{A}_{\mu}$ reads
\begin{align}
	\tilde{A}_{\mu} = -i\pi \frac{ {\rm e}^{+i\phi}\delta(k-\Omega) -  {\rm e}^{-i\phi}\delta(k+\Omega) }{\Omega}  \times (0,{\mathcal E}_1,{\mathcal E}_2,{\mathcal E}_3).
\end{align}
Therefore, we finally arrive at
\begin{align}
	n^{(\mp)}_{\pm{\bm p},s}
		&=	\frac{V}{(2\pi)^3}  {\rm e}^{- 2 \pi a_{\bm p}}
			\left[
					\left|
							1
							+ i\pi {\rm e}^{+i\phi} {\rm e}^{-i \frac{\Omega p_3}{e\bar{E}}}
								\Biggl\{
										\frac{ p_1 {\mathcal E}_1 + p_2 {\mathcal E}_2 }{\bar{E} \Omega} {\rm Re}\left[ {\rm e}^{-i\frac{\Omega^2}{4e\bar{E}}} {}_1\tilde{F}_1\left( 1 - ia_{\bm p}; 1; i \frac{\Omega^2}{2e\bar{E}} \right)  \right] \right. \right. \nonumber\\
										&\quad\quad\quad\quad\quad\quad\quad\quad\quad\quad\quad\quad\quad\quad\quad\quad\quad\quad - \sigma \frac{ - p_2 {\mathcal E}_1 + p_1 {\mathcal E}_2 }{\bar{E} \Omega} {\rm Im}\left[ {\rm e}^{-i\frac{\Omega^2}{4e\bar{E}}} {}_1\tilde{F}_1\left( 1 - ia_{\bm p}; 1; i \frac{\Omega^2}{2e\bar{E}} \right)  \right] \nonumber\\
										&\quad\quad\quad\quad\quad\quad\quad\quad\quad\quad\quad\quad\quad\quad\quad\quad\quad\quad - i \frac{{\mathcal E}}{\bar{E}}  a_{\bm p} {\rm Re}\left[ {\rm e}^{-i\frac{\Omega^2}{4e\bar{E}}} {}_1\tilde{F}_1\left( 1 - ia_{\bm p}; 2; i \frac{\Omega^2}{2e\bar{E}} \right)  \right] \left.
								\Biggl\}
					\right|^2 \nonumber\\
					&\quad\quad\quad\quad\quad\quad\quad\quad + \left. \pi^2 \frac{m^2}{\Omega^2} \frac{|{\mathcal E}_1|^2 + |{\mathcal E}_2|^2}{\bar{E}^2} \left|  {\rm Im}\left[ {\rm e}^{-i\frac{\Omega^2}{4e\bar{E}}} {}_1\tilde{F}_1\left( 1 - i a_{\bm p}; 1; i \frac{\Omega^2}{2e\bar{E}} \right)  \right] \right|^2
			\right].  \label{eqa13}
\end{align}
\end{widetext}

\section{Numerical results for momentum distribution} \label{appB}

In this appendix, we present some numerical results for the momentum distribution of, for simplicity, electrons with spin up $n_{{\bm p},\uparrow}$.

We consider the same parameter set as in the main text, i.e, $e\bar{E}/m^2 = 0.4, {\mathcal E}_3/\bar{E} = 0, \phi = 1, m\tau=100$, and discuss the other parameter dependences in the following.  Note that the numerical method used here to obtain the results is the same as in the main text (see Sec.~\ref{sec3b1}).

\subsubsection{$\Omega$-dependence}

\begin{figure*}[!t]
\begin{center}
\hspace*{-10mm}\includegraphics[clip, width=0.365\textwidth]{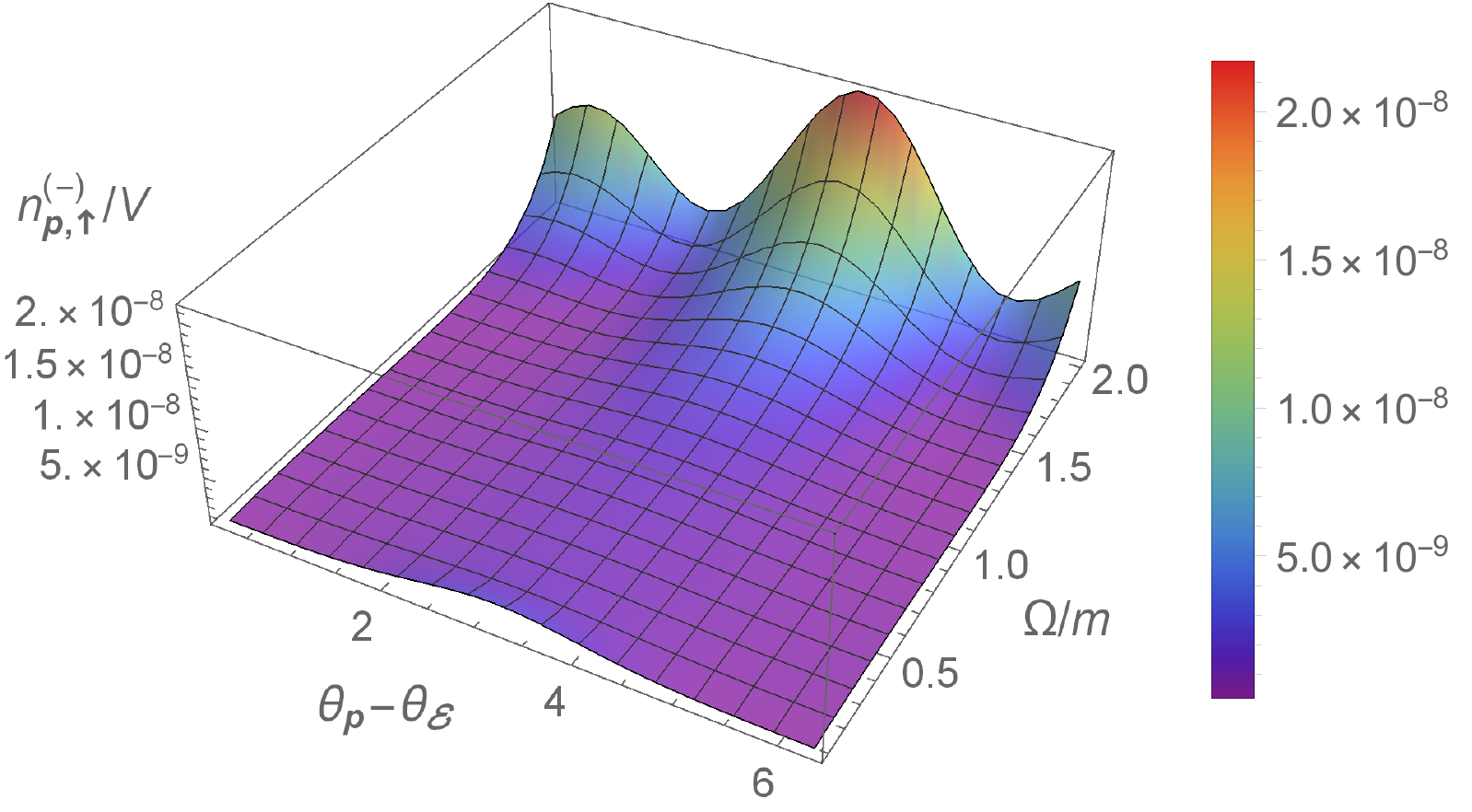}
\hspace*{-1mm}\includegraphics[clip, width=0.345\textwidth]{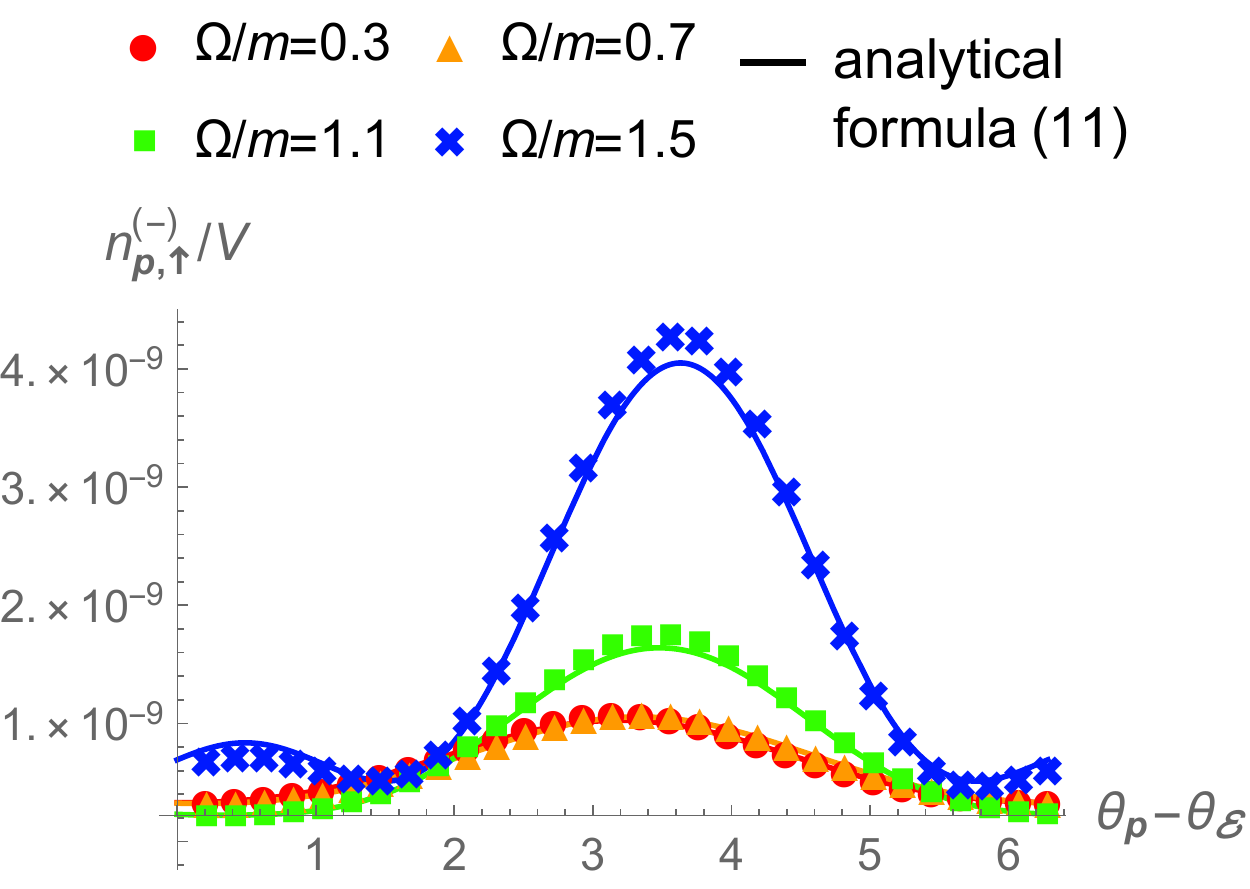}
\hspace*{-1mm}\includegraphics[clip, width=0.345\textwidth]{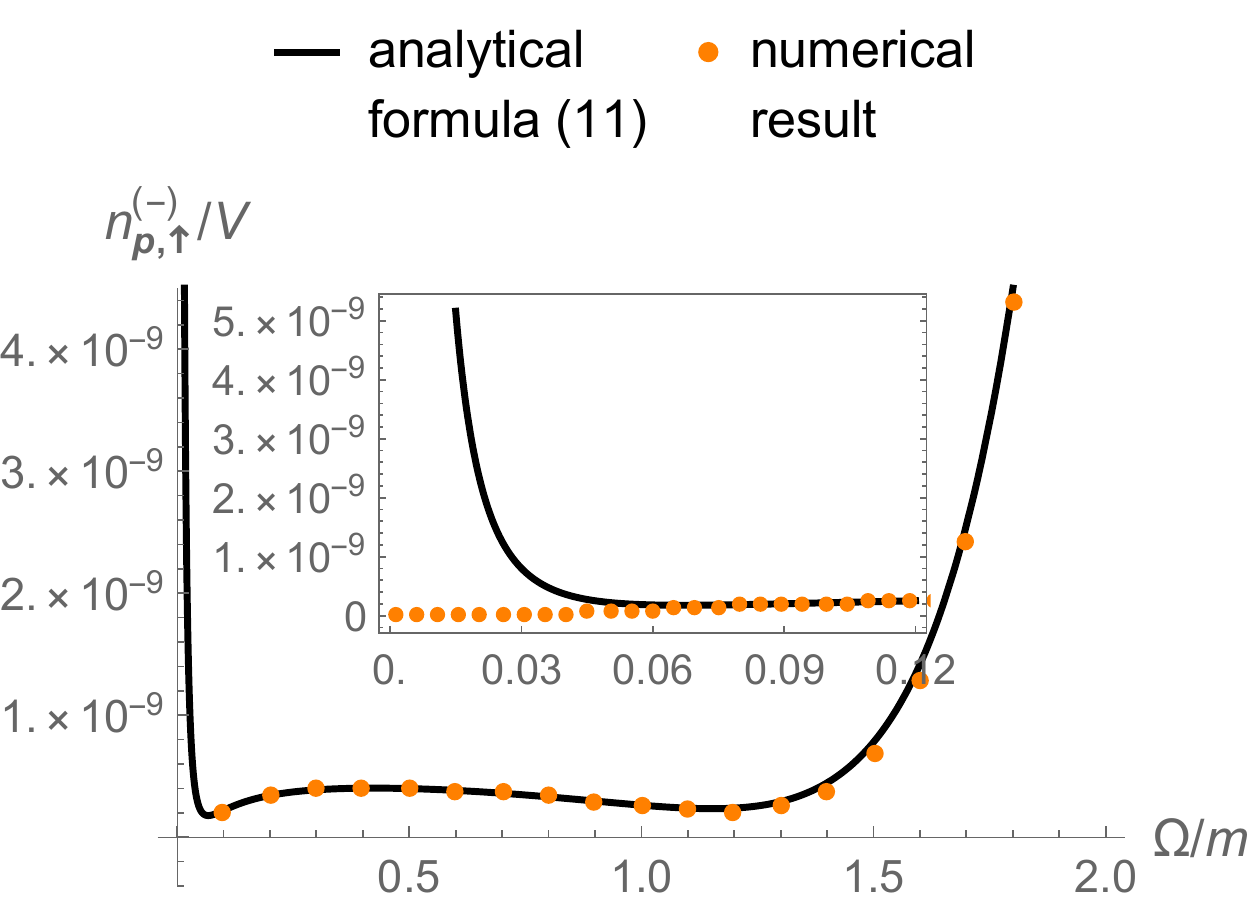}
\vspace*{3mm}\hfill \\
\mbox{(i) Small frequency $\Omega/m \in [0,2]$}\hfill \\
\vspace*{6mm}
\hspace*{-10mm}\includegraphics[clip, width=0.365\textwidth]{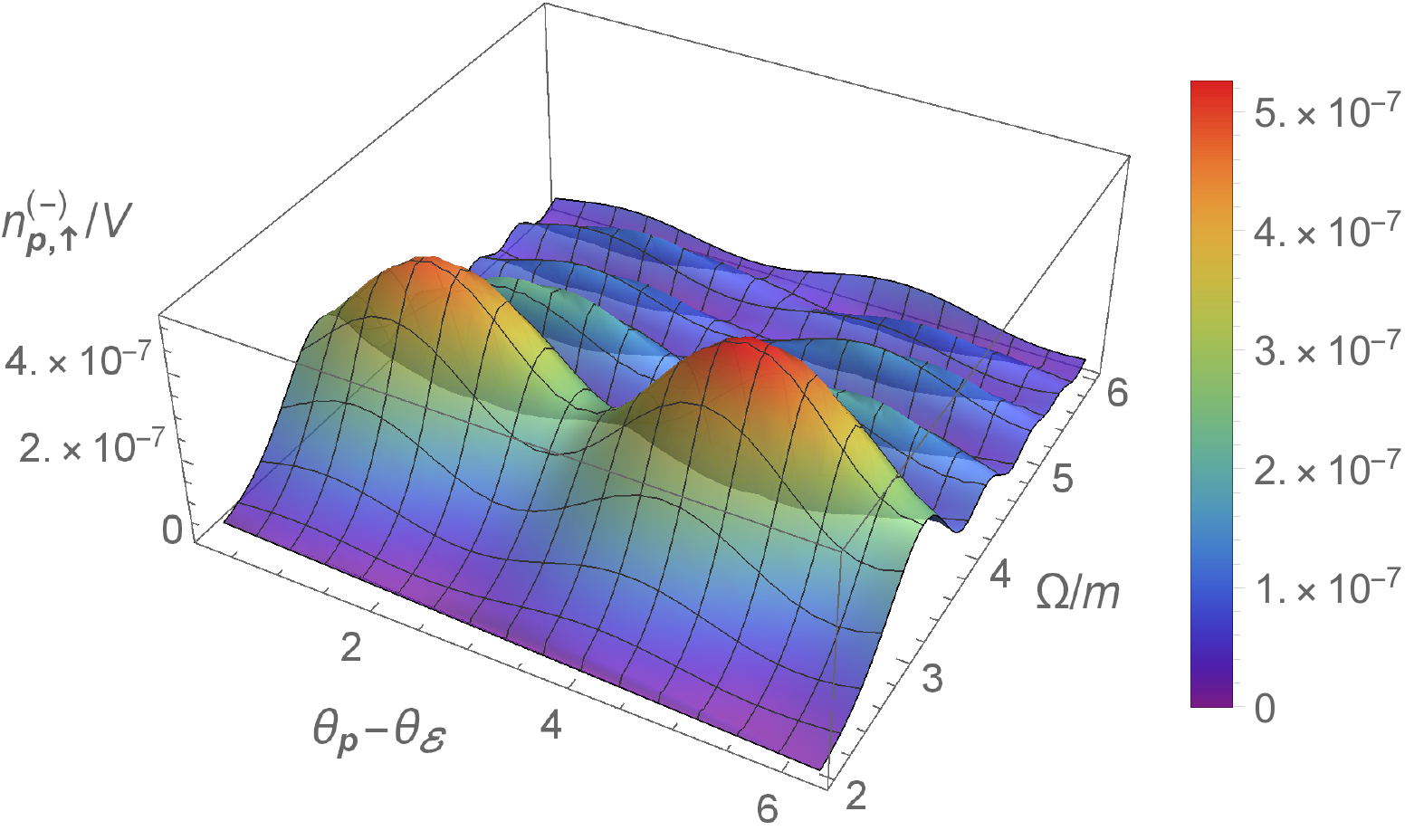}
\hspace*{-1mm}\includegraphics[clip, width=0.345\textwidth]{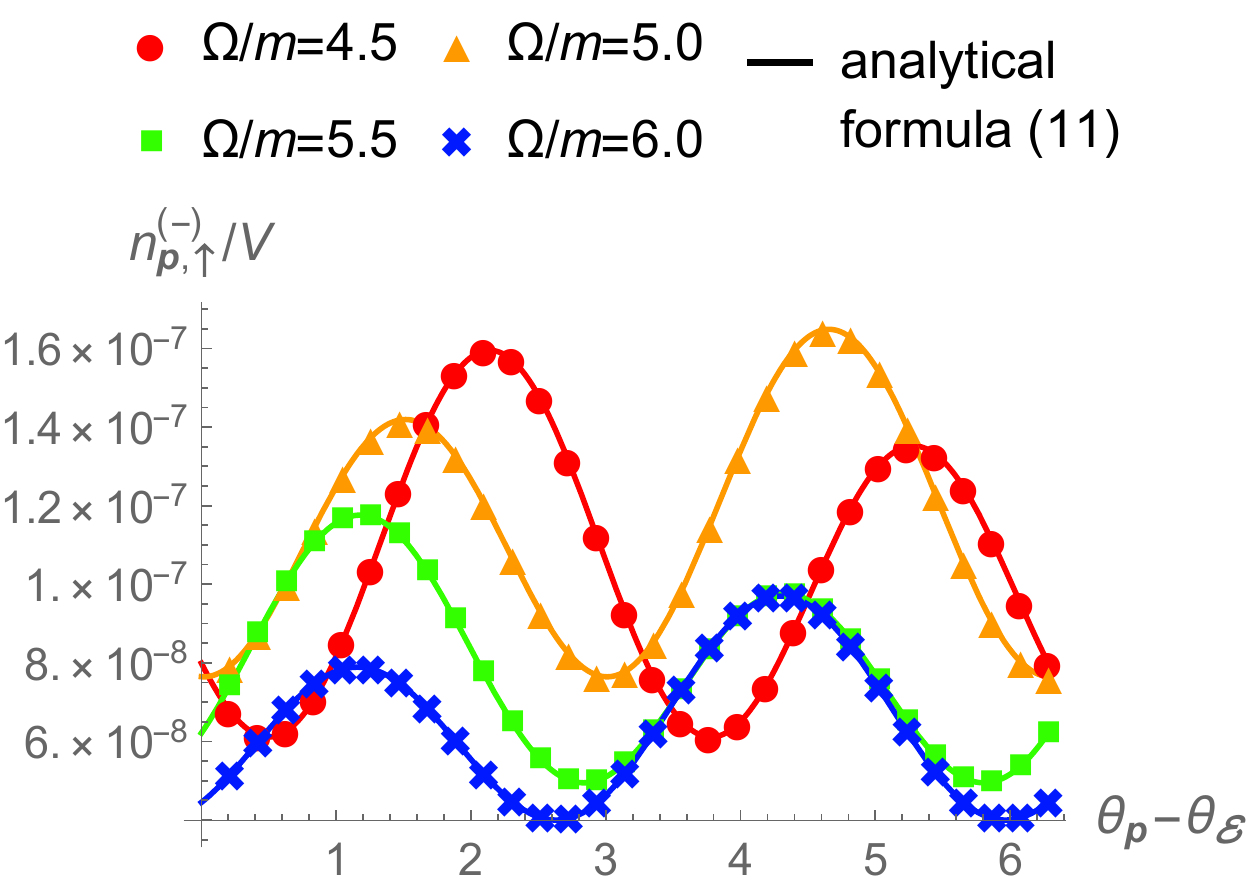}
\hspace*{-1mm}\includegraphics[clip, width=0.345\textwidth]{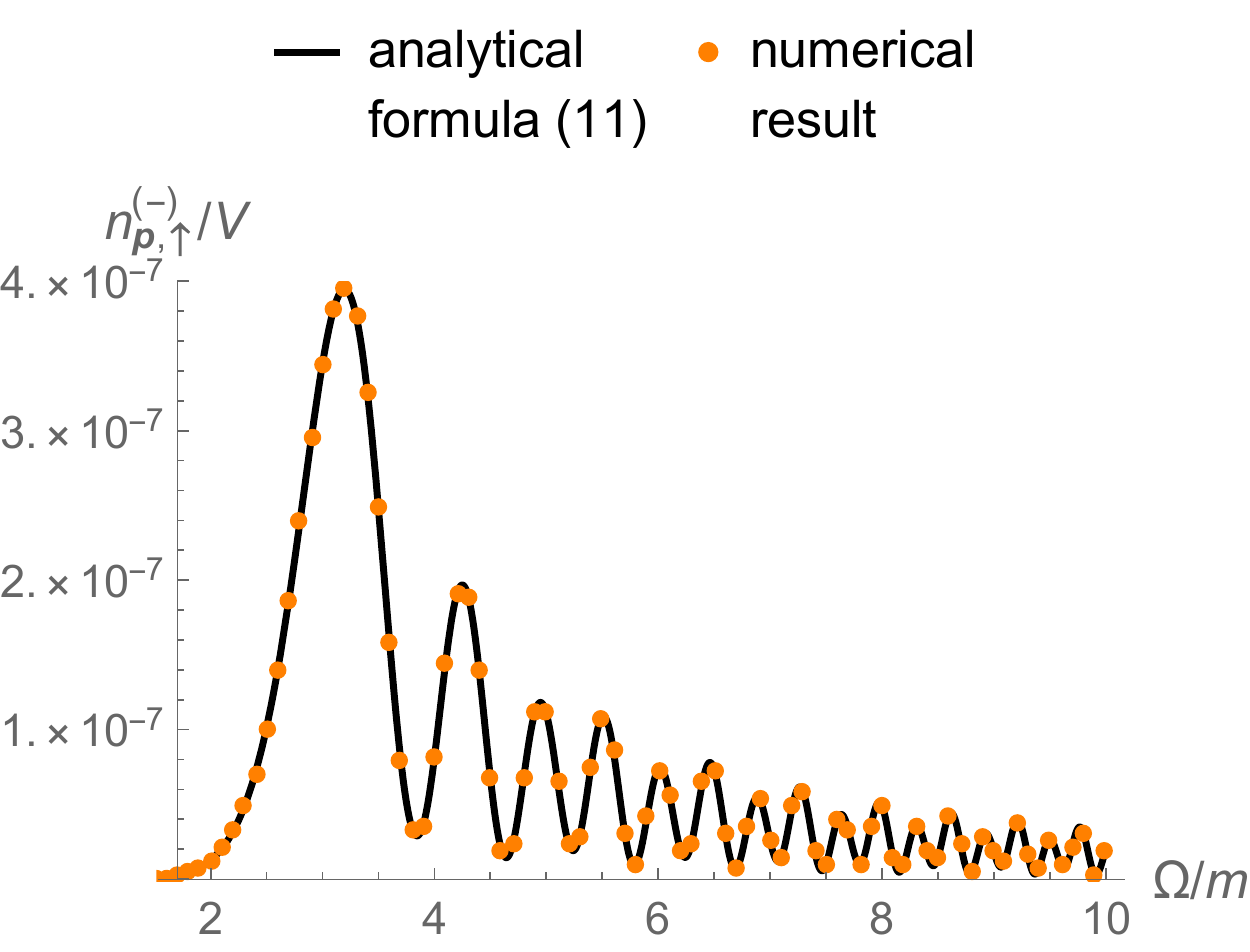}
\vspace*{3mm} \hfill \\
\mbox{(ii) Large frequency $\Omega/m \in [2,10]$}
\caption{\label{fig6} (color online) The numerical results of the momentum distribution $n^{(-)}_{{\bm p},\uparrow}$ as a function of $(\theta_{\bm p}, \Omega)$ (left); as a function of $\theta_{\bm p}$ for several values of $\Omega$ (center); and as a function of $\Omega$ for fixed $\theta_{\bm p}-\theta_{\bm {\mathcal E}} = \pi/4$ (right).  As a comparison, the analytical results (\ref{eq11}) are plotted as the lines in the center and right panels.  The top (i) and bottom (ii) panels distinguish the size of the frequency $\Omega$.  The parameters are the same as in Fig.~\ref{fig1}, i.e., $e\bar{E}/m^2 = 0.4, {\mathcal E}_{\perp}/\bar{E} = 0.025, {\mathcal E}_3/\bar{E} = 0, p_{\perp}/m = 1, p_3/m = 0, \phi = 1,\ {\rm and\ }m\tau=100$. .   }
\end{center}
\end{figure*}

Figure~\ref{fig6} shows the numerical results for the momentum distribution $n^{(-)}_{{\bm p},\uparrow}$ as a function of the azimuthal angle $\theta_{\bm p}$ and the frequency $\Omega$ (which corresponds to Fig.~\ref{fig1} in the main text).

The perturbation theory in the Furry picture, or the resulting formula for the momentum distribution (\ref{eq11}), reproduces the numerical results very well except for very small $\Omega$, which, for our parameter choices, reads $\Omega/m \lesssim 0.1$.  The formula (\ref{eq11}) exhibits a divergent behavior at $\Omega \to 0$ due to the collinear (infra-red) divergence, which may be eliminated by resumming higher order multi-photon scatterings.  Note that the formula for the spin-imbalance (\ref{eq16}) gives a constant (see Fig.~\ref{fig1}) at $\Omega \to 0$.  This is simply because the divergent contribution in Eq.~(\ref{eq11}) is spin-independent.

The azimuthal angle $\theta_{\bm p}$-dependence changes by changing the size of $\Omega$.  Basically, for small $\Omega \lesssim \sqrt{e\bar{E}}, \sqrt{m^2 + p_{\perp}^2}$, the usual Schwinger mechanism plus the interference effect between the two diagrams in Eq.~(\ref{eq7}) dominate the production.  The usual Schwinger mechanism is insensitive to $\theta_{\bm p}$, while the interference term gives $\sin (\theta_{\bm p} - \theta_{{\bm {\mathcal E}}}), \cos(\theta_{\bm p} - \theta_{{\bm {\mathcal E}}})$ (see Eq.~(\ref{eq11})), which determine the $\theta_{\bm p}$-dependence of the momentum distribution.  On the other hand, for large $\Omega \gtrsim \sqrt{e\bar{E}}, \sqrt{m^2 + p_{\perp}^2}$, the production becomes perturbative rather than non-perturbative, and the formula (\ref{eq11}) approaches the purely perturbative formula (\ref{eq14}).  Therefore, the momentum distribution acquires a higher order $\theta_{\bm p}$-dependence such as ($\sin (\theta_{\bm p} - \theta_{{\bm {\mathcal E}}}))^2, (\cos (\theta_{\bm p} - \theta_{{\bm {\mathcal E}}}))^2$, which results in the two-peak structure in the panel (ii) of Fig.~\ref{fig6}.

The production number is, basically, an increasing function of $\Omega$ below the mass gap $\Omega \lesssim 2m$.  This is nothing but the dynamically assisted Schwinger mechanism  \cite{sch08,dun09,piz09,mon10a,mon10b}.  Note that for very small $\Omega$ the production number does not necessarily increase.  This is due to the interference effect between the two diagrams in Eq.~(\ref{eq7}) \cite{gre17}, which can be controlled by changing the substructure of the weak field (e.g., the phase $\phi$).  For large $\Omega$ above the mass gap $\Omega \gtrsim 2m$, the momentum distribution exhibits an oscillating behavior.  This is a QED analog of the Franz-Keldysh oscillation \cite{tay19}, which occurs due to the quantum reflection of electrons by the tilted mass gap in the presence of strong electric field.

\begin{figure}[!t]
\begin{center}
\includegraphics[clip, width=0.365\textwidth]{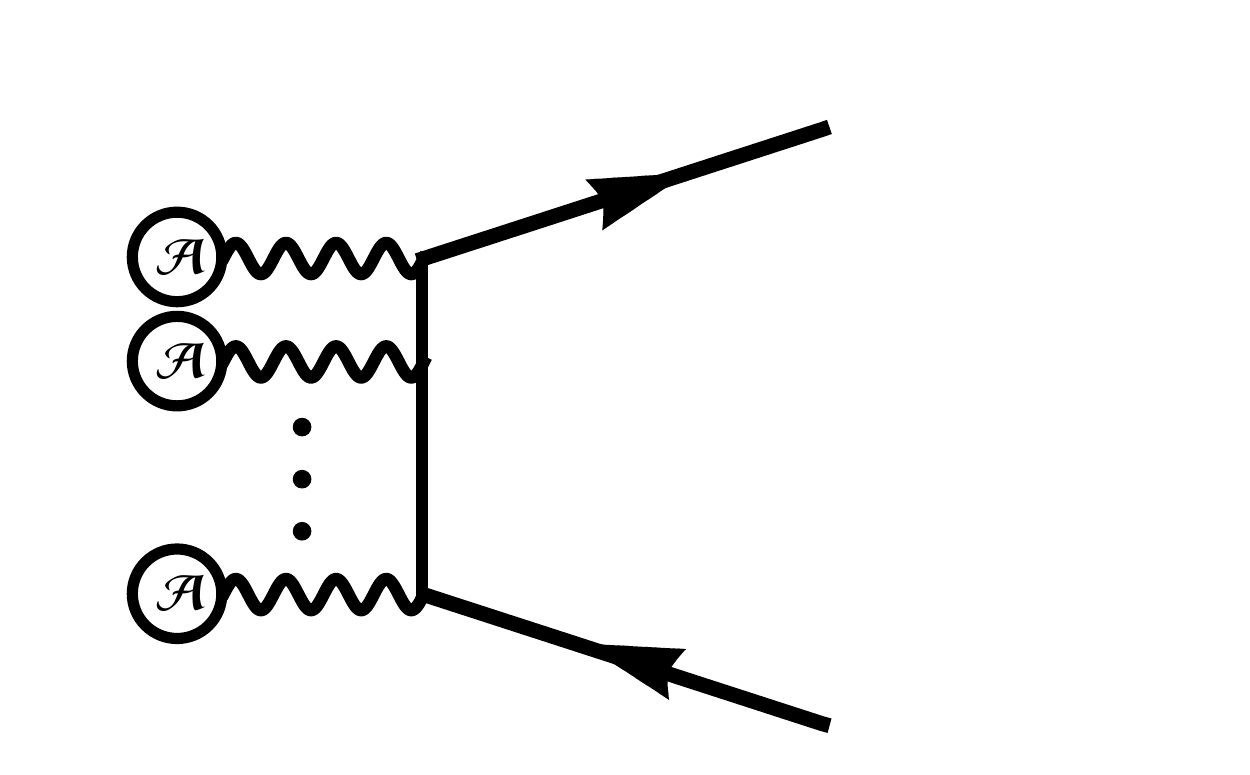}
\includegraphics[clip, width=0.365\textwidth]{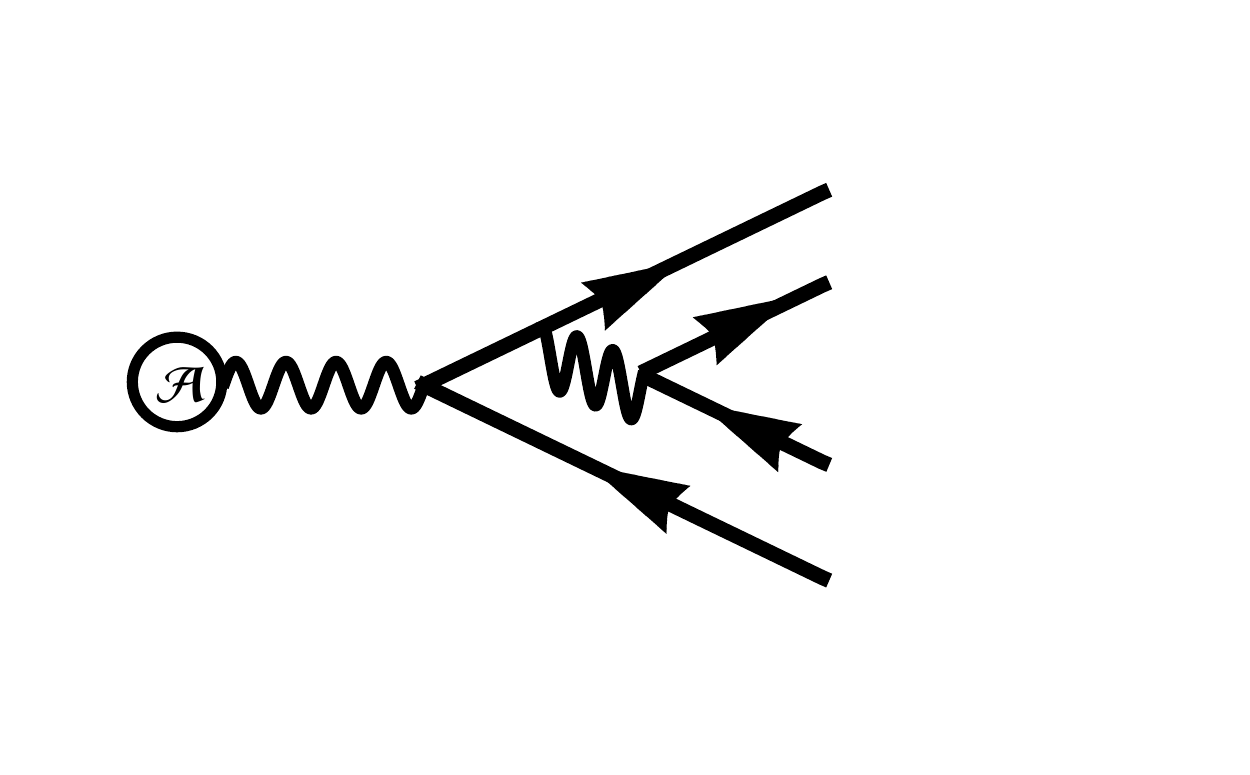}
\caption{\label{dia2} [Top] Tree-level diagram of the perturbation ${\mathcal A}_{\mu}$; [Bottom] The lowest order diagram to produce two pairs from one photon.  The thick lines are the full propagator dressed by the strong field $\bar{A}$; see Eq.~(\ref{eq_7}).  }
\end{center}
\end{figure}

Figure~\ref{fig6} does not seem to have peaks below the mass gap $\Omega = 2\omega_{\bm p}/n $ ($n \in {\mathbb N}$) originating from one pair production from $n$-photons.  This is because such multi-photon processes are strongly suppressed by the factor $(e{\mathcal E}_{\perp}/m^2)^n$, so that the peaks are buried in the exponential tail below the mass gap due to the dynamically assisted Schwinger mechanism.  One might naively expect that a very energetic photon with $\Omega = 2n\omega_{\bm p}$ is able to create $n$-pairs, which might result in sharp peaks on top of the Franz-Keldysh oscillation above the mass gap.  However, such multi pair production from one photon is not possible at the leading order in the coupling constant $e$ (or tree diagrams depicted in Fig.~\ref{dia2}), and is therefore strongly suppressed by the coupling constant $e$ (see, for example, the bottom diagram in Fig.~\ref{dia2} for the lowest order diagram).  Our framework based on the Dirac equation does not take into account higher order quantum interactions beyond the tree-level processes, and is justified in the limit of $e \to 0$.

\subsubsection{${\mathcal E}_{\perp}$-dependence}

\begin{figure*}[!t]
\begin{center}
\hspace*{-10mm}\includegraphics[clip, width=0.365\textwidth]{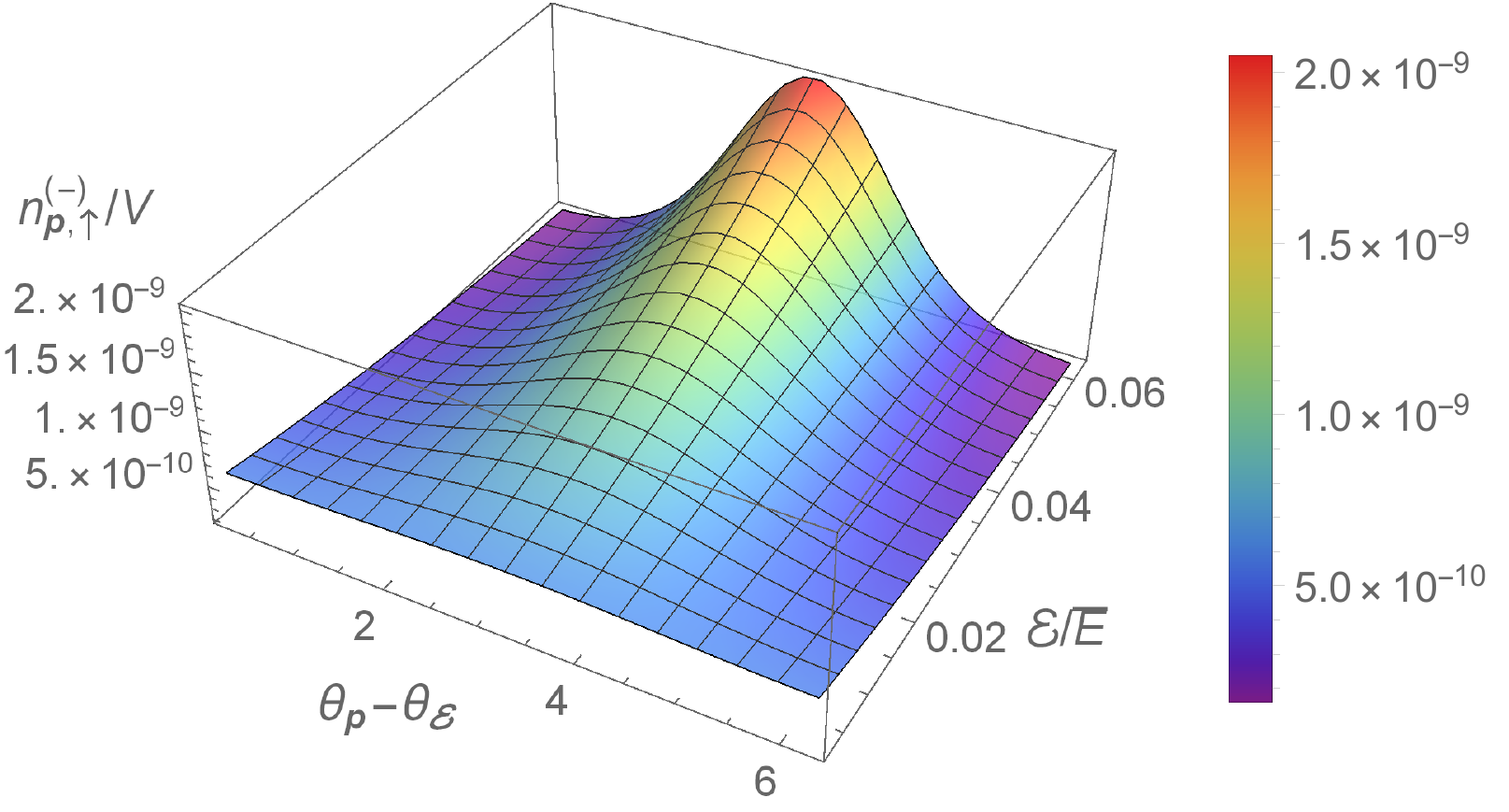}
\hspace*{-1mm}\includegraphics[clip, width=0.345\textwidth]{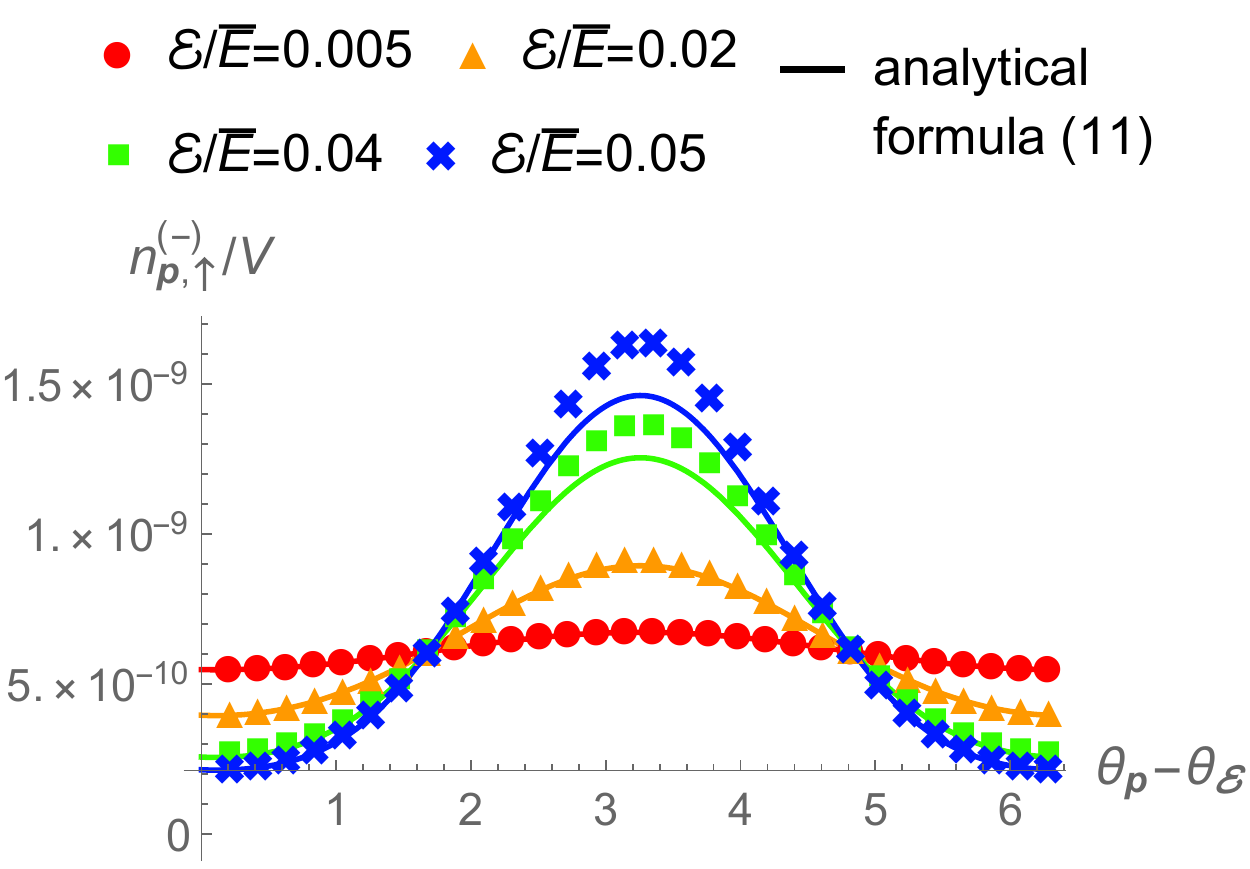}
\hspace*{-1mm}\includegraphics[clip, width=0.345\textwidth]{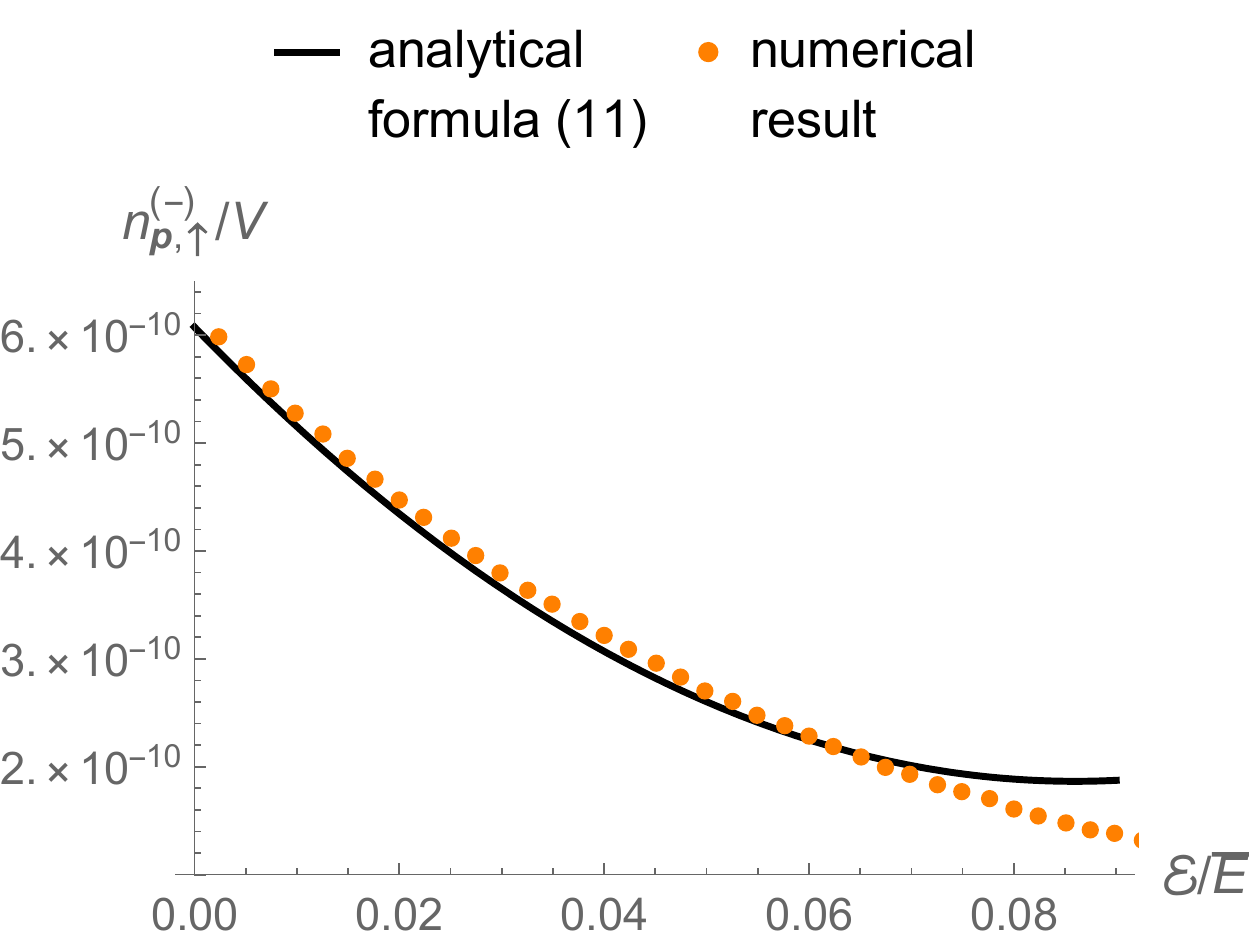}
\vspace*{3mm}\hfill \\
\mbox{(i) Weak perturbation ${\mathcal E}_{\perp}/\bar{E} \in [0,0.05]$}\hfill \\
\vspace*{6mm}
\hspace*{-10mm}\includegraphics[clip, width=0.365\textwidth]{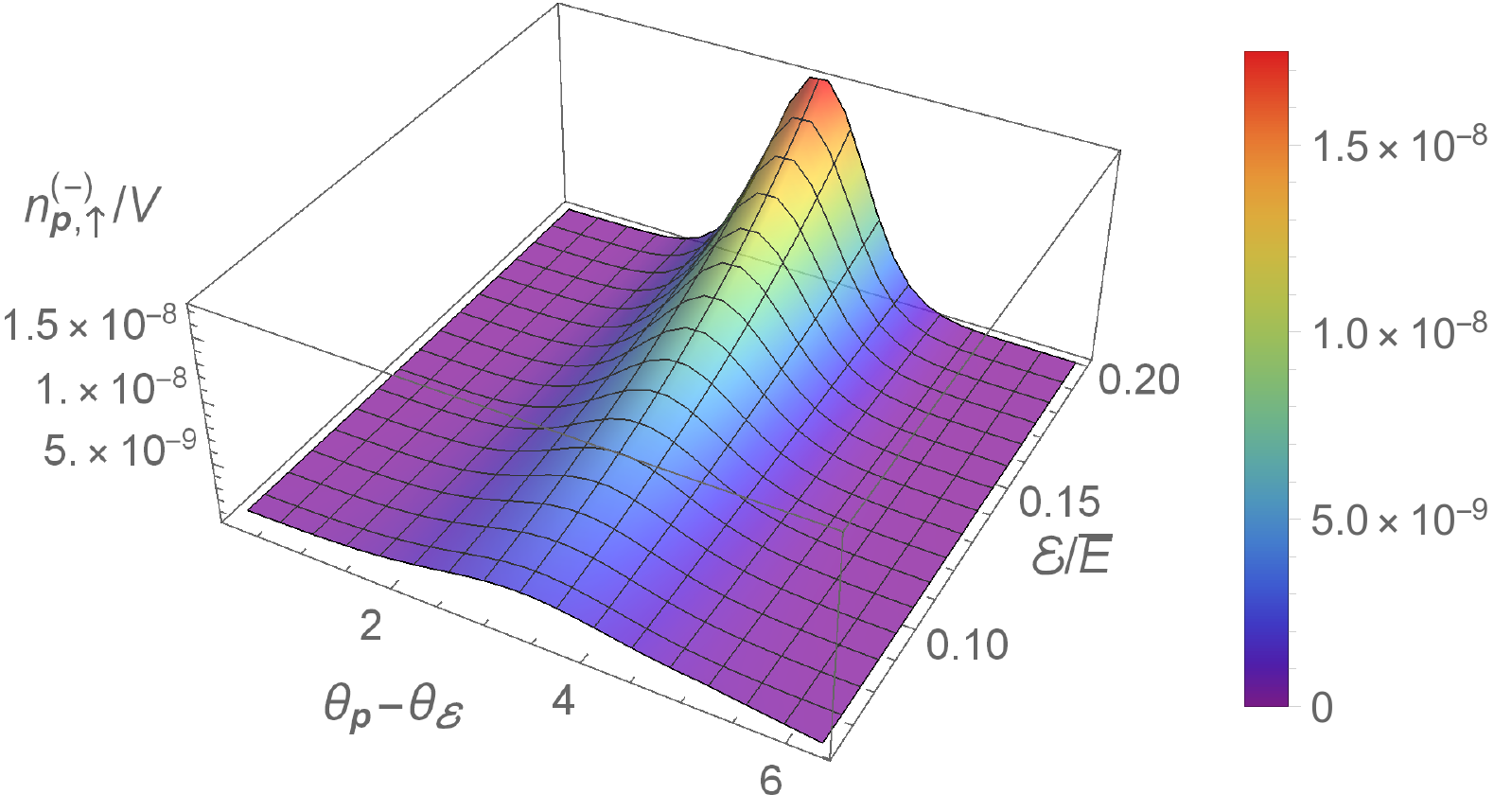}
\hspace*{-1mm}\includegraphics[clip, width=0.345\textwidth]{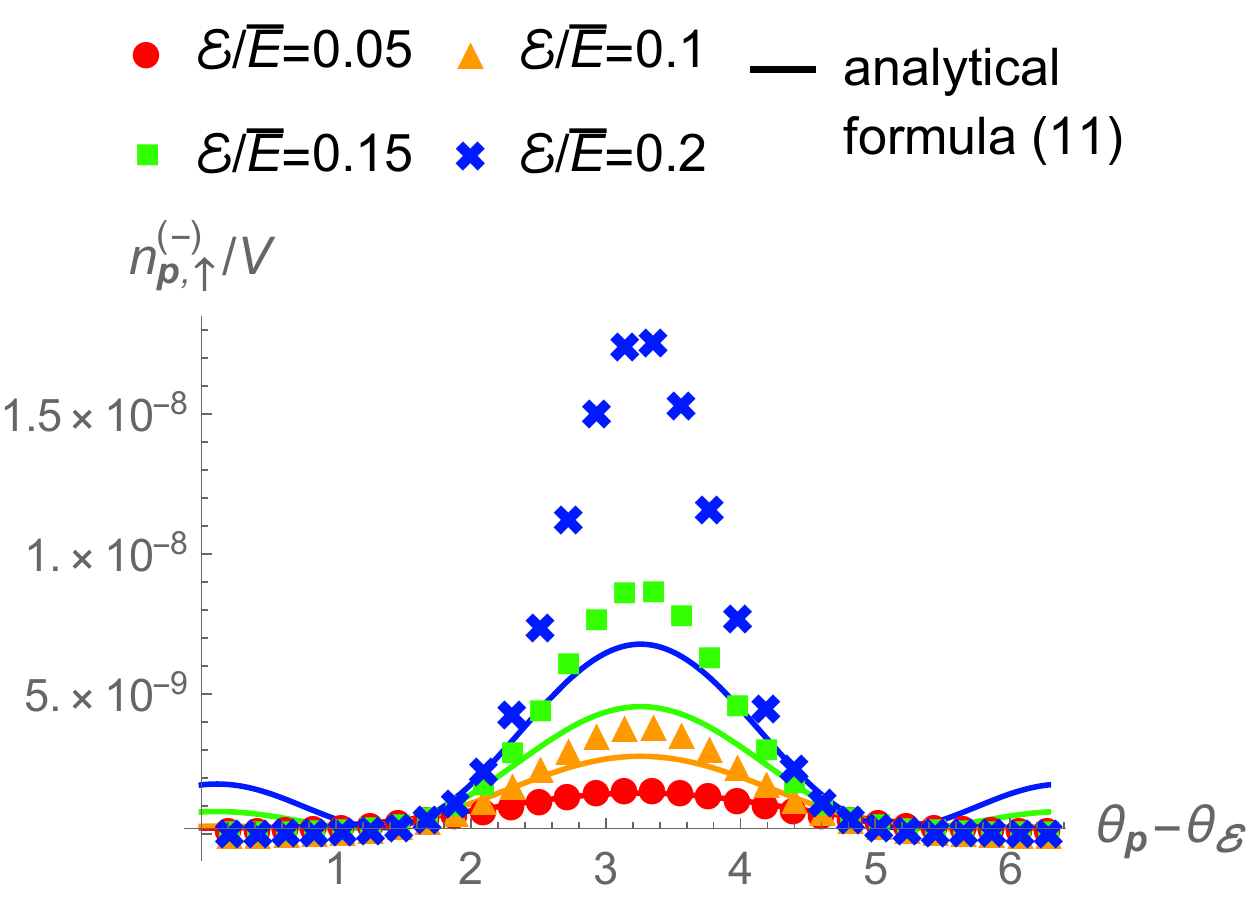}
\hspace*{-1mm}\includegraphics[clip, width=0.345\textwidth]{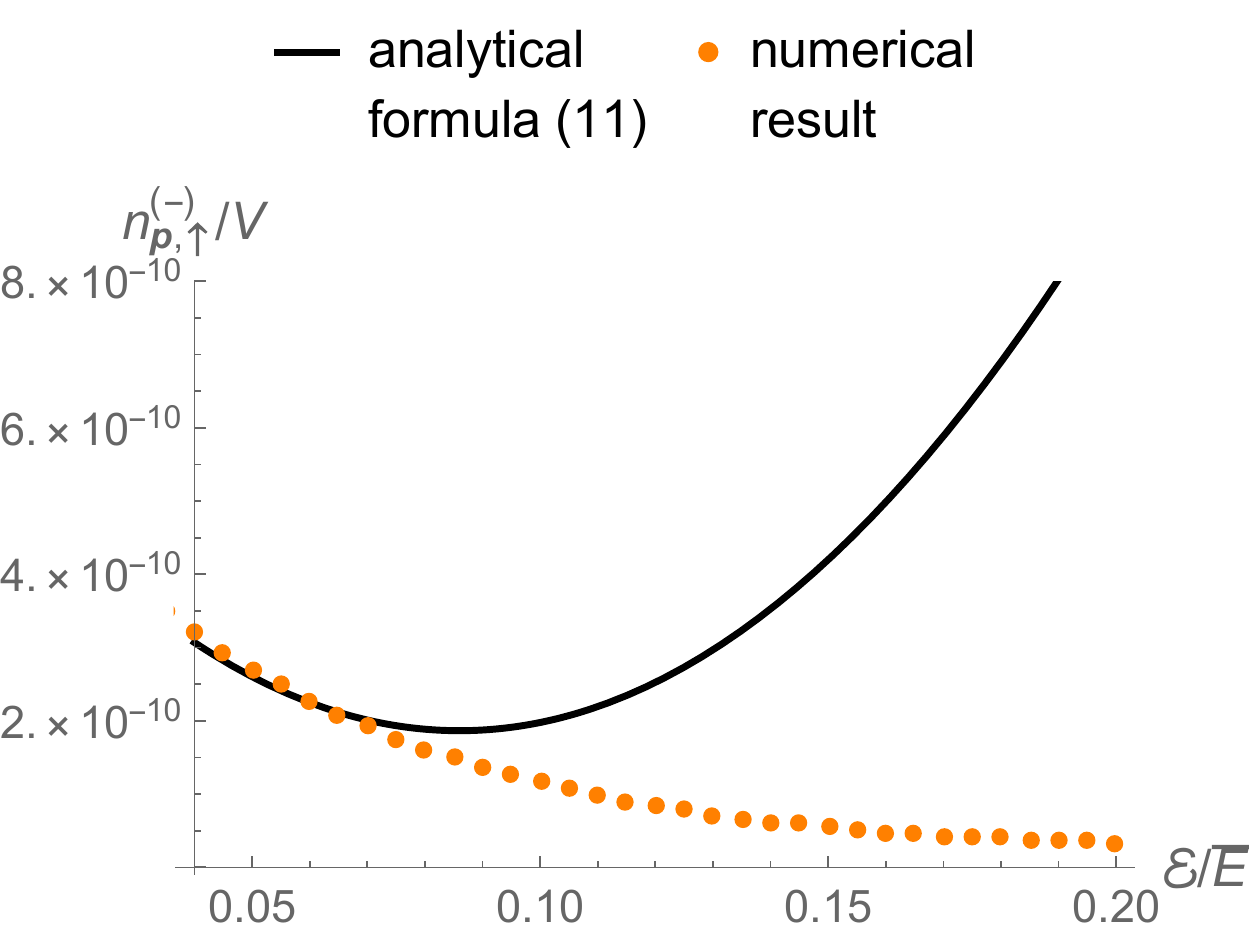}
\vspace*{3mm} \hfill \\
\mbox{(ii) Strong perturbation ${\mathcal E}_{\perp}/\bar{E} \in [0.05,0.2]$}
\caption{\label{fig7} (color online) The numerical results of the momentum distribution $ n^{(-)}_{{\bm p},\uparrow}$ as a function of $(\theta_{\bm p}, {\mathcal E}_{\perp})$ (left); as a function of $\theta_{\bm p}$ for several values of ${\mathcal E}_{\perp}$ (center); and as a function of ${\mathcal E}_{\perp}$ for fixed $\theta_{\bm p}-\theta_{\bm {\mathcal E}} = \pi/4$ (right).  As a comparison, the analytical results (\ref{eq11}) are plotted as the lines in the center and right panels.  The top (i) and bottom (ii) panels distinguish the size of the transverse perturbation ${\mathcal E}_{\perp}$.  The parameters are the same as in Fig.~\ref{fig2}, i.e., $e\bar{E}/m^2 = 0.4,  {\mathcal E}_3/\bar{E} = 0, \Omega/m = 0.5, p_{\perp}/m = 1, p_3/m = 0, \phi = 1,\ {\rm and\ }m\tau=100$.     }
\end{center}
\end{figure*}

\begin{figure}[!t]
\includegraphics[clip, width=0.4\textwidth]{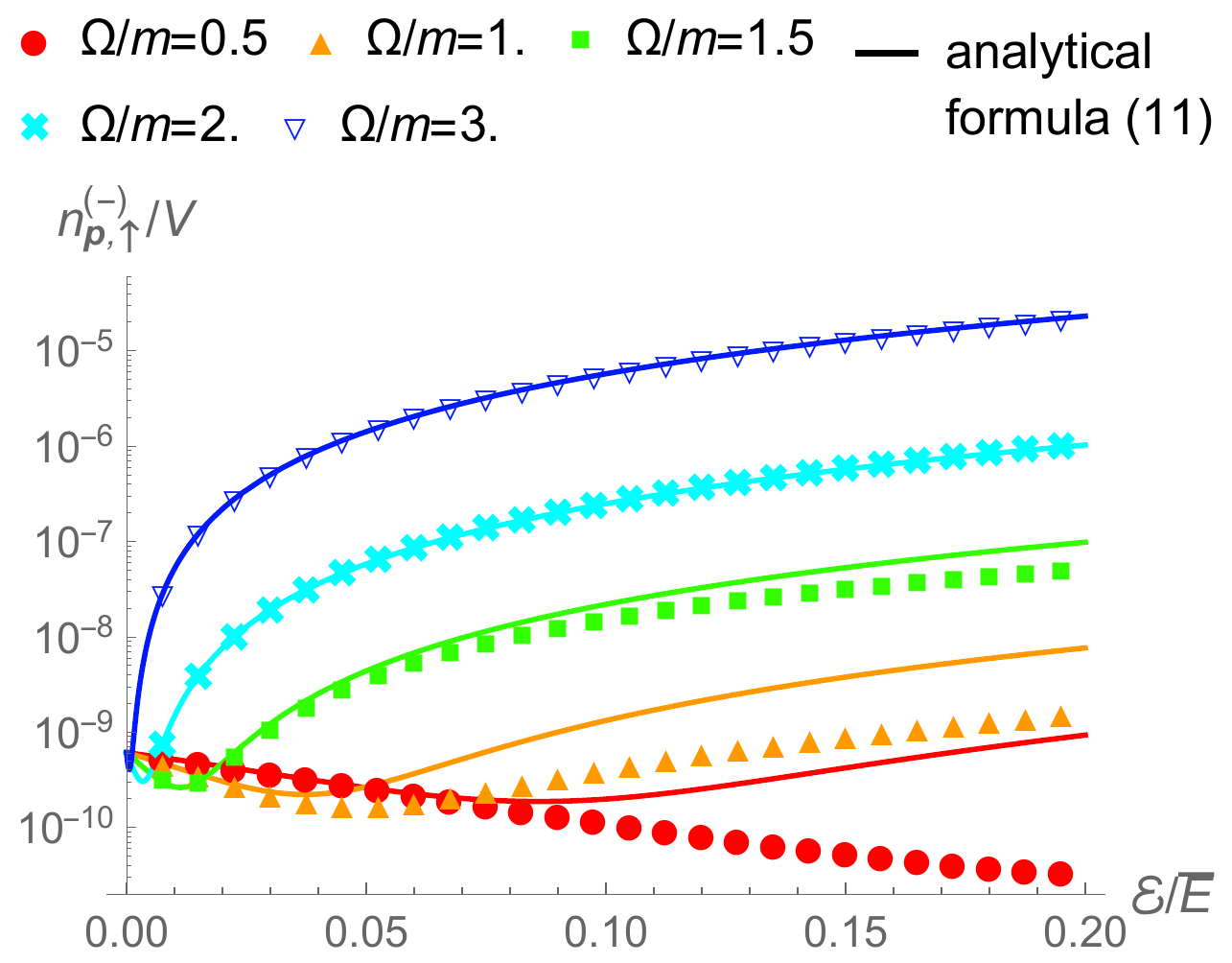}
\caption{\label{fig8} (color online) A comparison between the numerical results (points) and the analytical results (lines) for the momentum distribution $n^{(-)}_{{\bm p},\uparrow}$ as a function of the strength of the perturbation ${\mathcal E}_{\perp}$ for several values of the frequency $\Omega$.  The parameters are the same as in Fig.~\ref{fig3}, i.e., $e\bar{E}/m^2 = 0.4,  {\mathcal E}_3/\bar{E} = 0, p_{\perp}/m = 1, p_3/m = 0, \phi = 1,\ {\rm and\ }m\tau=100$.  }
\end{figure}

The ${\mathcal E}_{\perp}$-dependence of the momentum distribution $n_{{\bm p},\uparrow}$ is numerically investigated in Figs.~\ref{fig7} and \ref{fig8} (which correspond to Figs.~\ref{fig2} and \ref{fig3} in the main text, respectively).  

As discussed in the main text, the analytical formula (\ref{eq11}) is valid only when the production is dominated by the one-photon process in Eq.~(\ref{eq7}), so that the analytical formula (\ref{eq11}) deviates from the numerical results for large ${\mathcal E}_{\perp}$ if the weak field is less energetic $\Omega \lesssim 2m$.  Multi-photon processes beyond the formula (\ref{eq11}) enhance backward production $\theta_{\bm p} - \theta_{\bm {\mathcal E}} \sim \pi$, while collinear production $\theta_{\bm p} - \theta_{\bm {\mathcal E}} \sim 0$ is suppressed.  Note that the momentum distribution acquires $\theta_{\bm p}$-dependence only through the perturbative effect, i.e., the interaction with the weak field, so that the distribution becomes insensitive to $\theta_{\bm p}$ with ${\mathcal E}_{\perp} \to 0$.

\subsubsection{$p_{\perp}$-dependence}

\begin{figure*}[t!]
\begin{center}
\hspace*{-10mm}\includegraphics[clip, width=0.365\textwidth]{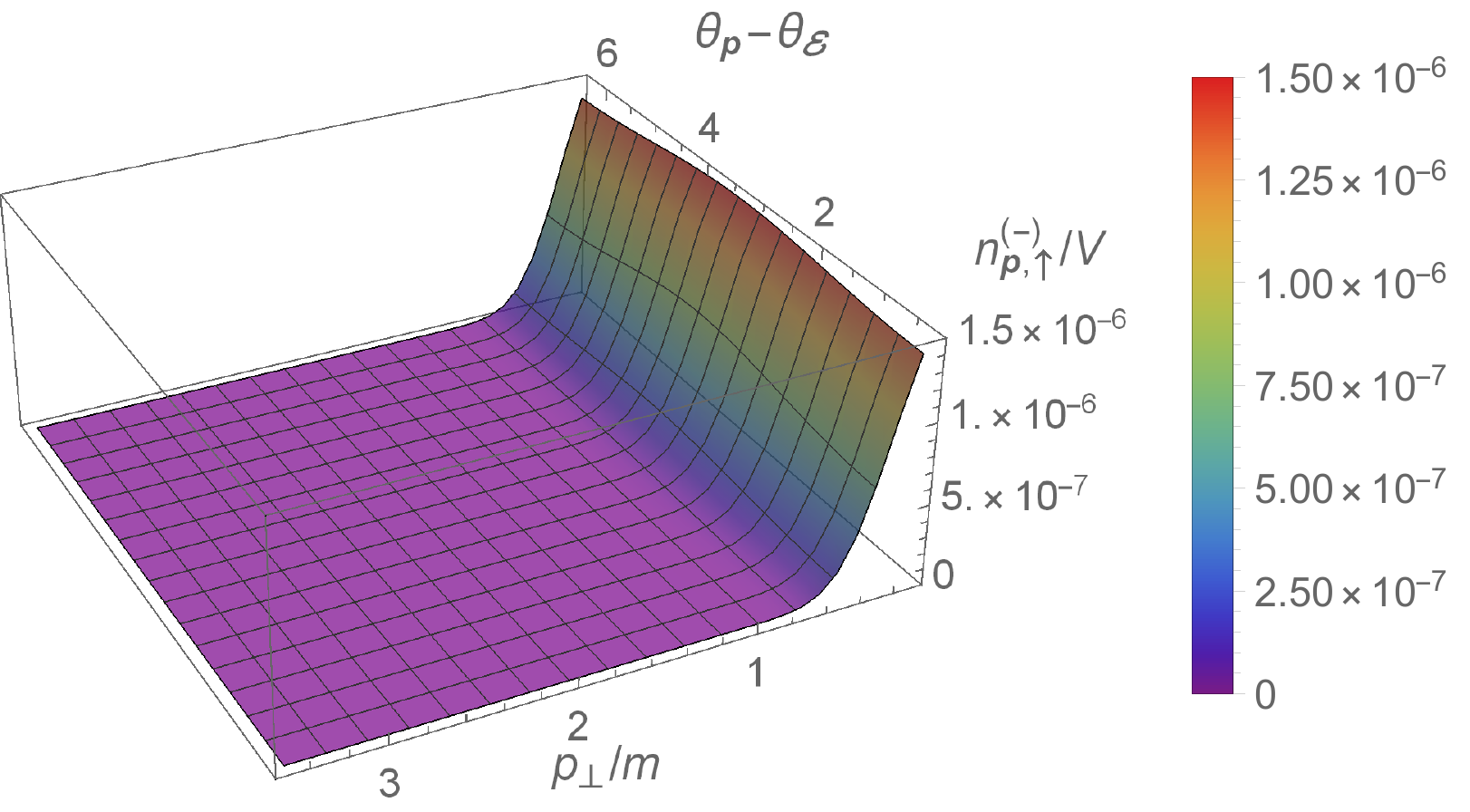}
\hspace*{-1mm}\includegraphics[clip, width=0.345\textwidth]{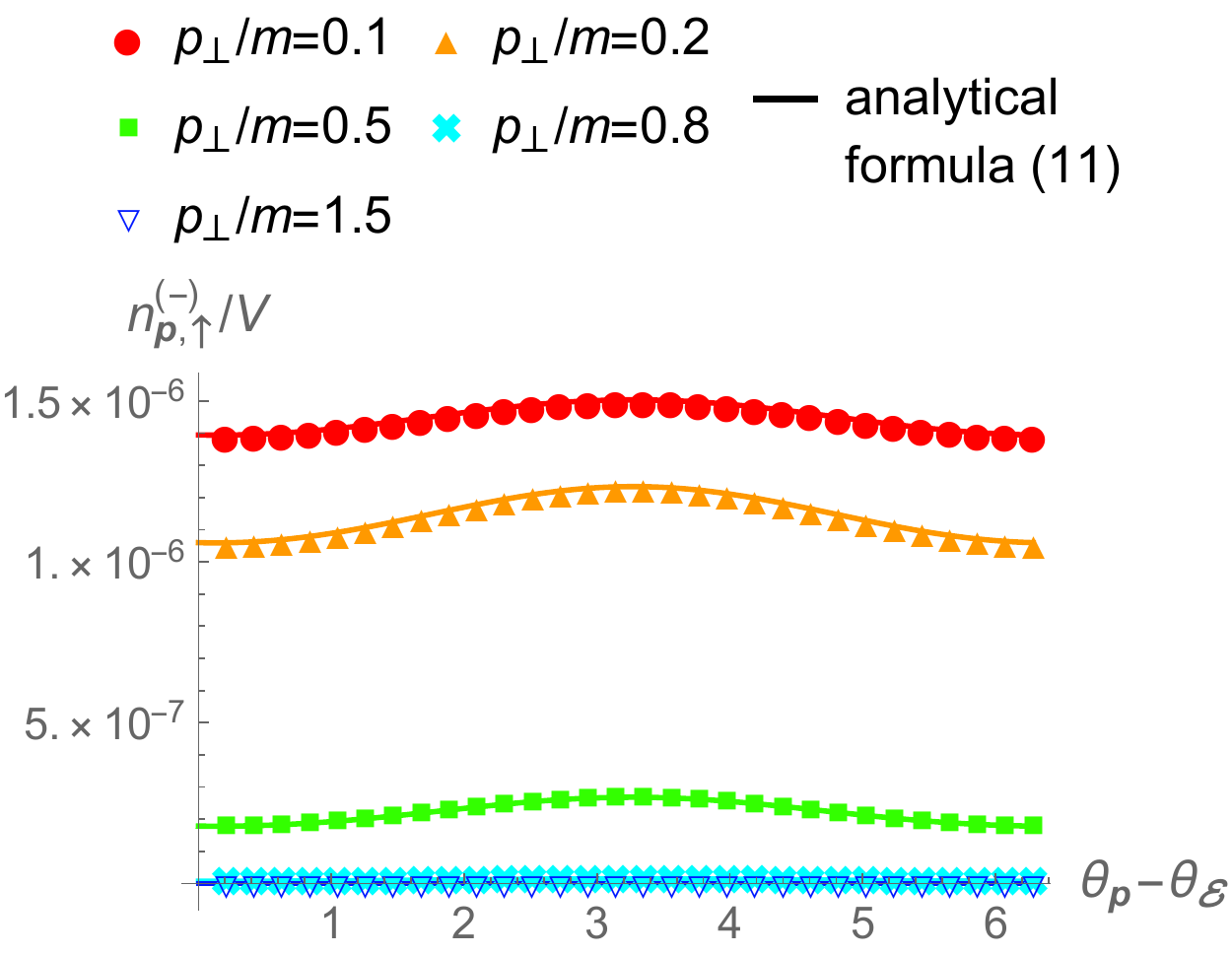}
\hspace*{-1mm}\includegraphics[clip, width=0.342\textwidth]{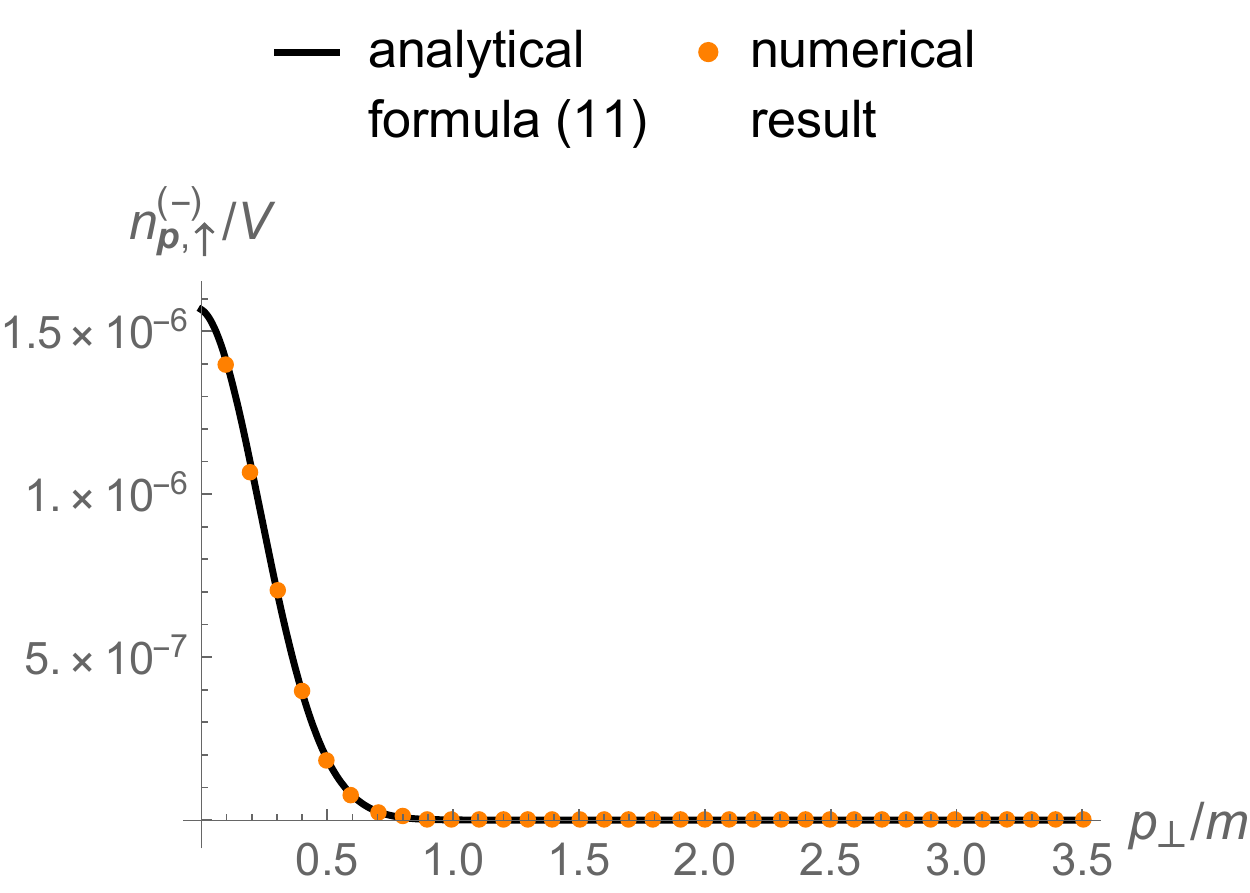}
\vspace*{-1mm}\hfill \\
\mbox{(i) Small frequency $\Omega/m =0.5 $ and weak perturbation ${\mathcal E}_{\perp}/\bar{E} = 0.025 $}\hfill \\
\vspace*{1mm}
\hspace*{-10mm}\includegraphics[clip, width=0.365\textwidth]{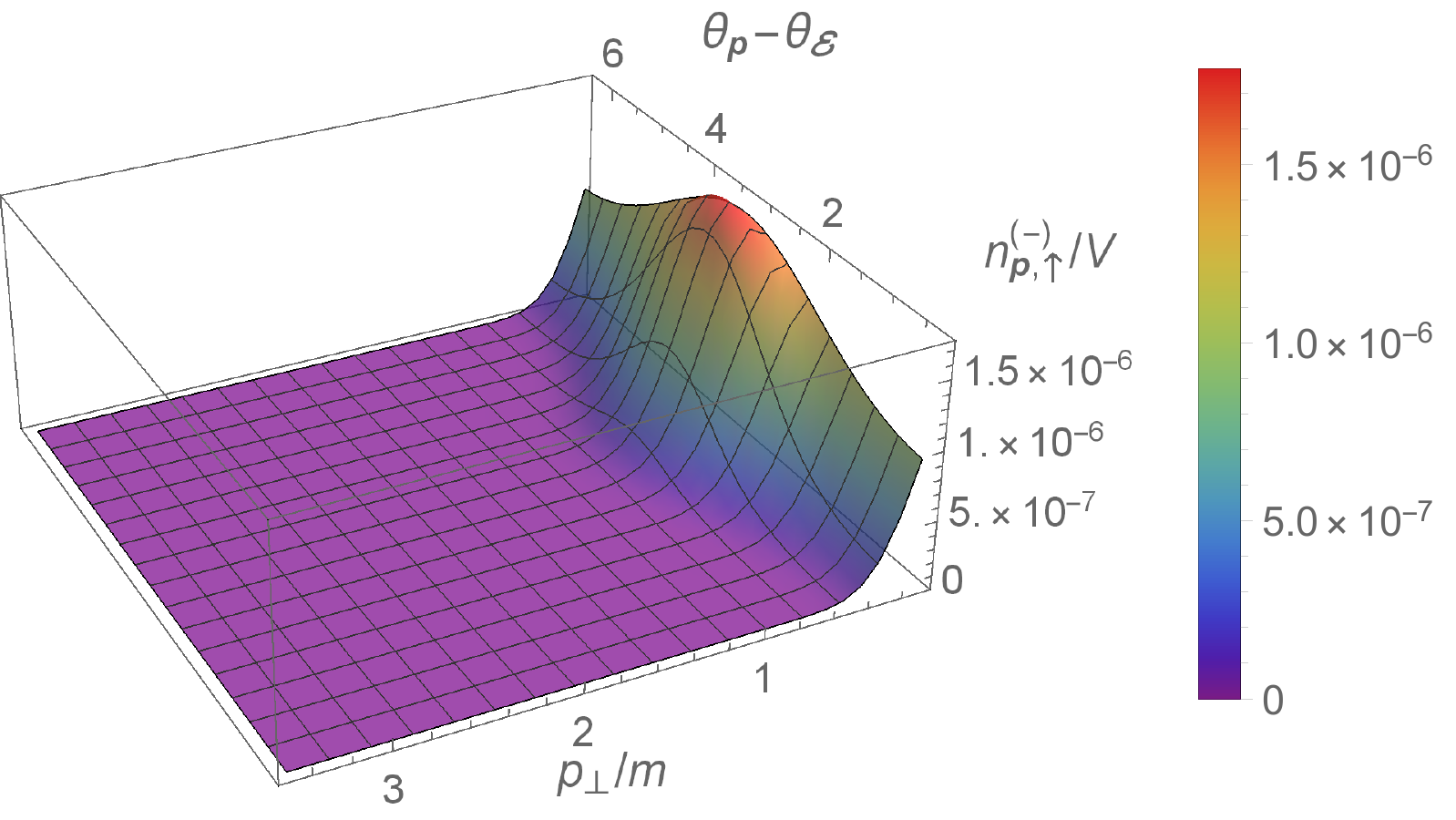}
\hspace*{-1mm}\includegraphics[clip, width=0.345\textwidth]{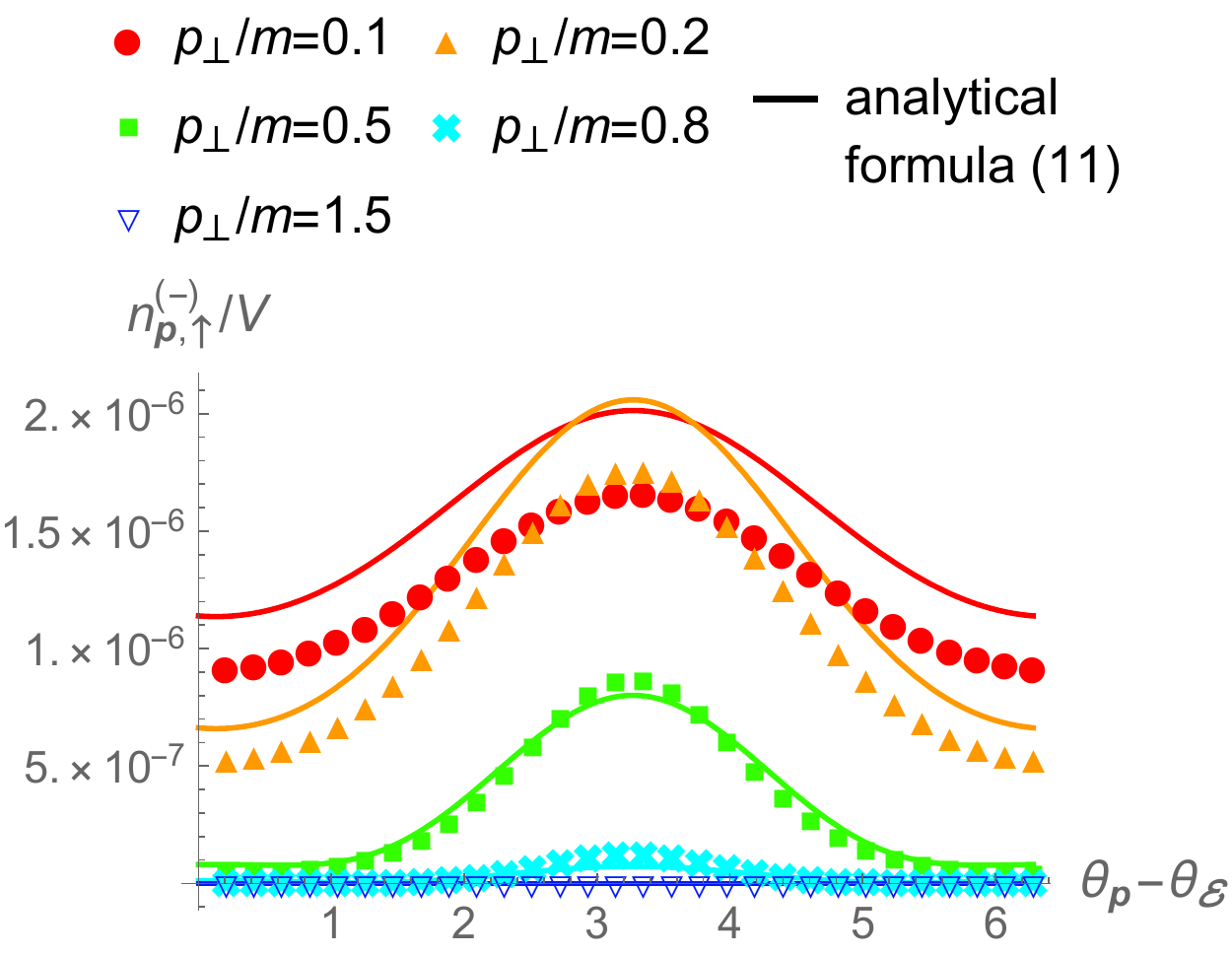}
\hspace*{-1mm}\includegraphics[clip, width=0.342\textwidth]{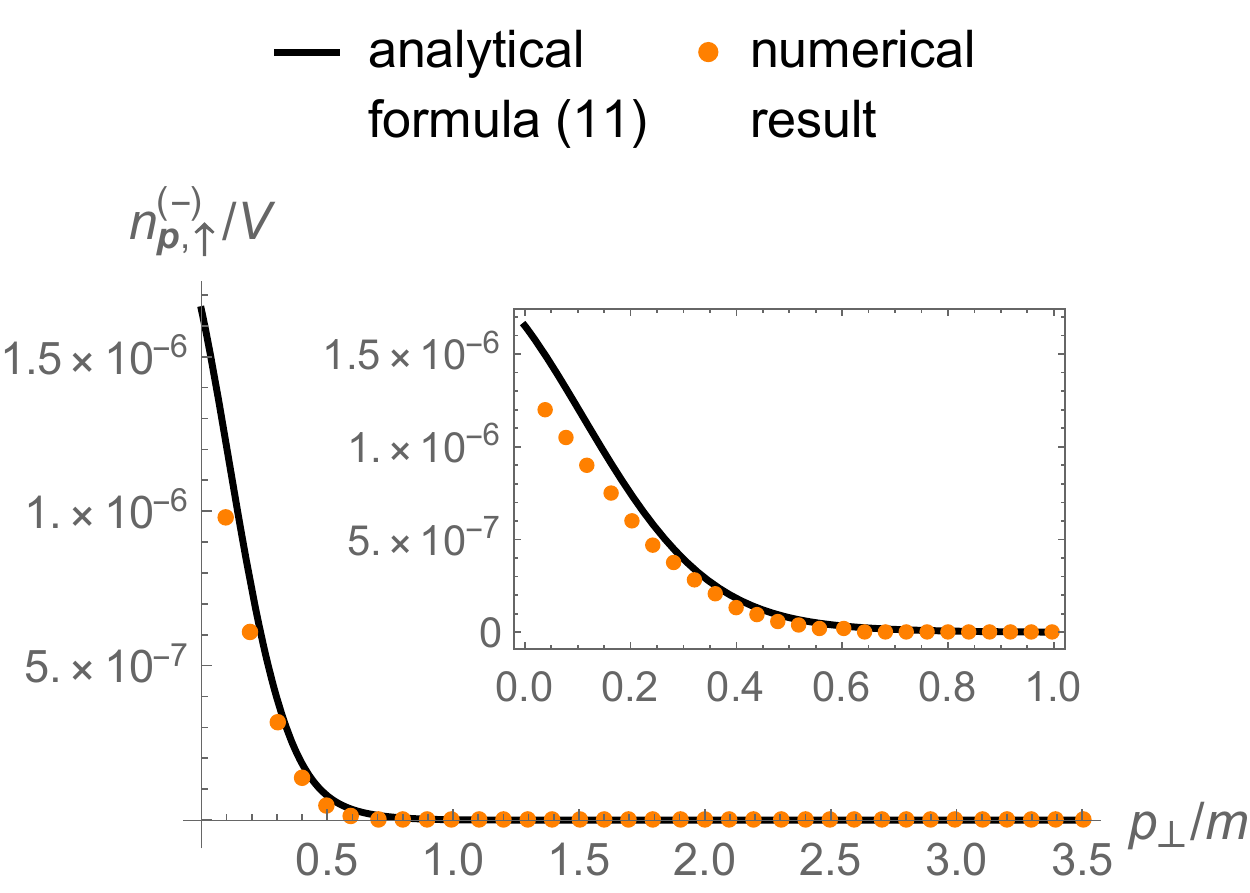}
\vspace*{-1mm} \hfill \\
\mbox{(ii) Small frequency $\Omega/m =0.5 $ and strong perturbation ${\mathcal E}_{\perp}/\bar{E} = 0.2 $}\hfill \\
\vspace*{1mm}
\hspace*{-10mm}\includegraphics[clip, width=0.365\textwidth]{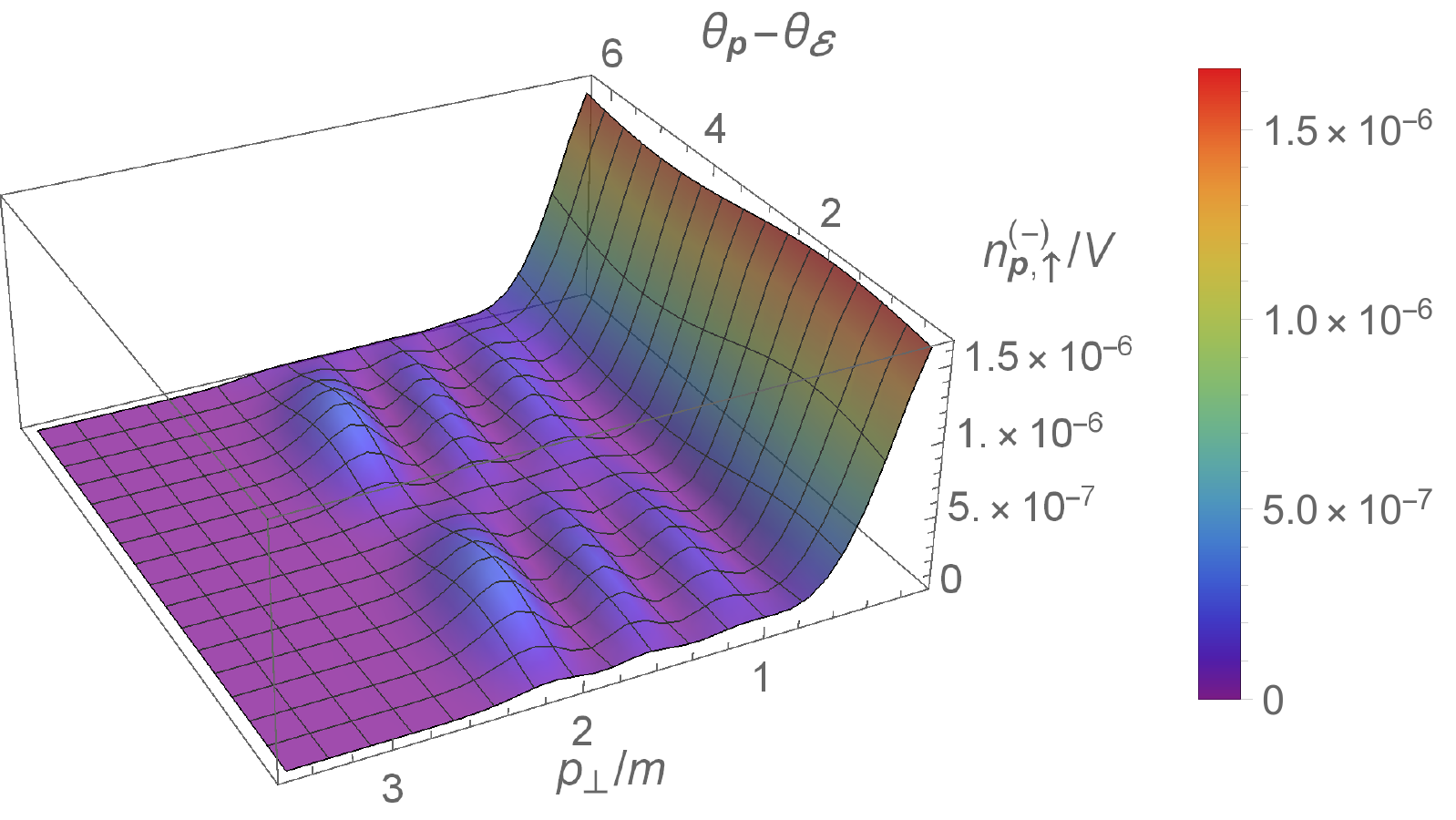}
\hspace*{-1mm}\includegraphics[clip, width=0.345\textwidth]{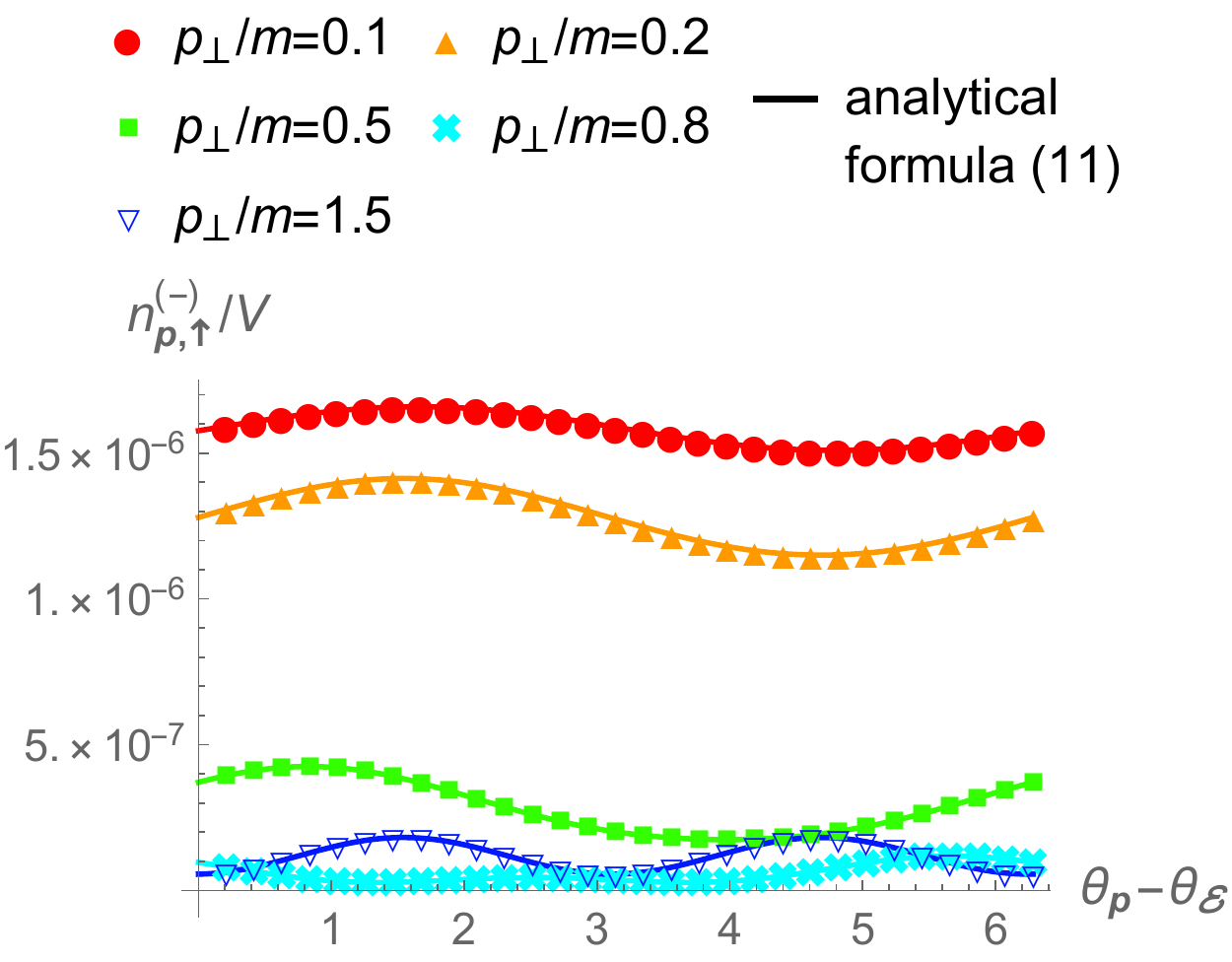}
\hspace*{-1mm}\includegraphics[clip, width=0.342\textwidth]{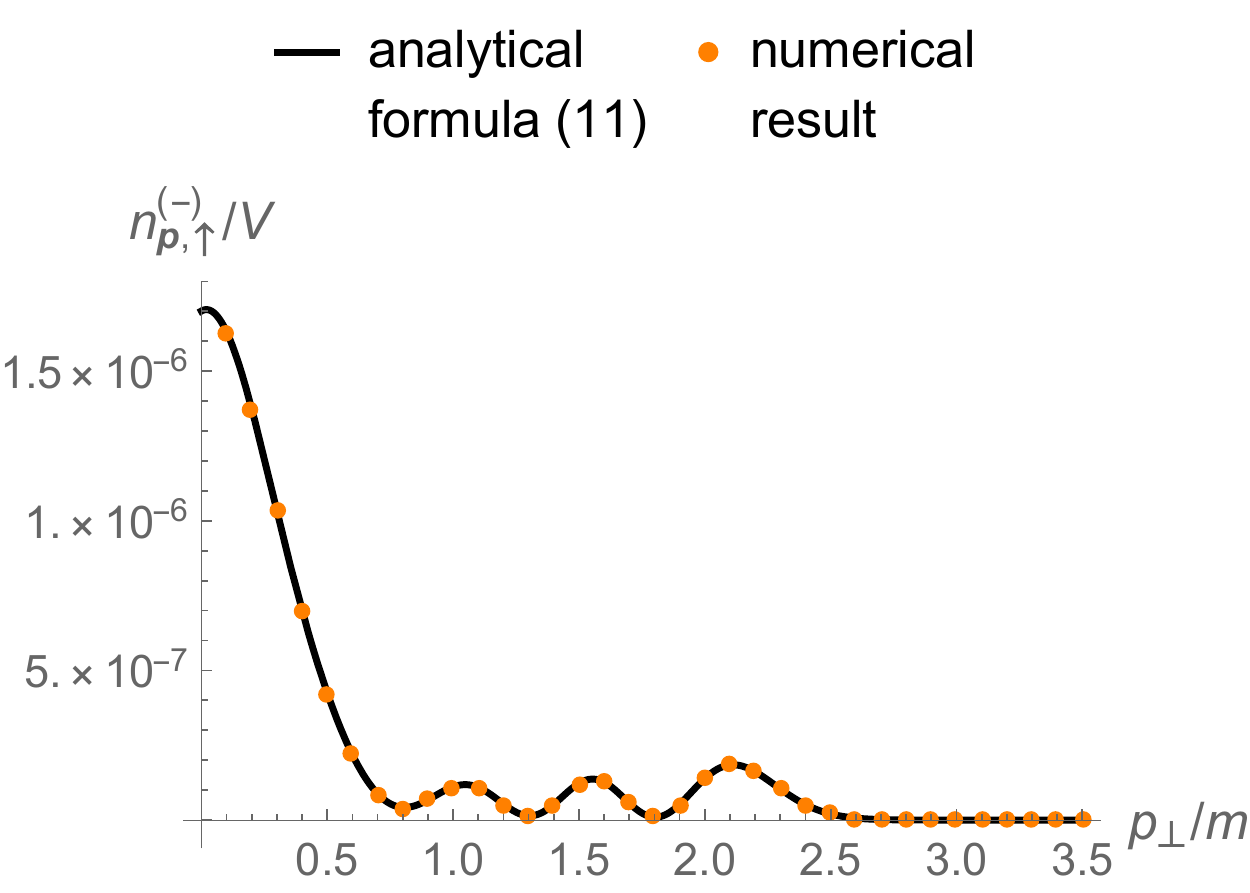}
\vspace*{-1mm} \hfill \\
\mbox{(iii) Large frequency $\Omega/m =5.0 $ and weak perturbation ${\mathcal E}_{\perp}/\bar{E} = 0.025 $}\hfill \\
\vspace*{1mm}
\hspace*{-10mm}\includegraphics[clip, width=0.365\textwidth]{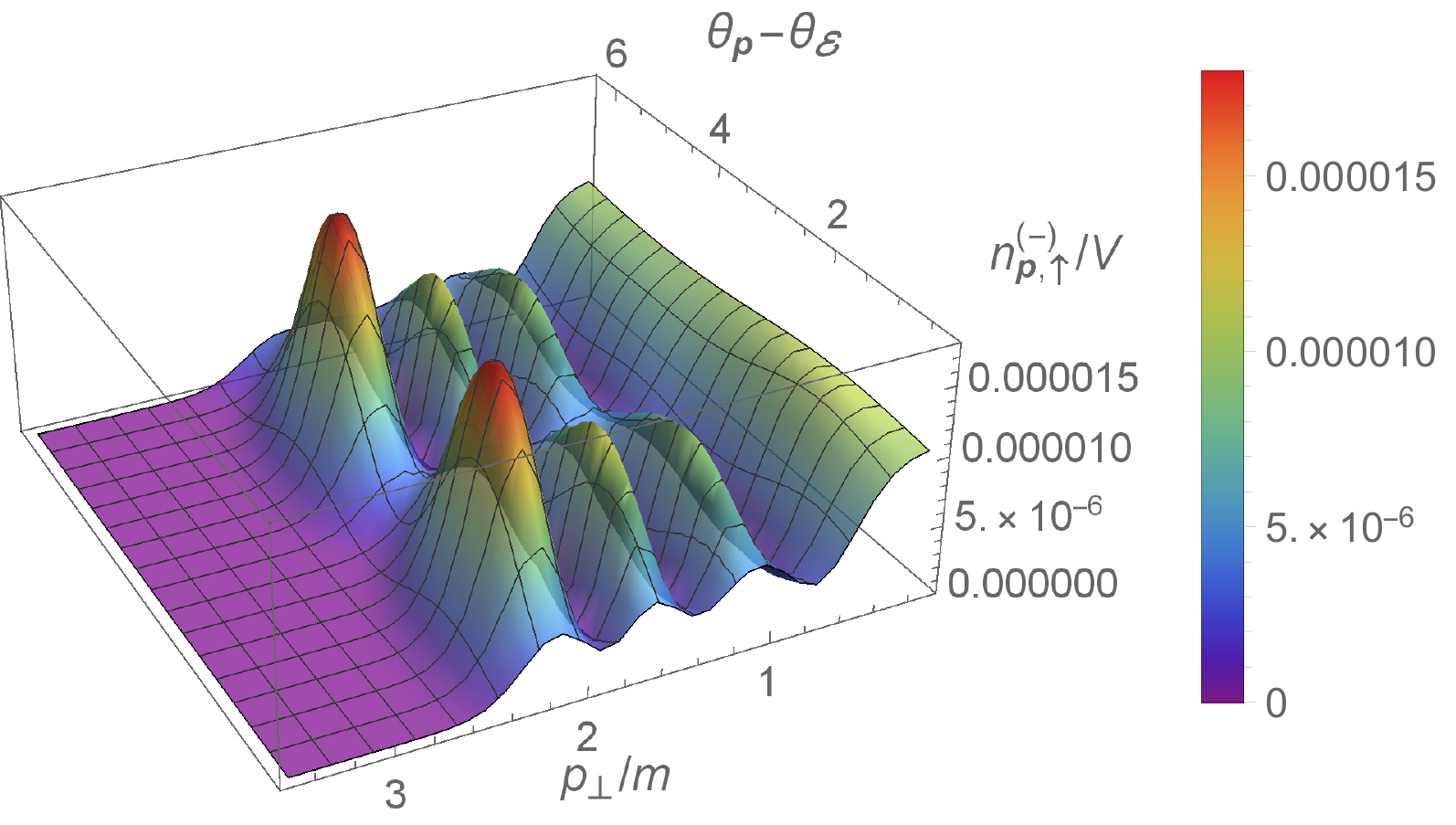}
\hspace*{-1mm}\includegraphics[clip, width=0.345\textwidth]{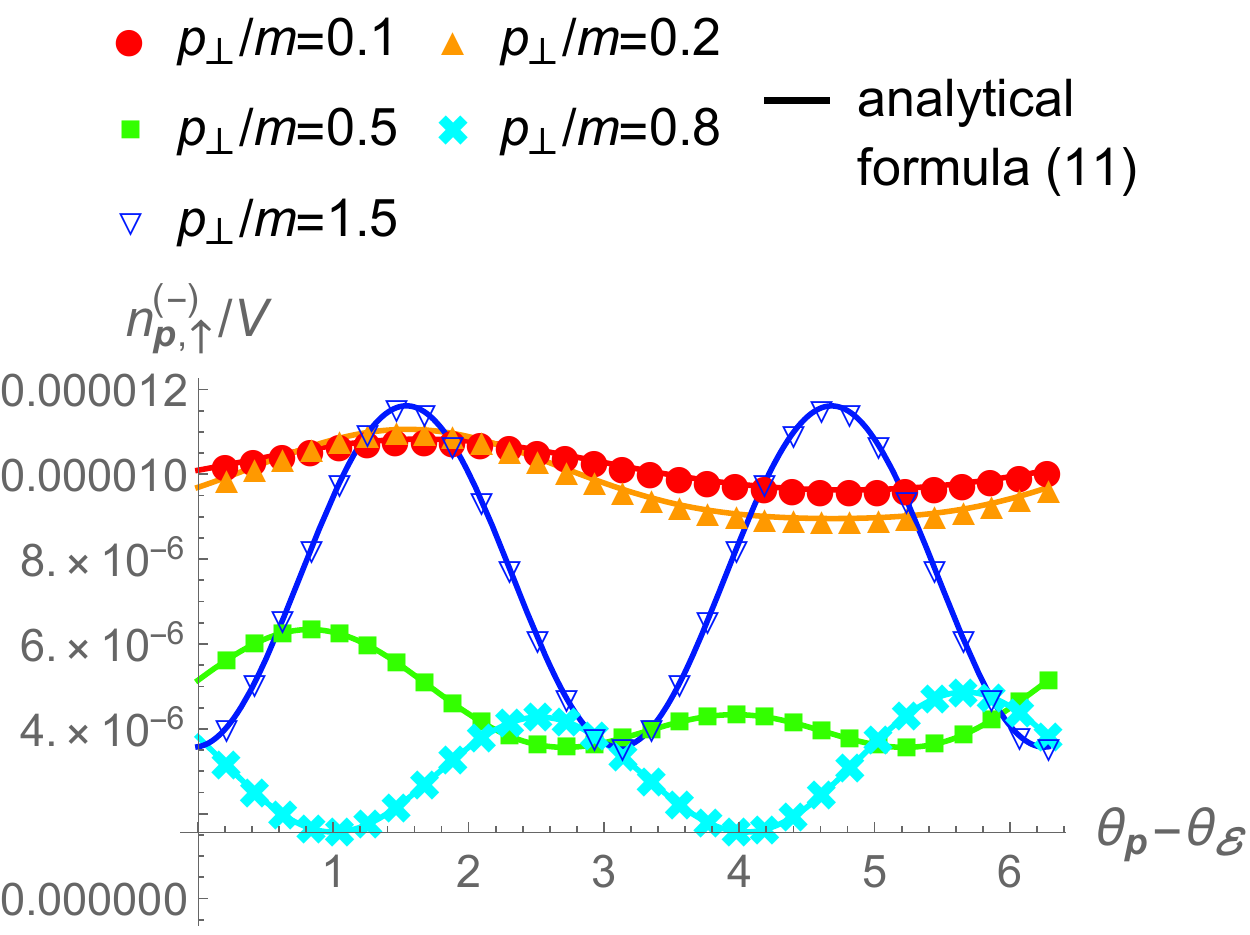}
\hspace*{-1mm}\includegraphics[clip, width=0.342\textwidth]{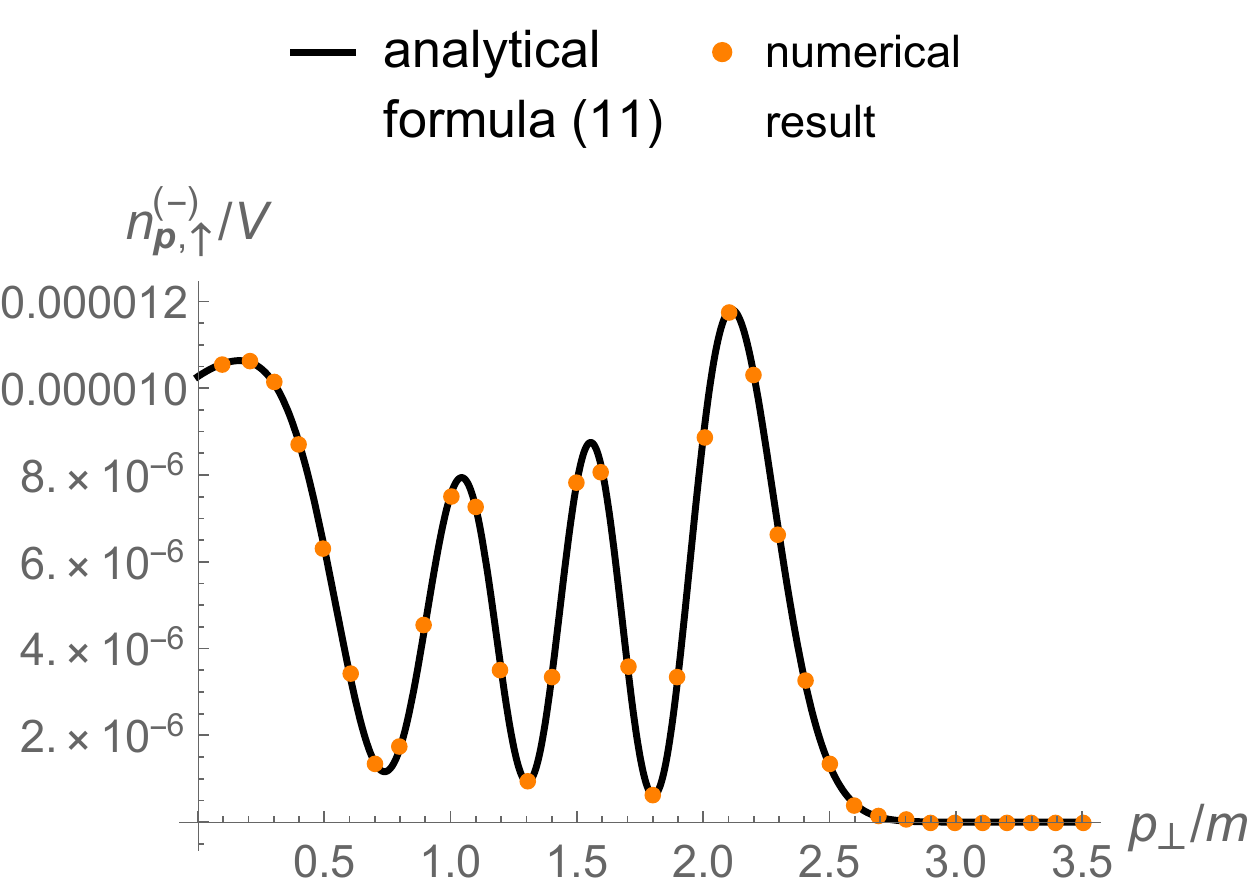}
\vspace*{-1mm} \hfill \\
\mbox{(iv) Large frequency $\Omega/m =5.0 $ and strong perturbation ${\mathcal E}_{\perp}/\bar{E} = 0.2 $}
\caption{\label{fig9} (color online) The numerical results of the momentum distribution $n^{(-)}_{{\bm p},\uparrow}$ as a function of $(\theta_{\bm p}, p_{\perp})$ (left); as a function of $\theta_{\bm p}$ for several values of $p_{\perp}$ (center); and as a function of $p_{\perp}$ for fixed $\theta_{\bm p}-\theta_{\bm {\mathcal E}} = \pi/4$ (right).  As a comparison, the analytical results (\ref{eq11}) are plotted as the lines in the center and right panels.  The upper (i) and lower (ii) panels distinguish the size of the frequency $\Omega$ and the perturbation ${\mathcal E}_{\perp}$.  The parameters are the same as Fig.~\ref{fig4}, i.e., $e\bar{E}/m^2 = 0.4, {\mathcal E}_3/\bar{E} = 0, p_3/m = 0, \phi = 1,\ {\rm and\ }m\tau=100$.   }
\end{center}
\end{figure*}

Figure~\ref{fig9} shows the numerical results for the momentum distribution as a function of the azimuthal angle $\theta_{\bm p}$ and the transverse momentum $p_{\perp}$ (which corresponds to Fig.~\ref{fig4} in the main text).

The basic features are the same as for the $p_{\perp}$-dependence for the spin-imbalance (see  Sec.~\ref{sec3b4}): For small $\Omega \lesssim \sqrt{e\bar{E}}, \sqrt{m^2 + p_{\perp}^2}$, the production is non-perturbatively suppressed by the exponential factor $\exp[-\pi (m^2+p_{\perp}^2)/e\bar{E}]$ (see Eq.~(\ref{eq11})).  Therefore, the momentum distribution can be non-negligible only for small values of $p_{\perp} \lesssim \sqrt{e\bar{E}}$.  For large $\Omega \gtrsim \sqrt{e\bar{E}}, \sqrt{m^2+p_{\perp}^2}$, the perturbative process dominates the production, so that the momentum distribution is no longer suppressed exponentially until $p_{\perp} \lesssim \Omega$.  The dominance of the perturbative effect implies that the momentum distribution depends on $\theta_{\bm p}$ as $(\sin (\theta_{\bm p} - \theta_{{\bm {\mathcal E}}}))^2, (\cos (\theta_{\bm p} - \theta_{{\bm {\mathcal E}}}))^2$ (see Eq.~(\ref{eq14})), which results in the two-peak structure at large $p_{\perp}$.  The distribution exhibits an oscillating behavior in terms of $p_{\perp}$ (see rightmost panels for large frequency (iii) and (iv) of Fig.~\ref{fig9}).  This is due to the Franz-Keldysh oscillation \cite{tay19}, and its frequency is determined by the phase factor $\varphi$ in Eq.~(\ref{eq19}).

\subsubsection{$p_{3}$-dependence}

\begin{figure*}[t!]
\begin{center}
\hspace*{-10mm}\includegraphics[clip, width=0.365\textwidth]{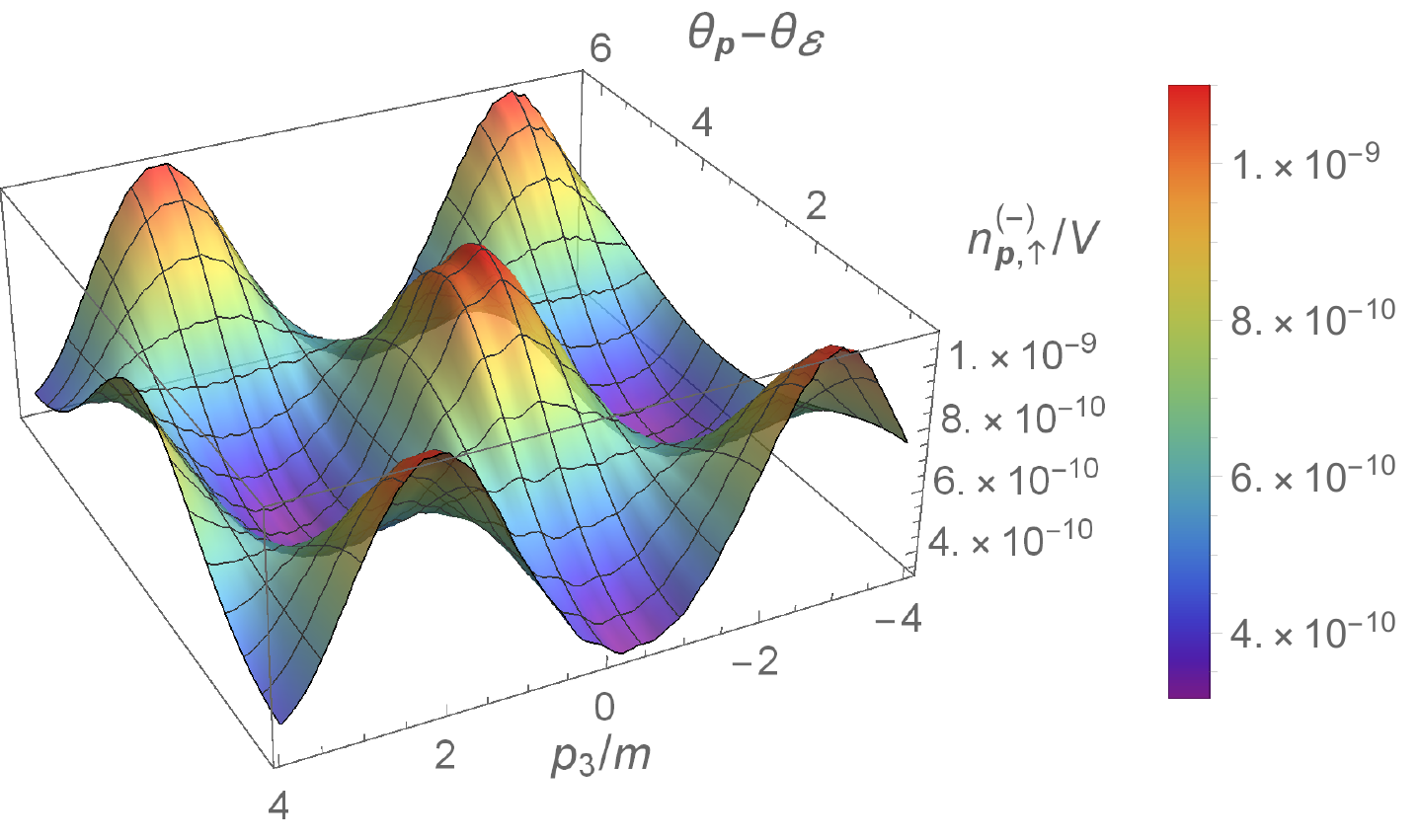}
\hspace*{-1mm}\includegraphics[clip, width=0.345\textwidth]{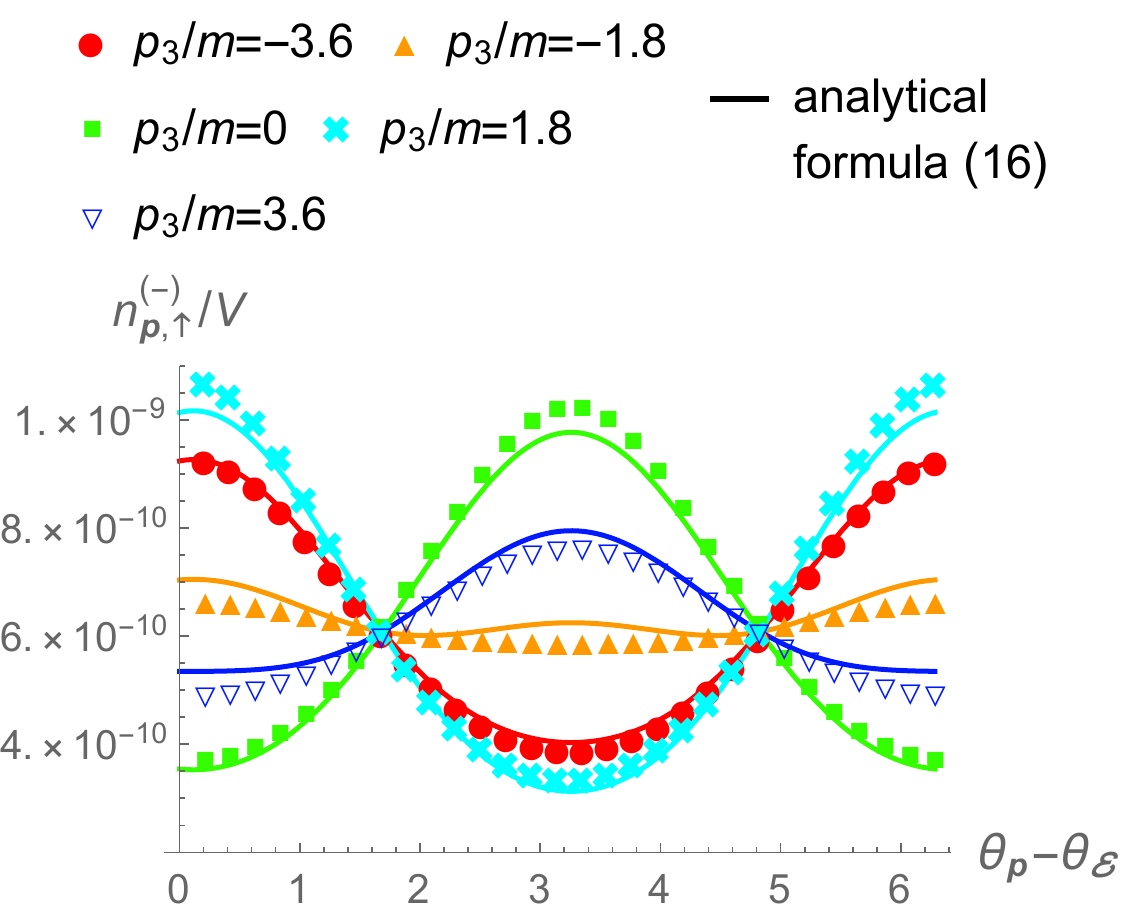}
\hspace*{-1mm}\includegraphics[clip, width=0.34\textwidth]{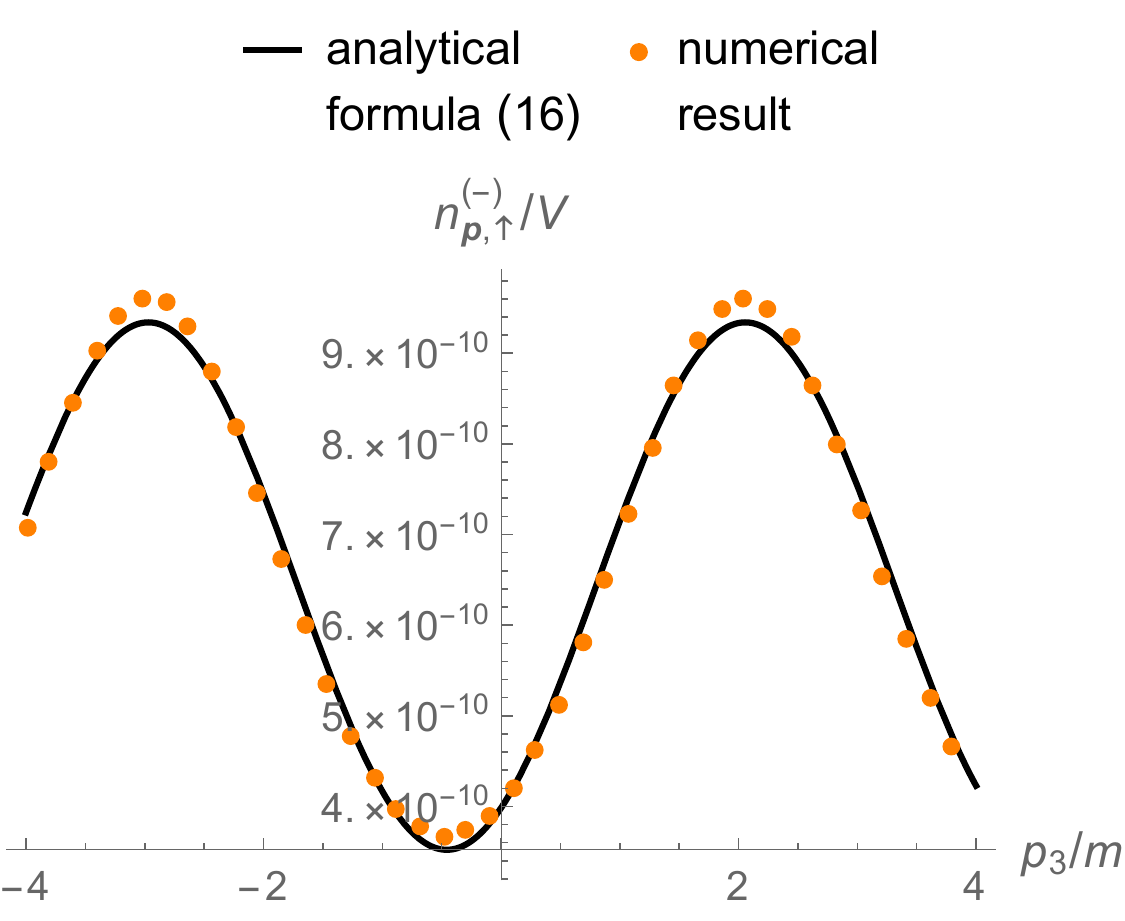}
\vspace*{-1mm}\hfill \\
\mbox{(i) Small frequency $\Omega/m =0.5 $ and weak perturbation ${\mathcal E}_{\perp}/\bar{E} = 0.025 $}\hfill \\
\vspace*{1mm}
\hspace*{-10mm}\includegraphics[clip, width=0.365\textwidth]{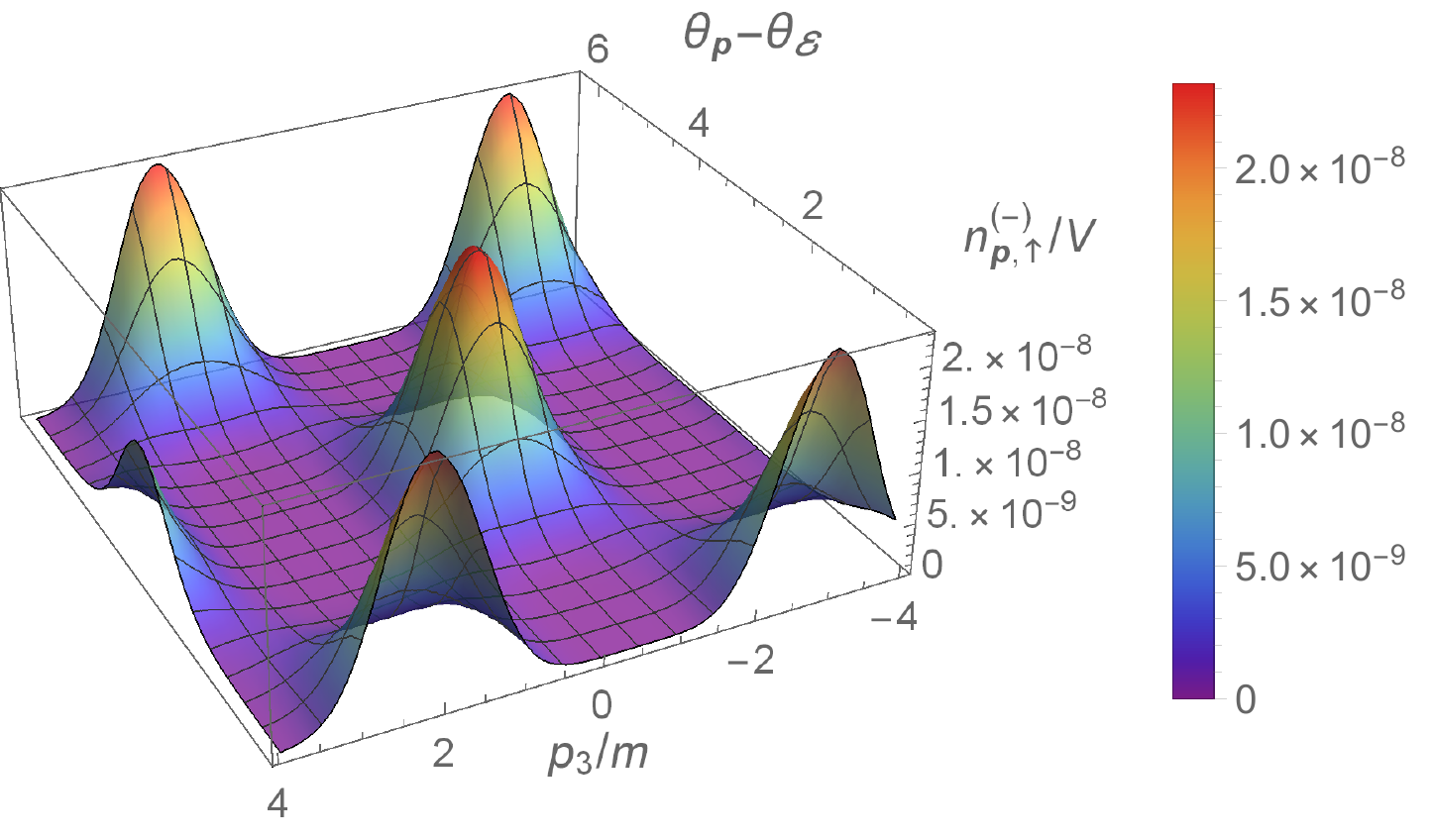}
\hspace*{-1mm}\includegraphics[clip, width=0.345\textwidth]{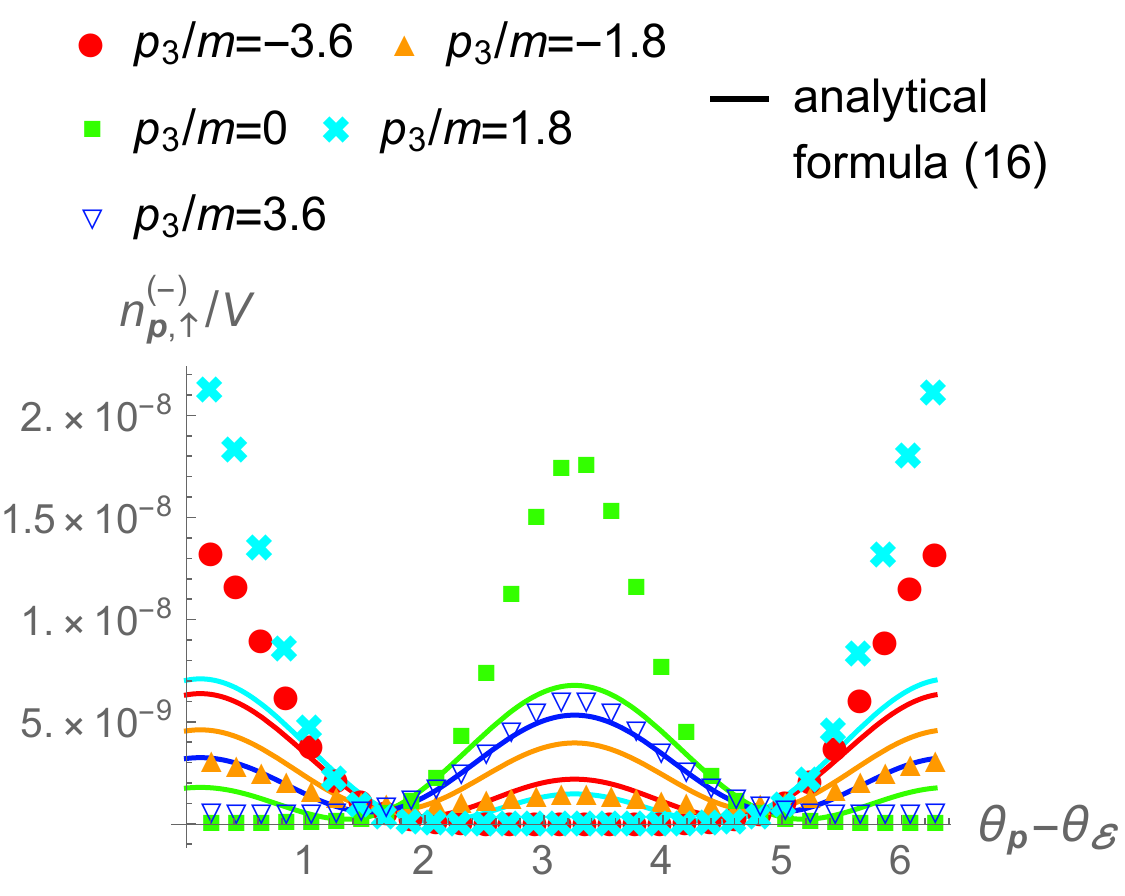}
\hspace*{-1mm}\includegraphics[clip, width=0.34\textwidth]{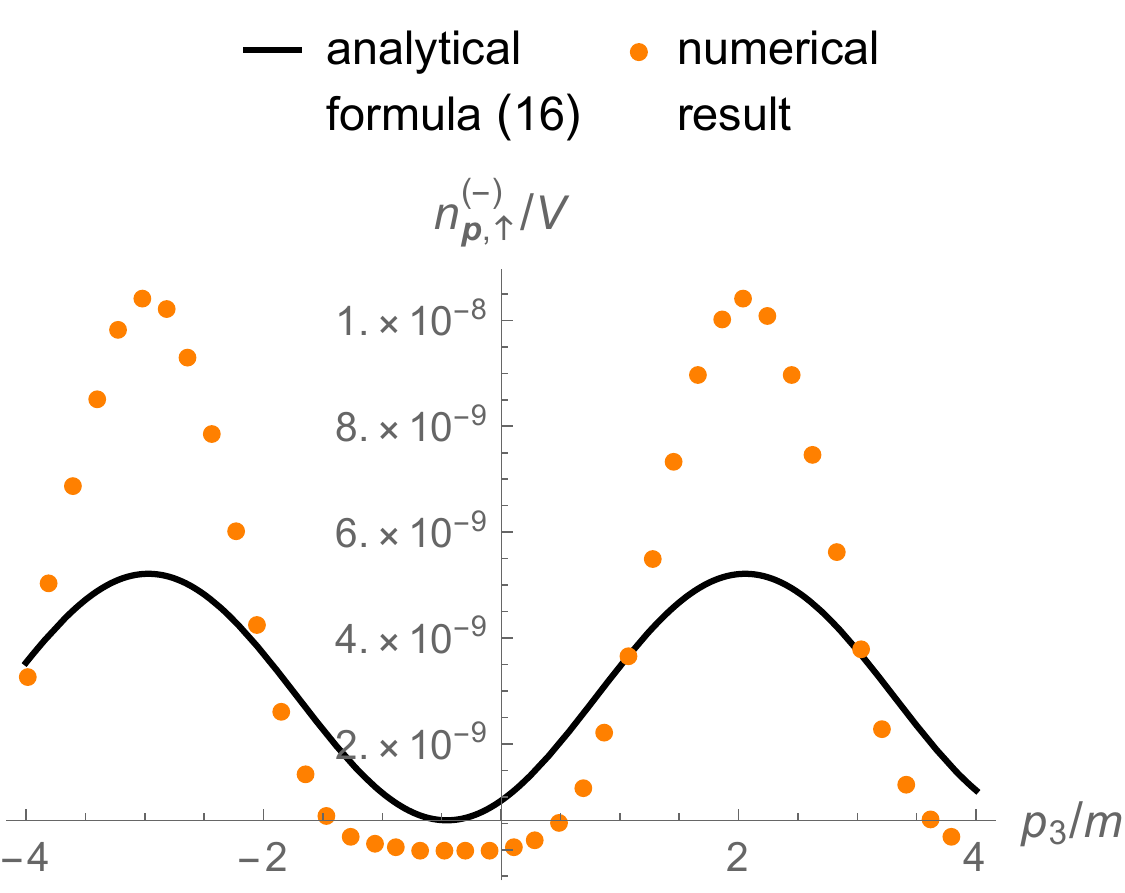}
\vspace*{-1mm} \hfill \\
\mbox{(ii) Small frequency $\Omega/m =0.5 $ and strong perturbation ${\mathcal E}_{\perp}/\bar{E} = 0.2 $}\hfill \\
\vspace*{1mm}
\hspace*{-10mm}\includegraphics[clip, width=0.365\textwidth]{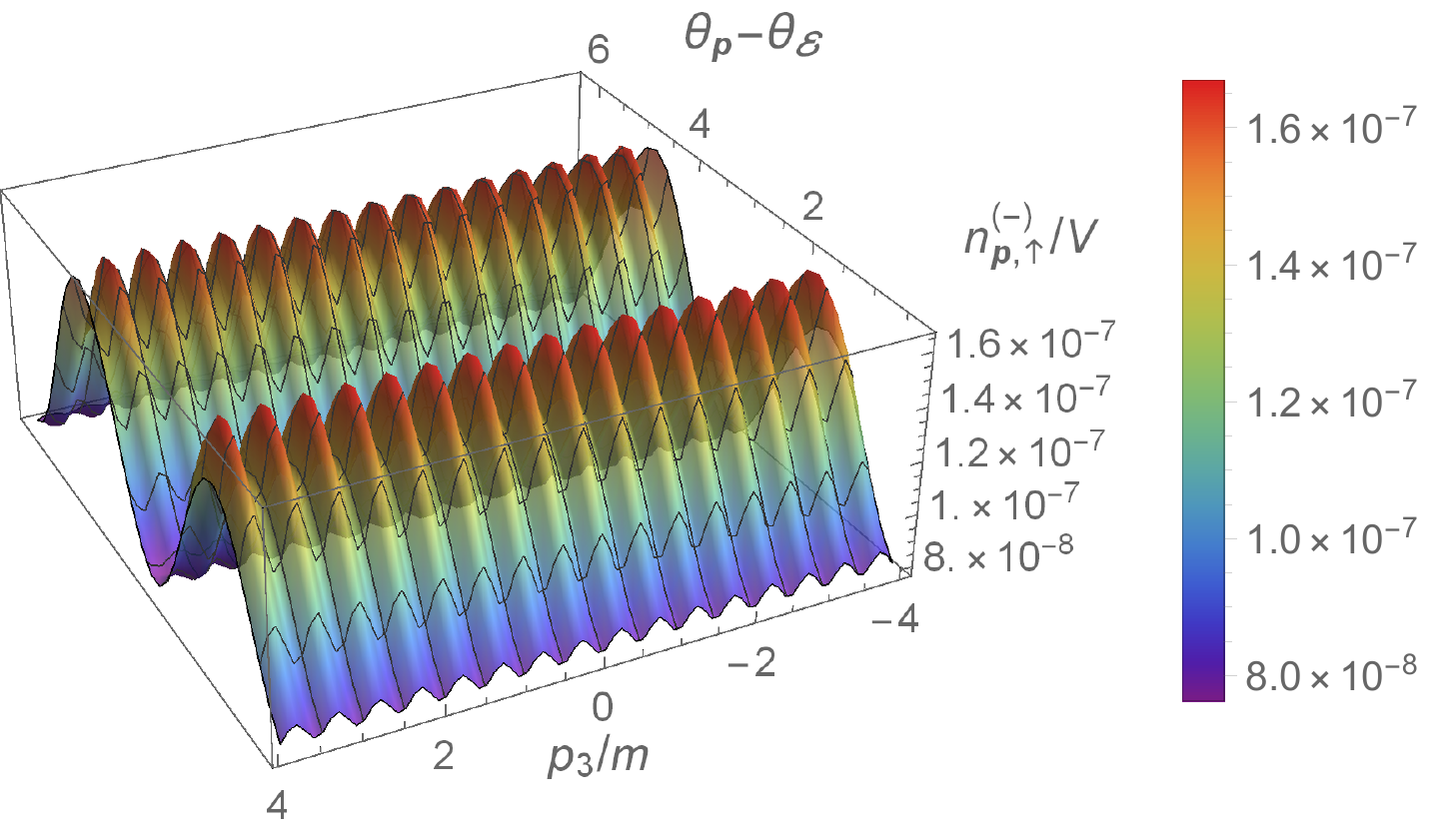}
\hspace*{-1mm}\includegraphics[clip, width=0.345\textwidth]{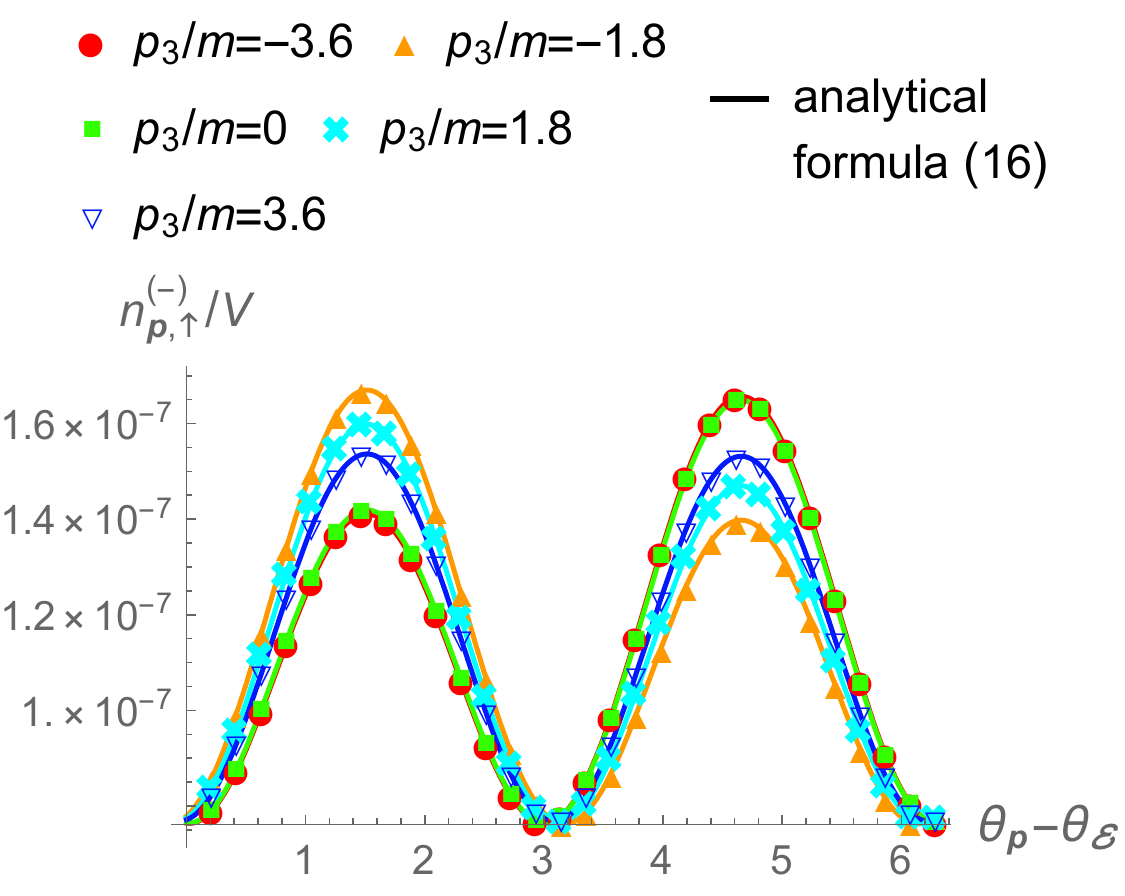}
\hspace*{-1mm}\includegraphics[clip, width=0.34\textwidth]{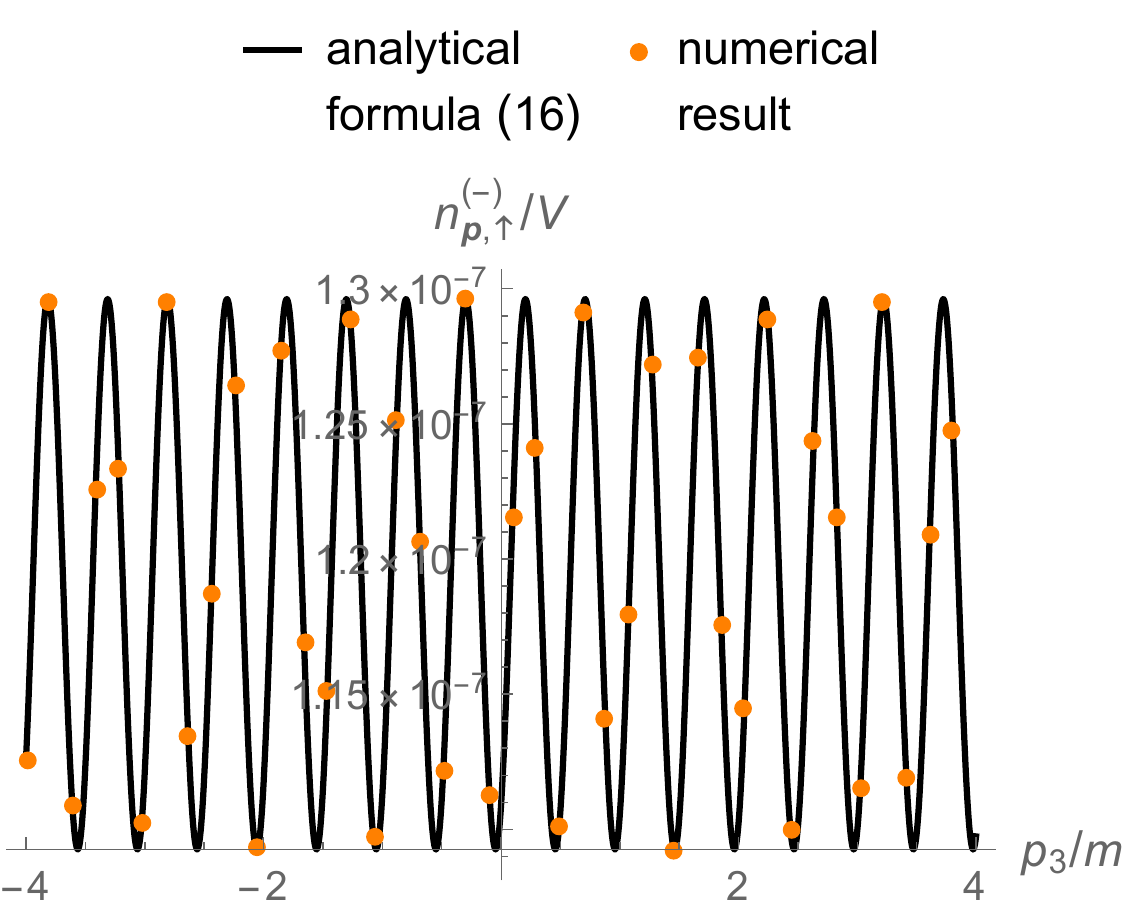}
\vspace*{-1mm} \hfill \\
\mbox{(iii) Large frequency $\Omega/m =5.0 $ and weak perturbation ${\mathcal E}_{\perp}/\bar{E} = 0.025 $}\hfill \\
\vspace*{1mm}
\hspace*{-10mm}\includegraphics[clip, width=0.365\textwidth]{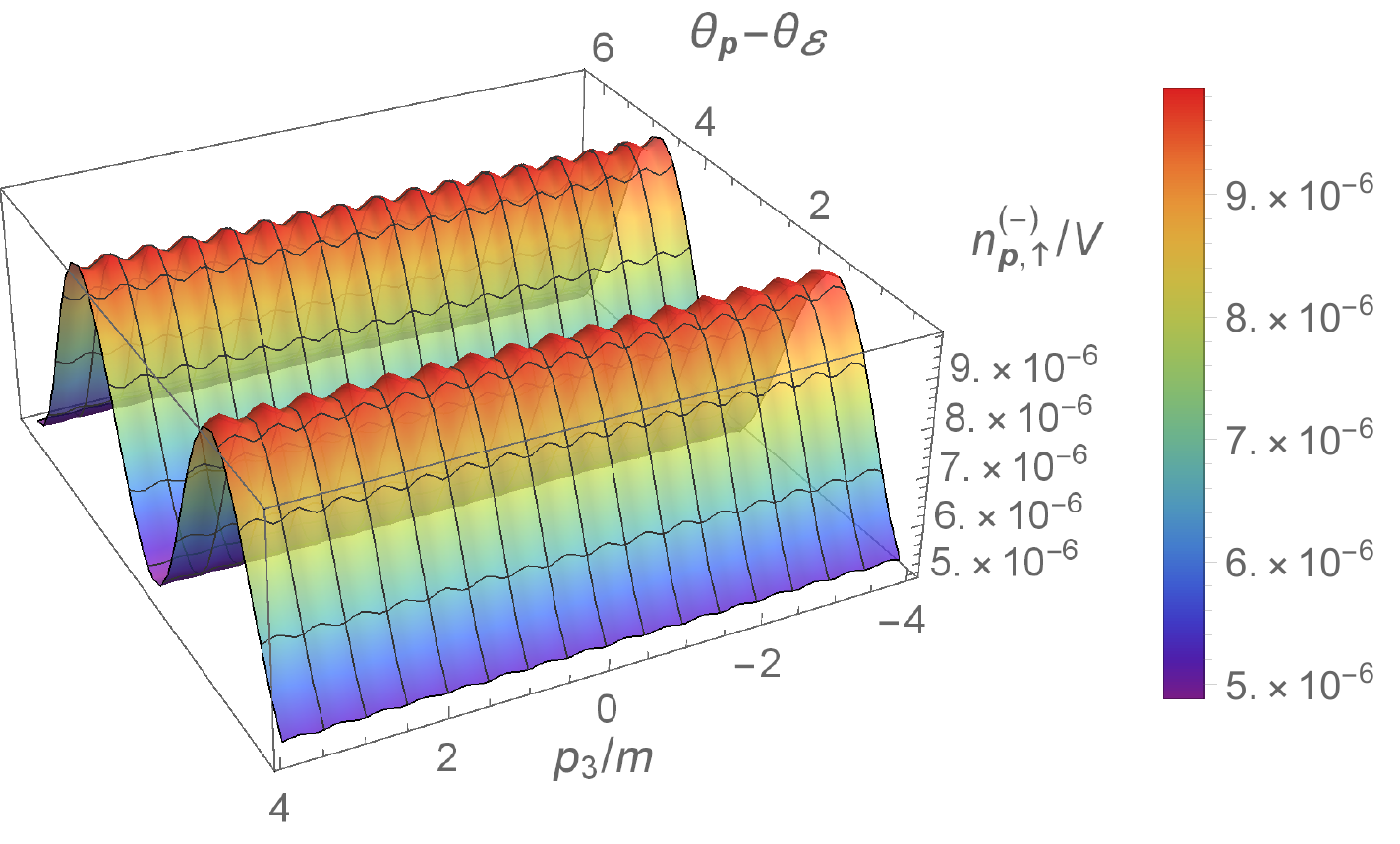}
\hspace*{-1mm}\includegraphics[clip, width=0.345\textwidth]{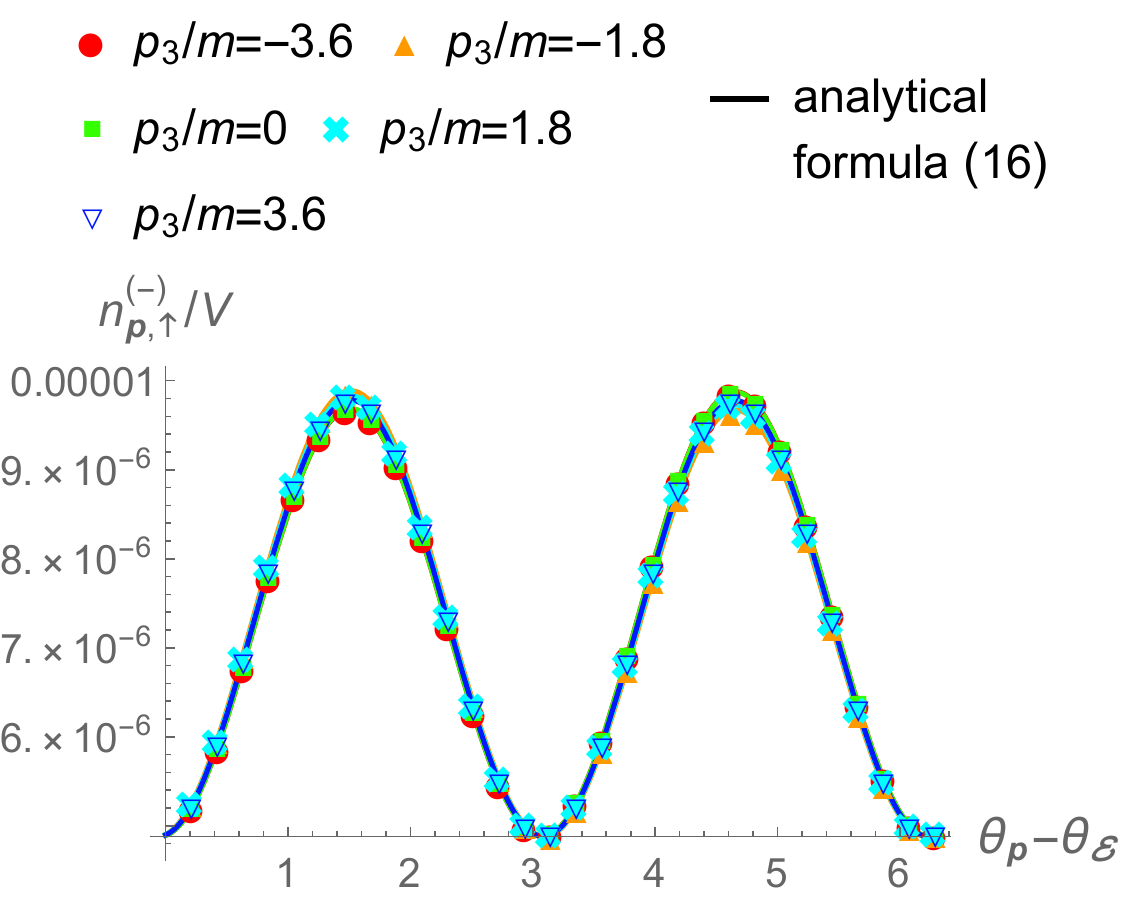}
\hspace*{-1mm}\includegraphics[clip, width=0.34\textwidth]{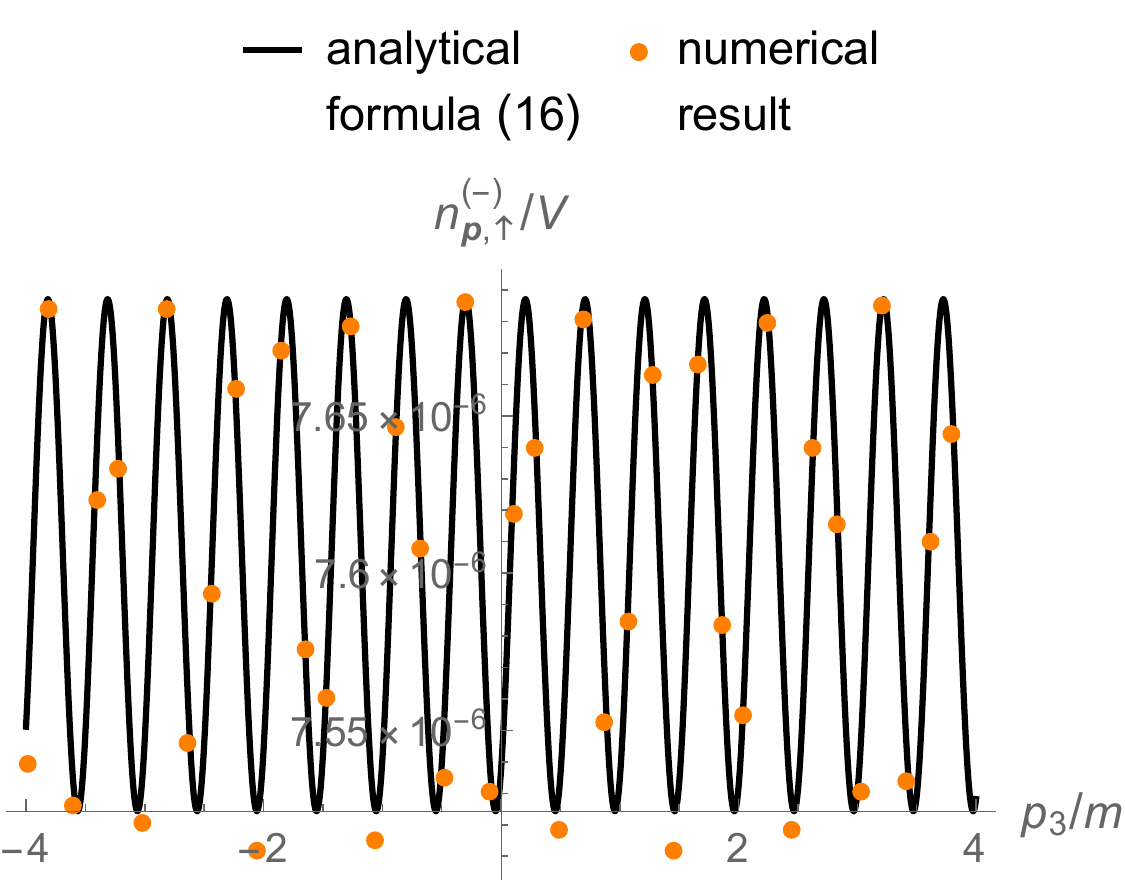}
\vspace*{-1mm} \hfill \\
\mbox{(iv) Large frequency $\Omega/m =5.0 $ and strong perturbation ${\mathcal E}_{\perp}/\bar{E} = 0.2 $}
\caption{\label{fig10} (color online) The numerical results of the momentum distribution $n^{(-)}_{{\bm p},\uparrow}$ as a function of $(\theta_{\bm p}, p_{3})$ (left); as a function of $\theta_{\bm p}$ for several values of $p_{3}$ (center); and as a function of $p_{3}$ for fixed $\theta_{\bm p}-\theta_{\bm {\mathcal E}} = \pi/4$ (right).  As a comparison, the analytical results (\ref{eq11}) are plotted as the lines in the center and right panels.  The upper (i) and lower (ii) panels distinguish the size of the frequency $\Omega$ and the perturbation ${\mathcal E}_{\perp}$.  The parameters are the same as Fig.~\ref{fig5}, i.e., $e\bar{E}/m^2 = 0.4, {\mathcal E}_3/\bar{E} = 0, p_{\perp}/m = 1, \phi = 1,\ {\rm and\ }m\tau=100$.   }
\end{center}
\end{figure*}

Figure~\ref{fig10} shows the numerical results for the $p_3$-dependence of the momentum distribution (which corresponds to Fig.~\ref{fig5} in the main text).  

As explained in Sec.~\ref{sec3b5}, $p_3$ is related to the production time of a pair as $t_{\rm prod} = -p_3/e\bar{E}$, so that the value of the weak field at $x^0 = t_{\rm prod}$ becomes important.  As a result, the momentum distribution becomes dependent on $p_3$, and exhibits an oscillating behavior in $p_3$ with frequency $\Omega/e\bar{E}$.

As shown in panel (ii) of Fig.~\ref{fig10}, the analytical formula (\ref{eq11}) deviates from the numerical results for large ${\mathcal E}_{\perp}$ with small $\Omega$.  As explained in the main text (see Sec.~\ref{sec3b3}), this is because the formula (\ref{eq11}) only takes into account the one-photon process in Eq.~(\ref{eq7}) and multi-photon processes become important for large ${\mathcal E}_{\perp}$ with small $\Omega$.  It is interesting to note that multi-photon processes strongly sharpen the $p_3$- and $\theta_{\bm p}$-distributions.  This is because the scattering amplitude for multi-photon processes with $n$-photons is always proportional to $|{\bm {\mathcal E}}|^n \sim |{\bm {\mathcal E}}(t_{\rm prod})|^n \propto (\cos (\phi -p_3/e\bar{E}))^n$, which is much sharper than that for the one-photon process $\propto \cos (\phi -p_3/e\bar{E})$.  Also, the weak field couples to ${\bm p}$ as ${\bm p} \cdot {\bm {\mathcal E}}, {\bm p} \times {\bm {\mathcal E}}$, and thus the scattering amplitude would contain higher order harmonics such as $({\bm p}\cdot {\bm {\mathcal E}})^n \propto (\cos(\theta_{\bm p}-\theta_{\bm {\mathcal E}}))^n$ and $ ({\bm p} \times {\bm {\mathcal E}})^n  \propto (\sin (\theta_{\bm p}-\theta_{\bm {\mathcal E}}))^n$.  These higher harmonics make the $\theta_{\bm p}$-distribution sharper.  It would be interesting to study the multi-photon effects further by extending our perturbative method, which we leave for future work.


\begin{thebibliography}{99}

\bibitem{sau31} F.~Sauter, ``{\it Ueber das Verhalten eines Elektrons im homogenen elektrischen Feld nach der relativistischen Theorie Diracs}," Z.~Phys. {\bf 69}, 742 (1931).
\bibitem{hei36} W.~Heisenberg and H.~Euler, ``{\it Folgerungen aus der Diracschen Theorie des Positrons}," Z.~Phys. {\bf 98}, 714 (1936).

\bibitem{sch51} J.~Schwinger, ``{\it On Gauge Invariance and Vacuum Polarization}," Phys.~Rev. {\bf 82}, 664 (1951).

\bibitem{lan32} L.~Landau, ``{\it Zur Theorie der Energieubertragung. II},'' Phys.~Z.~Sowjetunion {\bf 2}, 46 (1932).

\bibitem{zen32} C.~Zener, ``{\it Non-Adiabatic Crossing of Energy Levels},'' Proc.~R.~Soc.~Ser.~A {\bf 137}, 696 (1932).

\bibitem{stu32} E.~C.~G.~Stueckelberg, ``{\it Theorie der unelastischen Stosse zwischen Atomen},'' Helv.~Phys.~Acta. {\bf 5}, 369 (1932).

\bibitem{maj32} E. Majorana, ``{\it Atomi orientati in campo magnetico variabile},'' Nuovo~Cimento {\bf 9}, 43 (1932).

\bibitem{yan08} V.~Yanovsky {\it et al}., ``{\it Ultra-high intensity 300{\rm \;TW} laser at 0.1{\rm \;Hz} repetition rate},'' Opt.~Express {\bf 16}, 2109 (2008).

\bibitem{eli} \url{www.eli-beams.eu}

\bibitem{hiper} \url{www.hiper-laser.org}

\bibitem{sch08} R.~Schutzhold, H.~Gies, and G.~Dunne, ``{\it Dynamically Assisted Schwinger Mechanism}," Phys.~Rev.~Lett. {\bf 101}, 130404 (2008).

\bibitem{piz09} A.~Di~Piazza, E.~Lotstedt, A.~I.~Milstein, and C.~H.~Keitel, ``{\it Barrier control in tunneling $e^+$-$e^-$ photoproduction},'' Phys.~Rev.~Lett. {\bf 103}, 170403 (2009).

\bibitem{dun09} G.~V.~Dunne, H.~Gies, and R.~Schutzhold, ``{\it Catalysis of Schwinger Vacuum Pair Production},'' Phys.~Rev.~D {\bf 80}, 111301(R) (2009).

\bibitem{mon10a} A.~Monin, and M.~B.~Voloshin, ``{\it Photon-stimulated production of electron-positron pairs in electric field},'' Phys.~Rev.~D {\bf 81}, 025001 (2010).

\bibitem{mon10b} A.~Monin, and M.~B.~Voloshin, ``{\it Semiclassical Calculation of Photon-Stimulated Schwinger Pair Creation},'' Phys.~Rev.~D {\bf 81}, 085014 (2010).

\bibitem{fra58} V.~W.~Franz, ``{\it Einfluss eines elektrischen Feldes auf eine optische Absorptionskante},'' Z. Naturforsch. Teil A {\bf 13}, 484 (1958).

\bibitem{kel58} L.~V.~Keldysh, ``{\it The Effect of a Strong Electric Field on the Optical Properties of Insulating Crystals },'' Sov.~Phys.~JETP {\bf 7}, 788 (1958).

\bibitem{tah63} K.~Tharmalingam, ``{\it Optical Absorption in the Presence of a Uniform Field},'' Phys.~Rev. {\bf 130}, 2204 (1963).

\bibitem{cal63} J.~Callaway, ``{\it Optical Absorption in an Electric Field},'' Phys.~Rev. {\bf 130}, 549 (1963).

\bibitem{tay19} H. Taya, ``{\it Franz-Keldysh effect in strong-field QED},'' Phys.~Rev.~D {\bf 99}, 056006 (2019).  

\bibitem{gre17} G.~Torgrimsson, C.~Schneider, and R.~Schutzhold, ``{\it Dynamically assisted Sauter-Schwinger effect - non-perturbative versus perturbative aspects },'' JHEP {\bf 06}, 043 (2017).

\bibitem{gre19} G.~Torgrimsson, ``{\it Perturbative methods for assisted nonperturbative pair production},'' Phys.~Rev.~D {\bf 99}, 096002 (2019).  

\bibitem{fur51} W.~H.~Furry, ``{\it On Bound State and Scattering in Positron Theory},'' Phys.~Rev. {\bf 81}, 115 (1951).

\bibitem{fra81} E.~S.~Fradkin, and D.~M.~Gitman, ``{\it Furry Picture for Quantum Electrodynamics With Pair Creating External Field },'' Fortschr.~Phys. {\bf 29}, 381 (1981).

\bibitem{fra91} E.~S.~Fradkin, D.~M.~Gitman, and S.~M.~Shvartsman, ``{\it Quantum Electrodynamics with Unstable Vacuum},'' Springer-Verlag, Berlin (1991).

\bibitem{koh18} C.~Kohlfurst, ``{\it Spin-states in multiphoton pair production for circularly polarized light},'' Phys.~Rev.~D {\bf 99}, 096017 (2019).  

\bibitem{bre70} E.~Brezin and C.~Itzykson, ``{\it Pair Production in Vacuum by an Alternating Field}," Phys.~Rev.~D {\bf 2}, 1191 (1970).

\bibitem{pop72} V.~S.~Popov, ``{\it Pair production in a variable external field (quasiclassical approximation)}," JETP {\bf 34}, 709 (1972).

\bibitem{kel65} L.~V.~Keldysh, ``{\it Ionization in the field of a strong electromagnetic wave}," JETP {\bf 20}, 1307 (1965).

\bibitem{tay14} H.~Taya, H.~Fujii, and K.~Itakura, ``{\it Finite pulse effects on $e^+ e^-$ pair creation from strong electric fields}," Phys.~Rev.~D {\bf 90}, 014039 (2014).

\bibitem{gel16} F.~Gelis, and N.~Tanji, ``{\it Schwinger mechanism revisited},'' Prog.~Part.~Nucl.~Phys. {\bf 87}, 1 (2016).


\bibitem{gre18} G.~Torgrimsson, C.~Schneider, and R.~Schutzhold, ``{\it Sauter-Schwinger pair creation dynamically assisted by a plane wave},'' Phys.~Rev.~D {\bf 97}, 096004 (2018).

\bibitem{itz80} C.~Itzykson, J.-B.~Zuber, ``{\it Quantum Field Theory},'' McGraw-Hill (1980).

\bibitem{nik70} A.~I.~Nikishov, ``{\it Pair Production by a Constant External Field}," JETP {\bf 30}, 660 (1970).

\bibitem{tan09} N.~Tanji, ``{\it Dynamical view of pair creation in uniform electric and magnetic fields}," Ann.~Phys. {\bf 324}, 1691 (2009).

\bibitem{tay17} H.~Taya, ``{\it Quark and Gluon Production from a Boost-invariantly Expanding Color Electric Field },'' Phys.~Rev.~D {\bf 96}, 014033 (2017).

\bibitem{birrel} N.~D.~Birrell and P.~C.~W.~Davies, ``{\it Quantum Fields in Curved Space}," Cambridge University Press (1982).

\bibitem{pia12} A.~Di~Piazza, C.~Muller, K.~Z.~Hatsagortsyan, and C.~H.~Keitel, ``{\it Extremely high-intensity laser interactions with fundamental quantum systems},'' Rev.~Mod.~Phys. {\bf 84}, 1177 (2012).

\bibitem{koh13} C.~Kohlfurst, M.~Mitter, G.~von~Winckel, F.~Hebenstreit, and R.~Alkofer ``{\it Optimizing the pulse shape for Schwinger pair production},'' Phys.~Rev.~D {\bf 88}, 045028 (2013).

\bibitem{heb14} F.~Hebenstreit, and F.~Fillion-Gourdeau, ``{\it Optimization of Schwinger pair production in colliding laser pulses},'' Phys.~Lett.~B {\bf 739}, 189 (2014).

\bibitem{lin15} M.~F.~Linder, C.~Schneider, J.~Sicking, N.~Szpak, and R.~Schutzhold, ``{\it Pulse shape dependence in the dynamically assisted Sauter-Schwinger effect}," Phys.~Rev.~D {\bf 92}, 085009 (2015).

\bibitem{low75} F.~E.~Low, ``{\it Model of the bare Pomeron}," Phys.~Rev.~D {\bf 12}, 163 (1975).

\bibitem{nus75} S.~Nussinov, ``{\it Colored-Quark Version of Some Hadronic Puzzles}," Phys.~Rev.~Lett. {\bf 34}, 1286 (1975).

\bibitem{kov95a} A.~Kovner, L.~McLerran, and H.~Weigert, ``{\it Gluon Production at High Transverse Momentum in the McLerran-Venugopalan Model of Nuclear Structure Functions}," Phys.~Rev.~D {\bf 52}, 3809 (1995).

\bibitem{kov95b} A.~Kovner, L.~McLerran, and H.~Weigert, ``{\it Gluon Production from Non-Abelian Weizs\"{a}cker-Williams Fields in Nucleus-Nucleus Collisions}," Phys.~Rev.~D {\bf 52}, 6231 (1995).

\bibitem{lap06} T.~Lappi and L.~McLerran, ``{\it Some Features of the Glasma}," Nucl.~Phys.~A {\bf 772}, 200 (2006).

\bibitem{gle83} N.~K.~Glendenning, and T.~Matsui, ``{\it Creation of $q\bar{q}$ pairs in a chromoelectric flux tube},'' Phys.~Rev.~D {\bf 28}, 2890 (1983).

\bibitem{kaj85} K.~Kajantie and T.~Matsui, ``{\it Decay of strong color electric field and thermalization in ultra-relativistic nucleus-nucleus collisions},'' Phys.~Lett. {\bf 164B}, 373 (1985).

\bibitem{gat87} G.~Gatoff, A.~K.~Kerman, and T.~Matsui, ``{\it Flux-tube model for ultrarelativistic heavy-ion collisions: Electrohydrodynamics of a quark-gluon plasma},'' Phys.~Rev.~D {\bf 36}, 114 (1987).

\bibitem{gra15}  I.~S.~Gradshteyn, and I.~M.~Ryzhik,  ``{\it Table of Integrals, Series, and Products},'' 8th edition, Academic Press (2015).


\bibitem{bec2018} F. Becattini and Iu. Karpenko, ``{\it Collective Longitudinal Polarization in Relativistic Heavy-Ion Collisions at Very High Energy},'' Phys.~Rev. Lett. {\bf 120}, 012302 (2018).
    
\bibitem{nii2019} T. Niida (for STAR Collaboration), ``{\it Global and local polarization of $\Lambda$ hyperons in Au+Au collisions at 200 GeV from STAR},'' Nucl.~Phys. A {\bf 982}, 511 (2019).
    
\bibitem{deng2015} W. -T. Deng and X. -G. Huang, ``{\it Electric fields and chiral magnetic effect in Cu+Au collisions},'' Phys.~Lett. B {\bf 742}, 296 (2015).
    
\bibitem{hiro2014} Y.~Hirono, M.~Hongo, and T. Hirano, ``{\it Estimation of the electric conductivity of the quark gluon plasma via asymmetric heavy-ion collisions},'' Phys.~Rev. C {\bf 90}, 021903(R) (2014).
    
    
\end{thebibliography}
\end{document}